\documentclass[12pt]{article}

\AtBeginDocument{%
   \setlength\abovedisplayskip{0.05in}
   \setlength\belowdisplayskip{0.08in}}

\usepackage[onehalfspacing]{setspace}

\makeatletter
\renewcommand\@footnotetext[1]{%
  \insert\footins{%
    \reset@font\footnotesize
    \interlinepenalty\interfootnotelinepenalty
    \splittopskip\footnotesep
    \splitmaxdepth \dp\strutbox \floatingpenalty \@MM
    \hsize\columnwidth \@parboxrestore
    {\setstretch{1.0}\protect\@makefntext{%
      \rule{\z@}{\footnotesep}\ignorespaces#1}}}}
\makeatother

\usepackage{geometry}
\usepackage{lineno}
\usepackage{booktabs}
\usepackage{amssymb}
\usepackage{amsmath}
\usepackage{amsthm}
\usepackage{caption}
\usepackage{epsfig}
\usepackage{graphicx}
\usepackage{graphics}
\usepackage{float}
\usepackage{subfigure}
\usepackage{multirow}
\usepackage{xcolor}
\usepackage{lineno}
\usepackage{fullpage}
\usepackage[normalem]{ulem} 
\usepackage{makeidx}
\usepackage{xspace}
\usepackage{wrapfig}
\usepackage{lscape}
\usepackage{mathrsfs}
\usepackage{adjustbox}
\usepackage{multirow}
\usepackage{varwidth}
\usepackage{rotating}
\usepackage{lscape}
\usepackage{tikz}
\usetikzlibrary{automata, positioning}
\usepackage[bottom]{footmisc}
\usepackage[colorlinks,citecolor=darkblue,urlcolor=darkblue,linkcolor=darkblue]{hyperref} 
\graphicspath{{C:/Users/flori/Desktop/folders/minimumWagePaper/paper/}}
\makeindex
\newtheorem{theorem}{Theorem}
\newtheorem{assumption}{Assumption}
\newtheorem{corollary}{Corollary}

\newtheorem{lemma}{Lemma}
\newtheorem{ex}{Example}

\newtheorem{proposition}{Proposition}
\newtheorem{definition}{Definition}

\newtheorem{claim}{Claim}
\newtheorem{obs}{Observation}

\definecolor{purple}{rgb}{0.6, 0.4, 0.8}
\definecolor{darkred}{rgb}{1, 0.1, 0.3}
\definecolor{darkblue}{rgb}{0.0, 0.0, 0.55}
\definecolor{darkgreen}{rgb}{0,0.6,0.5}
\definecolor{forestgreen}{rgb}{0.0, 0.46, 0.37}
\definecolor{bittersweet}{rgb}{1.0, 0.44, 0.37}
\definecolor{navy}{rgb}{0.0, 0.0, 0.55}
\definecolor{brown}{rgb}{0.53, 0.18, 0.09}
\definecolor{Green}{rgb}{0.0, 0.47, 0.44}

\newcommand {\mm}[1] {\ifmmode{#1}\else{\mbox{\(#1\)}}\fi}

\newcommand\E{\mathbb{E}}
\newcommand\Proba{\mathbb{P}}
\newcommand\R{\mathbb{R}}

\usepackage{dsfont}
\newcommand{\m}[1]{\ensuremath{\boldsymbol {#1}}}

\usepackage{sectsty}
\sectionfont{\centering\Large}
\subsectionfont{\large}
\usepackage{fancyhdr}
\usepackage{natbib}

\newtheorem{remark}{Remark}

\usepackage{pst-node, pst-plot, pst-poly}
\usepackage{rotating}

\usepackage{tikz}
\usepackage{pstricks}
\usepackage{tikz-cd}
\usepackage{siunitx}
\usepackage{pgfplots}
\usepgfplotslibrary{fillbetween}
\usepackage{algorithm}
\usepackage{algorithmic}
\usepackage{bbm}
\usepackage{csquotes}

\usepackage{scalerel}
\usetikzlibrary{shapes,arrows,trees,calc}
\pgfplotsset{compat=1.12}

\usetikzlibrary{trees}
\allowdisplaybreaks 
\begin{document}

\title{\vspace{-3.95em}
\mbox{\small\bf\MakeUppercase{Aggregate Efficiency in Games}}\footnote{I want to express my gratitude to Drew Fudenberg for his guidance and support throughout this project. I also want to thank Ian Ball, Abhijit Banerjee, Lukas Bolte, Glenn Ellison, Matthew Jackson, Philippe Jehiel, Stephen Morris, Ariel Rubinstein, Ran Shorrer, Ran Spiegler, Alexander Wolitzky, Jiabin Wu, Muhamet Yildiz, and seminar participants at MIT Theory Lunch for their helpful comments. 
}
\vspace{-0.5em}
}

\author{
\textsc{Florian Mudekereza}\footnote{Department of Economics, MIT, \href{mailto:florianm@mit.edu}{\texttt{\footnotesize florianm@mit.edu}}.}
}

\date{}
\maketitle
\thispagestyle{empty}
\setcounter{page}{0}
\vspace{-0.38in}
\begin{abstract}
We show that, in large population games, decentralized information aggregation \textit{generically} corrects for individual-level biases. This establishes a new testable \textit{aggregate efficiency benchmark} where the behavior of boundedly rational agents mimics that of fully rational agents. However, we find that structural economic forces such as strategic network formation and profit-maximizing platforms can systematically select \textit{pathological} environments to exploit individuals' biases, thereby causing aggregate inefficiencies. We characterize these inefficiencies in monopoly and labor markets. Our findings therefore suggest that policy should shift focus from correcting individuals' behavior to monitoring and regulating information structures.

\end{abstract}

{\footnotesize   \noindent\textit{Households may be irrational and yet markets quite rational. If this were generally recognized, critics might be more receptive to models implying rational market responses, and economic theorists to models permitting erratic and other irrational household responses.} \hfill-- \citet[][p. 8]{becker62}}

\vspace{-0.1in}
\section{Introduction}
Consider the director of the World Health Organization (WHO), tasked with designing vaccination policy for an infectious disease. Individuals choose whether to get vaccinated based on private costs and benefits, and the aggregate vaccination rate. To inform their decisions, individuals estimate the prevalence of vaccination in the population by acquiring signals from their environment, e.g., observing neighbors, news stories, or social media discussions. The literature on ``correlation neglect'' reports that individuals typically treat these correlated signals as independent \citep{enke19}. The director's main policy question is determining whether  individual-level biases can distort aggregate vaccination rates, or if decentralized aggregation can correct them. This paper provides guidance by establishing an \textit{aggregate efficiency benchmark} for games played by boundedly rational agents: just as perfect competition generates allocative efficiency, decentralized information aggregation can systematically correct for individual-level biases.
 
\par We study large population games where agents infer the distribution of play from samples of others' actions. Our main result, Theorem \ref{thm:generic}, shows that equilibria in which action-correlations distort aggregate behavior are exceptionally rare with respect to a large family of probability measures over the information structure. Formally, in the set of all possible correlation structures, those that sustain welfare-altering biases have ``measure zero'' in large samples. Hence, in ``generic'' environments---those with ``full measure''---aggregate behavior of boundedly rational agents is indistinguishable from that of fully rational agents.\footnote{Online Appendix \hyperref[app:market]{A.II} microfounds generic environments as arising from competitive markets.} This finding complements recent experimental studies reporting that, in large markets, individual-level biases tend to ``cancel out'' in aggregate \citep[e.g.,][]{crockett21}.\footnote{In general-equilibrium experiments, \citet[][pp. 833--834]{crockett21} report: ``our individual subjects reveal preferences that can easily produce pathological economies [...] The heterogeneity in revealed preferences seen in our study (and in all previous studies of which we are aware) tends to cancel out the pathologies as we aggregate more and more different traders into the economy.''} 
 Thus, our aggregate efficiency benchmark  contributes to the literature on correlation neglect which has focused largely on its negative effects  but has not characterized where it does or not matter and the underlying mechanisms \citep[][]{rabin05,ed09,pol15,levy15,levy22,corr22,cont23,info21}.
\par Our benchmark serves primarily to identify environments where aggregate efficiency fails. Just as market power and externalities cause competitive markets to fail, aggregate efficiency breaks down when economic forces systematically select nongeneric, pathological equilibria. That is, while Theorem \ref{thm:generic} shows that nongeneric equilibria are mathematically rare, we find that they arise precisely in games where structural forces---such as engagement-maximizing algorithms or homophilic network formations---strategically select them to exploit biases. Thus, by establishing the benchmark where biases cancel out, we isolate the key mechanisms that require policy intervention: intervention is required when some agents or institutions have \textit{strategic control} over the information structure. The main policy question can therefore shift from ``do agents have biases?'' to ``are there information frictions?'' To support this shift and operationalize the benchmark, we provide  a simple statistical test that can distinguish between generic and nongeneric environments empirically using only a finite sample of agents' actions (Proposition \ref{thm:test}).

\par Table \ref{fig:invisible} provides a parallel comparison between classical market efficiency and aggregate efficiency.
\begin{table}[hbt!]
\centering
\caption{Classical Market Efficiency and Aggregate Efficiency}\label{fig:invisible}
\resizebox{\textwidth}{!}{
\begin{tabular}{ll}
\hline\hline
\multicolumn{1}{c}{\textbf{Classical Market Efficiency}} & \multicolumn{1}{c}{\textbf{Aggregate Efficiency}}              \\ \midrule
Baseline: Perfect competition $\rightarrow$ efficiency   & Baseline: Generic correlations $\rightarrow$ no aggregate bias \\
Failure: Market power, externalities                     & Failure: Strategic networks, Platform algorithms                     \\
Policy tool: Antitrust, Pigouvian taxes                  & Policy tool: Network regulation, Platform regulation           \\
Policy focus: Market structure                           & Policy focus: Information structure                            \\ \hline\hline
\end{tabular}}
\end{table}
\noindent Formally, Theorem \ref{thm:generic} identifies a \textit{null hypothesis} for the effect of biases in large population games: without knowledge of the information structure, a social planner should expect aggregate behavior to resemble rational expectations. Then, aggregate inefficiencies act as evidence of frictions in the information structure. Thus, just as the \textit{First Welfare Theorem} identifies when to look for market failures, Theorem \ref{thm:generic} identifies when to look for ``information failures'' in games played by boundedly rational agents. Thus, this may help social planners identify where to direct attention and allocate resources. 
\par Returning to WHO's director: Theorem \ref{thm:generic} provides guidance depending on whether the environment is generic or nongeneric, and our empirical test can help distinguish them. (1) In generic environments---where individuals learn about vaccines from decentralized news sources---no intervention is needed because aggregate efficiency will cancel out individuals' biases and produce rational expectations' outcomes. 
However, (2) in nongeneric environments---where information spreads through viral anti-vaccine campaigns, and platform-amplified misinformation---intervention is required because such campaigns and platforms are strategically designed to exploit individuals' biases. In these settings, correlation neglect  causes \textit{overprecision} bias: individuals will perceive their estimates of the vaccination rate to be more precise than they actually are. This bias will then cause the vaccination rate to be too low compared to rational expectations' prediction because this is a game of strategic substitutes. Propositions \ref{thm:noconv}--\ref{thm:compNE} show that this gap in vaccination rates will not vanish even as individuals acquire more signals, so aggregate inefficiencies will persist. Notably, when signals are highly correlated, there is \textit{no} sorting from individuals' types to their actions, i.e., individuals' actions depend only on their sample and not their type. This lack of sorting indicates that misinformation campaigns can fully \textit{manipulate} individuals. To prevent this, the director should focus on regulating  information campaigns because correcting individuals' psychology may be more challenging.\footnote{\citet[][Section 3.4, p. 329]{enke19} highlight the challenges of correcting the behavior of individuals who suffer from correlation neglect because policy intervention could either focus on ``teaching math'' or on raising awareness of correlations, which they argue is ``context-specific.''} 
\par We will now summarize our baseline framework. It consists of a unit mass of statistical agents who face the same binary choice.\footnote{Online Appendix \hyperref[app:multi]{B.I} extends our main results to games with more than two actions.}
Agents' payoffs depend on the distribution of actions in the population, which captures strategic externalities. Each agent acquires a sample of other agents' actions to determine the value of each choice. Our key novelty is that we allow sampled actions to be arbitrarily correlated, which reflects sampling or selection biases. The joint distribution of these actions encodes an exponential amount of correlation parameters, so it is not realistic to know it completely. Agents are boundedly rational in two ways: (1) they have a \textit{subjective model}---their best guess of the true joint distribution---which may be \textit{misspecified}, and (2) they use the empirical frequency of actions in their samples as a summary statistic. Then, they estimate the distribution of actions using inference procedures such as Bayesian inference and maximum likelihood estimation. Lastly, they best respond to their estimates by choosing the optimal action.
\par We propose a tractable unified equilibrium concept called correlated sampling equilibrium with statistical inference (CoSESI) for learning from correlated data in large population games. CoSESI is an equilibrium distribution of actions with the key property that sampling correlated signals from it and using subjective models to arrive at an optimal action results in the \textit{same} distribution of actions. CoSESI depends on agents' subjective models, their sample sizes, their inference procedures, and the true joint distribution of actions. CoSESI nests many existing solution concepts: Nash equilibrium (NE) where agents know the true distribution ex ante, \citeauthor{stat}'s (\citeyear{stat}) SESI where agents do not know the true distribution ex ante but can acquire independent signals, and \citeauthor{jehiel}'s (\citeyear{jehiel}) analogy-based expectation equilibrium (ABEE) where agents rely on coarse partitions of the state space. CoSESI also captures a wide range of misspecifications such as correlation neglect,  the hot-hand fallacy, and the gambler's fallacy.

\par  As \citet[][p. 330]{enke19} remark: ``Markets are an additional obvious candidate to study correlation neglect.'' We therefore explore applications of CoSESI in monopoly and labor markets in Section \ref{sec:equiapp} to illustrate how structural economic forces can engineer nongeneric correlation structures to exploit biases. Section \ref{sec:mono} introduces correlation neglect in monopoly pricing, where consumers have preference for uniqueness. Key examples are markets for luxury goods. Consumers value the good less whenever they encounter others who consume the good. These encounters are allowed to be correlated to reflect the idea of viral trends or fads. We show that the monopolist's profit increases as these encounters become more \textit{clustered}, so there are gains due to consumers having correlation neglect. However, these  gains vanish as consumers acquire more data and their encounters follow generic patterns. Thus, the monopolist has an incentive to design a nongeneric market environment (or platform) where  consumers' encounters are so clustered that they become \textit{uninformative} for the market demand. In practice, such clustering patterns can be achieved by making the good ``exclusive'' via club memberships or loyalty programs, which is consistent with  the so-called ``rarity principle'' used by many luxury brands such as Rolex and De Beers \citep[][]{lux95}. 
\par Section \ref{sec:twosided} considers a two-sided labor market where workers and firms are engaging, respectively, in a costly search for jobs and employees, and are matched subject to  some matching friction. 
Workers (resp., firms) acquire signals to estimate the market thickness on the firm (worker) side. Suppose homophilic \textit{job-referral} networks cause signals to be positively correlated  \citep{matt04}. We show that, in this nongeneric environment, correlation neglect can worsen the equilibrium participation and employment relative to rational expectations' predictions. In fact, the gap in employment between CoSESI and rational expectations can be proportional to the matching friction. Thus, when signals originate from job referrals, correlation neglect negatively impacts participation and employment  by  amplifying the fluctuations of the matching friction.

\par\noindent \textit{--- Related literature:} The closest paper to ours is \citet{stat}, who extend \citeauthor{osborne2}'s (\citeyear{osborne2}) (action) sampling equilibrium by introducing a general class of statistical inference procedures in games. However, both papers assume agents acquire independent signals. \citet{dynamic23} extend \citet{stat} to evolutionary dynamics but maintain the independent-sampling assumption.  
In contrast, modeling correlated sampling procedures allows us to study how some information structures  affect the welfare of  na{\"i}ve agents in a wide class of strategic environments.  Correlation neglect is very concerning in many settings because it causes other pervasive biases such as overconfidence \citep[][]{pol15}, polarization \citep{levy15}, persuasion bias \citep{dem03},  the winner’s curse \citep{rabin05}, and failures of contingent thinking \citep{cont23}. Moreover, our paper is also related to the growing literature on misspecified Bayesian learning pioneered by \citet[][]{esponda2},\footnote{\citet[][p. 1119]{esponda2} note a key difference between their framework and the sampling equilibrium framework: ``In the sampling equilibrium [...] beliefs may be incorrect due to learning from a limited sample, rather than from misspecified learning.''  } who also study some forms of correlation neglect.

\par More generally, CoSESI is a member of the class of boundedly rational solution concepts called ``sampling equilibrium,'' where agents sample signals from the equilibrium distribution. \citet{osborne1,osborne2}, \citet{stat}, \citet{sethi21}, and \citet[][]{dynamic23} interpret this as reflecting a steady state of a dynamic process in which new agents randomly sample the actions of past generations. We extend this dynamic interpretation to correlated sampling procedures in Online Appendix \hyperref[app:learning]{A.III}. 

\par\noindent \textit{--- Organization:}  Section \ref{sec:model} introduces our framework followed by the equilibrium concept in Section \ref{sec:cosesi}, and an analysis of genericity and aggregate efficiency in Section \ref{sec:gen}. Section \ref{sec:simplecosesi} characterizes the negative effect of correlation neglect in nongeneric environments. Sections \ref{sec:equiapp} and \ref{sec:conc} present some applications and some extensions, respectively. 

\section{Framework}\label{sec:model}
\subsection{Setup and Preliminaries}\label{sec:setup}
Our baseline framework consists of a unit mass of agents, where each agent is deciding whether to take one of two actions $\{A,B\}$ (e.g., get vaccinated or not).  
The utility of action $B$ is normalized to zero whereas the utility from taking action $A$ is 
\begin{align}\label{eq:util}
    u_{\gamma}(\xi,\theta)=\xi+\gamma c(\theta).
\end{align}
Here, $\xi\in\Xi$ is an agent's idiosyncratic preference for taking action $A$, which is distributed according to $\upsilon(\xi)$, where $\upsilon$ is absolutely continuous with respect to the Lebesgue measure on $\Xi$---a measurable subset of $[-1,1]$. Moreover, $c(\theta)\in[0,1]$ quantifies the strategic externality that agents face, i.e., the social incentive to take action $A$ given that a fraction $\theta\in[0,1]$ of agents are taking this action, and the constant $\gamma\in\{-1,1\}$ captures whether this externality is positive or negative. The function $c$ is continuous on $[0,1]$. 
\par If, ex ante, all agents knew the aggregate state, $\theta\in[0,1]$, then the \textit{objective} fraction of agents taking action $A$ would be a Nash equilibrium (NE) of the game, denoted $\theta_{\text{\tiny NE}}\in[0,1]$. Let's illustrate this in games of strategic substitutes and strategic complements. 
\par \noindent \textit{--- Strategic substitutes}: When $\Xi=[0,1]$, $\upsilon$ is uniform, i.e., $\upsilon(\xi)=\mathcal{U}[0,1]$, $\gamma=-1$, and $c$ is increasing, we obtain a game of strategic substitutes, where $\xi$ and $c$ can be interpreted, respectively, as the idiosyncratic benefit and cost of taking action $A$. Here, there exists a unique NE as the solution to $1-\theta=c(\theta)$, i.e., agents with preference $\xi$ above (below) $ c(\theta_{\text{\tiny NE}})$ strictly prefer to take action $A$ ($B$). This special case coincides with the setting studied in \citet[][Section 2.1]{stat} and \citet[][Section 2.1]{dynamic23}.
\par \noindent \textit{--- Strategic complements}: When $\Xi=[-1,0]$, $\upsilon$ is uniform, $\gamma=1$, and $c$ is increasing, we obtain a game of strategic complements (e.g., positive network externalities), where  $\xi$ and $c$ can be interpreted, respectively, as the idiosyncratic cost and benefit of taking action $A$. Here, NE solves $\theta=c(\theta)$, and depending on $c$, there may be multiple NEs.  
\par In this paper, agents do not know $\theta$ ex ante. Thus, they have to estimate it using statistical inference given samples of others' actions. This process is described below. 

\subsection{Sampling Process}\label{sec:sampling}
 To determine whether to take action $A$ or $B$, each agent acts as a statistician and estimates the fraction $\theta\in[0,1]$ of agents taking action $A$ after observing a (private) sample of $n$ Bernoulli signals from their environment. Signals can originate from neighbors, social media, or can even be retrieved from memory. All agents understand that $\theta$ is the success probability in the population. A success is interpreted as observing an agent taking action $A$, and a failure as observing action $B$.  Let $\Delta_n$ and $\bar{\Delta}_n$ denote, respectively, the space of all joint distributions of arbitrary Bernoulli random variables, and the subset containing all joint distributions of identically distributed Bernoulli random variables.
\par \noindent\textit{--- Objective Model}: Let $p(.|\theta)\in\bar{\Delta}_n$ denote the \textit{true} joint distribution of $n$ Bernoulli signals with success probability $\theta\in[0,1]$. We often write $p:=p(.|\theta)$, but it is understood that $p$ is always a function of $\theta$ and $n$. Notably, $p$ represents the \textit{information structure} in the game. The total number of correlation parameters that characterizes $p$ is $2^n-n-1$ \citep[see,][]{baha61}. Let $\vartheta_{n,p}\in\R^{2^n-n-1}$ denote the vector whose components are all the correlation parameters that characterize $p$. For example, when all $n$ signals are (mutually) independent, all the correlations parameters are zero, so $\vartheta_{n,p}$ becomes the zero vector. Otherwise, $\vartheta_{n,p}$ contains some nonzero entries capturing the correlation structure. The distribution of the sum of $n$ Bernoulli random variables induced by $p$ is denoted $\mathscr{L}_n(p)$.
\par\noindent \textit{--- Subjective Model}: The main novelty of this paper is that we do not assume agents know the true joint distribution $p$, i.e., they do not know $\vartheta_{n,p}$. Instead, agents are endowed with a \textit{subjective} model, denoted $q\in\bar{\Delta}_n$,\footnote{It is straightforward to allow agents to have different subjective models (see, Online Appendix \hyperref[app:hetero]{B.IV}).} or equivalently, a vector of subjective correlation parameters $\vartheta_{n,q}$ that the agents ``perceive'' ex ante. Notice that there are two types of subjective correlation parameters: (1) $N_1$ correlation parameters, $\vartheta_{1,n,q}$, that agents believe they know ex ante and therefore will not estimate, and (2) $N_2$ correlation parameters, $\vartheta_{2,n,q}$, that agents believe they do not know ex ante and therefore will estimate them using signals. Thus, $\vartheta_{1,n,q}$ and $\vartheta_{2,n,q}$ partition the vector $\vartheta_{n,q}$, so $N_1+N_2=2^{n}-n-1$. If agents know all $2^n-n-1$ correlation parameters, i.e., $\vartheta_{1,n,q}=\vartheta_{n,p}$ and $\vartheta_{2,n,q}=\varnothing$, then they are \textit{correctly specified}, so $q=p$. To ease notation, let $\vartheta_1:=\vartheta_{1,n,q}$ and $\vartheta_2:=\vartheta_{2,n,q}$.
\par\noindent \textit{--- Relating Objective and Subjective Models}:  We maintain the assumption that the true distribution $p$ is absolutely continuous with respect to the subjective model $q$. This is a minimal necessary condition which appears in various forms in the literature on misspecified learning  \citep[e.g.,][Assumption (iii)]{esponda2}. It rules out settings where agents could observe signal realizations they deemed impossible. Without this assumption, most statistical methods (e.g., Bayes rule) may not be well-defined. 

\subsection{Statistical Inference}\label{sec:equidgp} 
After acquiring $n$ Bernoulli signals, each agent wishes to estimate $\theta\in[0,1]$. This is a very challenging task, however, because observed signals are arbitrarily correlated, and agents may not know the true joint distribution $p$. Thus, agents have to jointly estimate  $(\theta,\vartheta_2)$---the success probability $\theta$, and the vector of $N_2$ correlation parameters that they deemed unknown ex ante $\vartheta_2$. For example, if $N_2=2^n-n-1$, then given $n$ signals, an agent has to estimate a total of $2^n-n$ unknown parameters.
Since correlation parameters are not payoff relevant, they are what statisticians call ``nuisance'' parameters.\footnote{As \citet[][p. 3079]{schmid05} notes: ``Nuisance parameters are parameters for which inference is not desired, but which need to be taken account of in model estimation.''}
\subsubsection{Inference Procedures}
\par Given an arbitrary sample of Bernoulli signals $(x_i)_{i=1}^n$ satisfying $x_i\sim\text{Bern}(\theta)$ for all $i$, define $y:=\sum_{i=1}^nx_i$, sample mean $z:=y/n$, and let the pair $(n,z)$ denote a sample. 
Below is the statistical inference method agents use to jointly estimate $(\theta,\vartheta_2)\in[0,1]\times\R^{N_2}$.
\begin{definition}[Inference Procedure]\label{def:infproc}\normalfont
 \begin{enumerate}
  \hfill \item Given a subjective model $q\in\bar{\Delta}_n$, a joint \textit{inference procedure}, $\pi^q = \{\pi^q_{n,y}\}$, assigns to every pair $(n,z)$ a cumulative joint distribution function $\pi^q_{n,z}(.,.|\vartheta_1)$ on $[0,1]\times \R^{N_2}$ conditional on $\vartheta_1\in\R^{N_1}$, called an \textit{estimate}.
  \vspace{-0.1in}
    \item The marginal inference procedure on $\theta$ is $F^{q}=\{F^q_{n,z}\}$, where the estimate is the cumulative marginal distribution function $F^q_{n,z}(.|\vartheta_1):=\int_{\R^{N_2}}d\pi^q_{n,z}(.,\vartheta_2|\vartheta_1)$ on $[0,1]$.
\end{enumerate} 
\end{definition}
\par  An inference procedure is the analogue of an estimator in  the statistics literature. 
It captures most methods used by frequentists, Bayesians, and even those used by inexperienced statisticians. Standard examples include the maximum likelihood estimation, Bayesian inference, whereas a nonstandard 
example is ``overweighting low probabilities'' \citep[][Example 2]{dynamic23} inspired by the biases in \citet{tversky92}.
\par Thus far, inference procedures are arbitrary estimators of the pair $(\theta,\vartheta_2)$. Below is a key special case that imposes a monotonic structure on the marginal over $\theta$. 
\begin{definition}[Monotone Inference Procedure]\label{def:ninf}\normalfont
Given a subjective model $q\in\bar{\Delta}_n$, a \textit{monotone inference procedure} $G^q = \big\{G^q_{n,z}:=F^{q}_{n,z}\big\}$ is an inference procedure on $\theta$ such that estimate $G^q_{n,\widetilde{z}}$ strictly
first-order stochastically dominates  estimate $G^q_{n,z}$ when $\widetilde{z}>z$. 
\end{definition}
For notation, let $p_0:=\theta^{nz}(1-\theta)^{n(1-z)}$, i.e., the joint distribution of iid Bernoulli variables. When $q=p_0$, we write $G:=G^{p_0}$ as the monotone inference procedure when agents believe signals are independent. In this sense, a monotone inference procedure $G$ captures the idea that, for an agent who believes signals are independent, observing a higher mean in sample $(n,\widetilde{z})$ than in sample $(n,z)$ implies that the value of $\theta$ that generated these samples must be higher. \citet{stat} and \citet[][]{dynamic23} use this concept in their setting where signals are independent. We can now define what it means for an agent to suffer from complete \textit{correlation neglect} in our framework.
\begin{definition}[Correlation Neglect]\label{def:corrneglect}
    \normalfont An agent is said to suffer from complete \textit{correlation neglect} if $q=p_0$ and $F^q=G$, but $p\neq p_0$.
\end{definition}
An agent suffers from complete correlation neglect if their subjective model $q$ is misspecified in the following sense: $q$ treats signals as independent ($q=p_0$), whereas, in reality, the true joint distribution $p$ encodes some correlation structure ($p\neq p_0$). Correlation neglect is often associated with complexity \citep[][]{enke19}.

\subsubsection{Discussion: Inference Procedures}
Notice that the inference procedure depends only on the sample mean $z$ and not on the vector $(x_i)_{i=1}^n$. This is a key feature of sampling equilibrium models \citep{osborne1,osborne2}---agents use the empirical frequency of actions as a summary statistic. This modeling choice is consistent with the theoretical and empirical literature:
\begin{itemize}
    \item Normatively, \citet[][p. 79]{jehiel22} argues that: ``When there are many players such as in trading environments or in complex network environments, it is somehow natural for players to summarize the state of the economy by some moments of the distribution of plays.''
    \item Positively, \citet[][p. 1363]{enke20} reports: ``A substantial fraction of experimental participants follows a simple `what you see is all there is' heuristic, according to which participants exclusively consider information that is right in front of them, and directly use the sample mean to estimate the population mean.''
\end{itemize}
  Section \ref{sec:mle} highlights that inference with Bernoulli data can be very challenging even under simple correlation structures. This is consistent with \citet[][p. 109]{samp18}: ``Anecdotal and experimental evidence suggests that humans are prone to systematic errors when estimating the probability of a Bernoulli distribution by means of finite samples.''

\subsection{Examples of Inference Procedures}\label{sec:ex}
We will illustrate two examples of inference procedures. To this end, suppose the true joint distribution $p\in\bar{\Delta}_n$ is such that the signals $(x_i)_{i=1}^n$ that have a common pairwise correlation parameter $\rho\in[0,1]$, i.e., $\text{corr}(x_i,x_j)=\rho$ for all $i\neq j$. This joint distribution will be useful because it has the following  closed-form \citep[][eq. (10)]{stra20}:\footnote{Here, $\m{1}_{s}$ (resp. $\m{0}_s$) denotes the $s$-dimensional vector of all ones (resp. zeros), and $(\m{1}_s,\m{0}_{r})$ is an $s+r$-dimensional vector of $s$ ones and $r$ zeros in any ordering.}
\begin{align}\label{eq:prho}
 p_{\rho}\big((x_i)_{i=1}^n\big|\theta\big):=\begin{cases}
     \theta\rho+\theta^n(1-\rho) &\text{ if }(x_i)_{i=1}^n=\m{1}_n,\\
        (1-\theta)\rho+(1-\theta)^n(1-\rho) &\text{ if }(x_i)_{i=1}^n=\m{0}_n,\\    
        \theta^k(1-\theta)^{n-k}(1-\rho) &\text{ if }(x_i)_{i=1}^n=(\m{1}_k,\m{0}_{n-k}), 0<k<n.\\
    \end{cases}
\end{align}
To ease notation, let $p_{\rho}:=p_{\rho}((x_i)_{i=1}^n|\theta)$ and notice also that the induced distribution of the sum $y=\sum_{i=1}^nx_i$ has the following closed-form \citep[][Theorem 1]{latent10}: 
\begin{align}\label{eq:distpositive}
    \mu^{\rho}_n(y|\theta):=\mathscr{L}_n(p_{\rho})=(1-\rho)\underbrace{\binom{n}{y}\theta^y(1-\theta)^{n-y}}_{\mu^{0}_n(y|\theta)}+\rho\underbrace{\Big( \theta\mathds{1}_{y=n}+(1-\theta)\mathds{1}_{y=0}\Big)\vphantom{\binom{n}{y}}}_{\mu^{1}_n(y|\theta)}.
\end{align}
This is the distribution of the sum of identically distributed Bernoulli random variables with common pairwise correlation. It is known as the ``correlated binomial distribution'' in statistics \citep[][Section 2.2]{luce95}.\footnote{In economics, \citet{stra20} use $\mu^{\rho}_n$ to study strategic transmission of correlated
information.} 
When $\rho=0$, signals are independent, so eq. (\ref{eq:distpositive}) is the binomial distribution, $\mu^{0}_n(.|\theta)$ (or $\mathscr{B}_n(\theta))$. When $\rho>0$, there is  \textit{sampling bias}: signals are positively correlated, which captures a degree of informational redundancy or double counting in the sampling process. In the context of memory retrieval, this sampling process can capture \citeauthor{kan73}'s (\citeyear{kan73}) \textit{availability heuristic}, where $\rho$ would quantify how easily certain memories come to agents' minds. Online Appendix \hyperref[app:network]{A.I} provides a network microfoundation for $\mu^{\rho}_n$ in eq. (\ref{eq:distpositive}) by showing that it is a reduced-form representation of a standard strategic network formation model.

\subsubsection{Bayesian Inference}\label{sec:ex1}
The first inference procedure that we consider is Bayesian inference, so agents are Bayesians. Here, the subjective model $q$ is the ``perceived'' likelihood function. Suppose the true joint distribution is $p_{\rho}$, so the vector of true correlation parameters is $\vartheta_{n,p_{\rho}}=\rho\in[0,1]$. Consider a common prior on $\theta$, $\pi(\theta)=\text{Beta}(\alpha,\beta)$, with known constants $\alpha,\beta>0$, and for each $\rho\in[0,1]$, define $\eta_z(\rho):=
\mathds{1}_{0<z<1}
+
\mathds{1}_{z\in\{0,1\}}\hspace{0.02in}
\frac{(1-\rho)\hspace{0.02in}B(\alpha+nz,\beta+n(1-z))}
     {(1-\rho)\hspace{0.02in}B(\alpha+nz,\beta+n(1-z))
       +\rho\hspace{0.02in}B(\alpha+z,\beta+1-z)}$, where $B(a,b):=\int_0^1 r^{a-1}(1-r)^{b-1}\hspace{0.02in}dr$ denotes the Beta function. There are three relevant cases:
\begin{itemize}
    \item[(1)] Agents know the correct value of $\rho$, so they do not estimate it ($\vartheta_1=\rho,\vartheta_2=\varnothing$).
    \item[(2)] Agents \textit{believe} they know that $\rho=0$, so they do not estimate it ($\vartheta_1=0,\vartheta_2=\varnothing$).   
    \item[(3)] Agents do not know the value of $\rho$, so they will estimate it as $\hat{\rho}$ ($\vartheta_1=\varnothing,\vartheta_2=\hat{\rho}$).
\end{itemize}
\par \textit{--- Case (1)}: Here, since agents know the value of $\rho$, they are correctly specified, so $q=p_{\rho}$ and $(\vartheta_1=\rho, \vartheta_2=\varnothing)$. Thus, the posterior on $\theta$, $F^{q}_{n,z}(.|\rho)$, is conditional on $\vartheta_1=\rho$.

\begin{obs}\label{thm:beta2}
Given the subjective model $q= p_{\rho}$ and marginal prior $\pi(\theta)=\text{\normalfont Beta}(\alpha,\beta)$,
\[
F^{q}_{n,z}(\theta|\rho)
=
\eta_z(\rho)\hspace{0.02in}\text{\normalfont Beta}(\alpha+nz,\beta+n(1-z))
+
(1-\eta_z(\rho))\hspace{0.02in}\text{\normalfont Beta}(\alpha+z,\beta+1-z).
\]
\end{obs}

\par \textit{--- Case (2)}: Here, agents assume signals are independent, so $q=p_0$ and $(\vartheta_1=0, \vartheta_2=\varnothing)$. The posterior on $\theta$ is   $F^{q}_{n,z}(\theta|\rho=0)=\text{Beta}\big(\alpha+nz,\beta+n(1-z)\big)$ (Observation \ref{thm:beta2}). Notice also that this is a monotone inference procedure, so $G_{n,z}(.):=F^{q}_{n,z}(.|\rho=0)$. 
\par \textit{--- Case (3)}: Here, the subjective model is $q=p_{\hat{\rho}}$, so $(\vartheta_1=\varnothing,\vartheta_2=\hat{\rho})$, and hence agents will jointly estimate $(\theta,\hat{\rho})$. Let the joint prior on $(\theta,\hat{\rho})\in[0,1]\times[0,1]$ be $\pi(\theta,\hat{\rho})\propto \theta^{\alpha-1}(1-\theta)^{\beta-1}$, i.e., $\pi(\hat{\rho})=\mathcal{U}[0,1]$ is the marginal prior on $\hat{\rho}$ independent of $\theta$. 
\begin{obs}\label{thm:beta}
Given a subjective model $q=p_{\hat{\rho}}$ and joint prior
$\pi(\theta,\hat{\rho})\propto \theta^{\alpha-1}(1-\theta)^{\beta-1}$,
\[
F^{q}_{n,z}(\theta)
=
\eta_z(1/2)\hspace{0.02in}\text{\normalfont Beta}(\alpha+nz,\beta+n(1-z))
+
(1-\eta_z(1/2))\hspace{0.02in}\text{\normalfont Beta}(\alpha+z,\beta+1-z).
\]
\end{obs}

Thus, a Bayesian agent with a uniform prior on $\rho\in[0,1]$ will estimate it to be the midpoint $\hat{\rho}=1/2$. Below, we compare the posteriors $G_{n,z}$ (Case (2)) and $F^{p_{\rho}}_{n,z}$ (Case (1)) to describe the effect of correlations on Bayesian beliefs. Let $\alpha=\beta$, so given $n$ signals,
\begin{itemize}
       \item Case (2): The posterior mean under $G_{n,z}$ is $\big(\frac{2\alpha}{2\alpha+n}\big)\frac{1}{2}+\big(\frac{n}{2\alpha+n}\big)z$, so Bayesian agents who have correlation neglect na{\"i}vely give equal weight $1/n$ to every signal. Thus, as $n\rightarrow\infty$, these agents' posteriors converge to the sample mean $z$. However, when $\rho>0$,  $z$ is not a consistent estimator of $\theta$ \citep[][p. 1642]{luce95}.
    \item Case (1): If $0<z<1$, then $F^{q}_{n,z}$ in Observation \ref{thm:beta2} coincides with $G_{n,z}$ in Case (2).  However, if $z\in\{0,1\}$, then the posterior mean under  $F^{q}_{n,z}$ in Observation \ref{thm:beta2} is $(1-w_n(\alpha,\rho))\frac{1}{2}+w_n(\alpha,\rho)z$, where $w_n(\alpha,\rho):=(1-\eta_z(\rho))\frac{1}{2\alpha+1}+\eta_z(\rho)\frac{n}{2\alpha+n}$. When $\rho>0$, Bayesians who know $\rho$ put less weight on the sample mean than those who na{\"i}vely assume $\rho=0$ because $\frac{1}{2\alpha+1}\leq w_n(\alpha,\rho)\leq \frac{n}{2\alpha+n}$ for all $n\geq1$ and $\rho\geq0$. Intuitively, since the sample contains less information due to correlations, agents should put less weight on the sample mean. In fact, for any $\rho>0$, the sample mean receives lower weight as $n$ grows, i.e., $w_n(\alpha,\rho)\rightarrow\frac{1}{2\alpha+1}$ as $n\rightarrow\infty$ for any $\rho>0$. 
\end{itemize}

\subsubsection{Maximum Likelihood Estimation}\label{sec:mle}
We now consider maximum likelihood estimation (MLE), so agents are frequentists. Here, the subjective 
model $q$ is also the perceived likelihood function. Unlike Bayesian inference, this section highlights that some popular inference methods perform poorly with correlated data. Let the true joint distribution be $p_{\rho}$. Since MLE is a point estimator, the inference procedure on $\theta$, $F^{p_{\rho}}_{n,z}$, is always a point mass on the sample mean $z$, denoted  $\mathds{1}_{\theta\geq z}$ for all $\rho\geq0$. This is a monotone inference procedure, so this MLE is $G_{n,z}(\theta)=\mathds{1}_{\theta\geq z}$. 
\par Thus, whether or not agents know $\rho$, if they estimate $\theta$ using MLE,  then their belief about $\theta$ will always be a point mass on $z$, so they will fail to account for $\rho$. This inference contrasts the Bayesian inference in Section \ref{sec:ex1}, where Bayesians accounted for the distributional effect of $\rho$. Focusing on the sample mean $z$ is problematic for MLE because  when $\rho>0$, $z$ is not a consistent estimator for $\theta$. Moreover, estimating $\rho$ using MLE is also problematic because,
\citet[][p. 1642]{luce95} make the following remark regarding $p_{\rho}$: ``The log likelihood function provided by this model yields an absolute maximum with non vanishing first order derivative with respect to $\rho$. This affects the large sample properties of the maximum likelihood estimator of $\rho$.'' 
 
\subsection{Decision-Making Process}
\par Let $C^{q}_{n,z}:=\int_0^1c(\theta)\hspace{0.03in}dF^{q}_{n,z}(\theta)$ denote the expected value of the social incentive of taking action $A$ given a sample $(n,z)$ and inference procedure $F^q$. Then, after acquiring $n$ signals and forming $C^{q}_{n,z}$ based on $F^{q}$, an agent of type $\xi$ takes action $A$ if $\xi\geq -\gamma C^{q}_{n,z}$.\footnote{We assume agents choose action $B$ when there is a tie. For each $\theta\in[0,1]$, the set of agents who are indifferent between actions has measure zero. Thus, our results do not depend on the tie-breaking rule.}
\par Given all the above, we define two types of game based on agents' subjective models.
\begin{definition}
 \normalfont   A \textit{subjective game}, $\langle u_{\gamma},\upsilon,p,F^{q}\rangle_n$, is a game of incomplete information where agents of type $\xi\sim\upsilon$ get utility $u_{\gamma}(\xi,\theta)=\xi+\gamma c(\theta)$ from action $A$, acquire $n$ signals from $p\in\bar{\Delta}_n$, and infer $\theta$ using a subjective model $q\in\bar{\Delta}_n$ and inference procedure $F^{q}$.
\end{definition}

An \textit{objective game}, $\langle u_{\gamma},\upsilon,p,F^{p}\rangle_n$, is a special case of a subjective game, where $q=p$, i.e., when agents know the true distribution $p$. When $q\neq p$, a subjective game can also be called a ``misspecified'' game because agents' subjective models are incorrectly specified.

\section{Equilibrium Concept}\label{sec:cosesi}

Every agent in a subjective game acquires correlated signals from a joint distribution of actions based on other agents' decision making. A correlated sampling equilibrium with statistical inference
(CoSESI) describes the resulting solution concept.
\begin{definition}[CoSESI]\label{def:CoSESI}\normalfont
A  \textit{correlated sampling equilibrium with statistical inference} is a number $\theta_{n,F^q}(p)\in[0,1]$ such that
a $\theta_{n,F^q}(p)$ fraction of agents take action $A$ when the following hold. (1) \textit{Sampling}: Each agent obtains $n$ Bernoulli signals from a joint distribution $p\in\bar{\Delta}_n$ with success probability $\theta_{n,F^q}(p)$; (2) \textit{Inference}: Given a subjective model $q\in\bar{\Delta}_n$, each agent forms an estimate according to the inference procedure $F^q$; and (3) \textit{Optimality}: Each agent best responds to their estimate by choosing an  action.
\end{definition}
In CoSESI, the joint distribution from which agents acquire correlated signals is representative of the distribution of agents' actions. This solution concept, summarized by a number $\theta_{n,F^q}(p)$, depends on all primitives of the game: the sample size $n$, the subjective model $q$, the inference procedure $F^q$, the true joint distribution $p$, and it also depends on the type distribution $\upsilon$, but we suppress that dependence to ease notation.

\subsection{Characterization of CoSESI}

For each sample size $n\geq 1$ and subjective game $\langle u_{\gamma},\upsilon,p,F^{q}\rangle_n$, fix $q$ and $F^{q}$ and consider the following best-response map.
For every possible sample mean $z\in\{0,1/n,\dots,1\}$, let
\[
\sigma(z\mid q,F^{q})
:= \upsilon\big(\{\xi\in\Xi : \xi \geq -\gamma C_{n,z}^q\}\big)
\]
denote the fraction of agents who take action $A$ conditional on observing $z$.
To ease notation, we write $\sigma(z)$ whenever $q$ and $F^{q}$ are clear from the context. Given the true joint distribution $p(.|\theta)\in\bar{\Delta}_n$ and success probability $\theta\in[0,1]$, let $\mathscr{L}_n(p)(y|\theta)$ denote the probability that the sum of the $n$ Bernoulli signals equals $y$ when signals are drawn from $p(.|\theta)$. The next result uses the functions $\sigma(.)$ and $\mathscr{L}_n(p)(.|\theta)$ to characterize CoSESI.\footnote{Online Appendix \hyperref[app:hetero]{B.IV} shows that it is straightforward to extend CoSESI to settings where agents have different subjective models $q$, different inference procedures $F^q$, and different sample sizes $n$.}

\begin{obs}[Characterization]\label{thm:cosesichar}
Fix $n\geq 1$ and a subjective game $\langle u_{\gamma},\upsilon,p,F^{q}\rangle_n$.
A number, $\theta_{n,F^{q}}(p)\in[0,1]$, is a CoSESI  if and only if it solves the equation
\begin{align*}
    \theta = \sum_{y=0}^n \mathscr{L}_n(p)(y|\theta)\hspace{0.02in} \sigma(y/n).
\end{align*}
\end{obs}

The right-hand-side of the equation above is the \textit{aggregate best-response function} in CoSESI. It will take a very simple form in games of strategic substitutes in Section \ref{sec:exist}. Proposition \ref{thm:coscont} shows that a sufficient condition for existence of CoSESI is the continuity of the true joint distribution $p(.|\theta)$ in the success probability $\theta$. Note that this is the most standard assumption used in statistics because all popular inference methods (e.g, MLE, Bayesian inference) require that the joint distribution of the data is differentiable in the parameter of interest \citep[e.g.,][Miscellanea 10.6.2, (A3)]{casella02}. 

\begin{proposition}\label{thm:coscont}
  A  CoSESI, $\theta_{n,F^q}(p)\in[0,1]$,  exists in every subjective game $\langle u_{\gamma},\upsilon,p,F^{q}\rangle_n$, where the mapping from the success probability to the joint distribution $p$ is continuous.
\end{proposition}
Let us illustrate an example where continuity of the mapping $\theta\mapsto p(.|\theta)$ is violated. Consider a population in which vaccination is socially stigmatized. If the true vaccination rate is below, say 10\%, people hide it, so the probability of encountering a person who reveals being vaccinated is near zero. However, if the vaccination rate hits a threshold, say 10.01\%, the stigma breaks, and suddenly everyone can now reveal if they are vaccinated. Notice that this creates an \textit{informational shock}: A marginal increase in the vaccination rate causes a large increase in the number of vaccinated people that are observable, which then causes beliefs to swing wildly. Thus, such shocks may prevent CoSESI from existing. 
\par A key special case of Proposition \ref{thm:coscont} arises when signals are independent because in this case $p=p_0:=\theta^{nz}(1-\theta)^{n(1-z)}$, which is continuous in $\theta$ for all $(n,z)$. Importantly, this case nests \citeauthor{stat}'s (\citeyear{stat}) SESI. Specifically, SESI, denoted $\theta_{n,G}(p_0)$, is  based on the objective game $\langle u_{\gamma},\mathcal{U}[0,1],p_0,G^{p_0}\rangle_n$, i.e., the special case of CoSESI where signals are independent ($p=p_0$), agents' subjective model is correctly specified ($q=p_0$), and the inference procedure is monotone ($F^{p_0}=G$). This is formalize below.
\begin{corollary}[SESI]\label{thm:sesi1}
   In Proposition \ref{thm:coscont}, let $q=p=p_{0}$, $F^q=G$, and $\upsilon=\mathcal{U}[0,1]$. Then, CoSESI coincides with SESI $\theta_{n,G}(p_0)$. If $\gamma=-1$ and $c$ is increasing, then SESI is unique.
\end{corollary}
SESI induces a binomial distribution as the equilibrium distribution of actions, $\mathscr{B}_n(\theta_{n,G}(p_0))$, where $\theta_{n,G}(p_0)\in[0,1]$ is the equilibrium probability of action $A$. Online Appendix \hyperref[app:ABEE]{B.V} shows how CoSESI nests a large-game version of \citeauthor{jehiel}'s (\citeyear{jehiel}) ABEE. 

\subsubsection{Examples}

Let us illustrate how correlations affect strategic behavior in CoSESI. Consider the following game of strategic substitutes. In eq. (\ref{eq:util}), set $u_{\gamma}(\xi,\theta)=\xi-\theta^4$, where $\gamma=-1$, $\upsilon=\mathcal{U}[0,1]$, and $c(\theta)=\theta^4$. As showed in Section \ref{sec:setup}, a unique NE exists and solves $1-\theta=\theta^4$, which is about $0.72$. Now, suppose agents have to infer $\theta$ from $n=2$ signals $(x_1,x_2)$ with  $z=\frac{1}{2}(x_1+x_2)$. Recall $p_{\rho}$ in eq. (\ref{eq:prho}), which is summarized in Table \ref{tab:joint}. 
\begin{table}[hbt!]
\centering
\caption{Joint Distribution, $p_{\rho}$, of $n=2$ Signals}
\begin{tabular}{ccccc}
\hline
\hline
        &  & $x_2=0$                               &  & $x_2=1$                       \\\hline
$x_1=0$ &  & $(1-\theta)\rho+(1-\theta)^2(1-\rho)$ &  & $\theta(1-\theta)(1-\rho)$    \\
$x_1=1$ &  & $\theta(1-\theta)(1-\rho)$            &  & $\theta\rho+\theta^2(1-\rho)$ \\ 
                       \hline\hline   
\end{tabular}\label{tab:joint}
\end{table}

\par\noindent--- \textit{Correlation Neglect}: Agents have correlation neglect, i.e., $q=p_0$ and $F^q=G$ but $p=p_{\rho}$ with $\rho>0$.  Let $G=\text{MLE}$, so $C^{q}_{n,z}=c(z)$. From Table \ref{tab:joint}, there are three cases:
\begin{itemize}
    \item With probability $(1-\theta)\rho+(1-\theta)^2(1-\rho)$, an agent observes two failures. She (na{\"i}vely) estimates that no one takes action $A$, and chooses $A$ if $\xi \geq c(0)$;
    \item With probability $2\theta(1-\theta)(1-\rho)$, an agent observes a success and a failure. She estimates that half of the population takes action $A$, and chooses $A$ if $\xi \geq c(1/2)$;
    \item With probability $\theta\rho+\theta^2(1-\rho)$, an agent observes two successes. She estimates that everybody takes action $A$, and chooses $A$ if $\xi\geq c(1)$.
\end{itemize}    Then, the fraction of agents taking action $A$ in equilibrium is a fixed point of the expected value of $1-c(z)$ with respect to the joint distribution $p_{\rho}$ in Table \ref{tab:joint}, i.e.,
\begin{align*}
    1-\theta=\big((1-\theta)\rho+(1-\theta)^2(1-\rho)\big)c(0)+2\theta(1-\theta)(1-\rho)c(1/2)+\big(\theta\rho+\theta^2(1-\rho)\big)c(1),
\end{align*}
which is a quadratic equation in $\theta$ and has a unique solution on $[0,1]$, for all $\rho\in[0,1]$. When $\rho=0$, the solution is $0.6$, which is the SESI. The difference between NE and SESI is due to \textit{sampling error}---the fact that the sample mean is noisy in small sample. Now, when $\rho>0$, the solution differs from SESI, e.g.,  if $\rho=1$, the solution is $0.5$. The difference between SESI and CoSESI is due to \textit{sampling bias}---the fact that signals are less \textit{informative} for $\theta$ due to positive correlations. Thus, the difference between NE and CoSESI is due to both sampling error and sampling bias. These two errors then cause agents to \textit{overreact} to their signals and therefore overestimate the cost of taking action $A$, so too few agents end up taking action $A$. We formalize these insights in Section \ref{sec:simplecosesi}. 
\par Although this paper focuses mainly on correlation neglect, let us illustrate how our framework captures the hot-hand fallacy \citep[e.g.,][]{rabin10}.\footnote{The gambler's fallacy is the opposite of the hot-hand fallacy in that agents think signals are negatively correlated when they are independent. The correlated binomial distribution in eq. (\ref{eq:distpositive}) requires nonnegative correlations $\rho\in[0,1]$, but this can  be easily extended to allow negative correlations $\rho\in[-1,1]$. The resulting distribution is known as the ``additive binomial model'' \citep[see,][eq. (3)]{corr16}. 
} It is worth noting that the interpretation here is that agents sample actions in a sequential fashion.

\par\noindent\textit{--- Hot-Hand Fallacy}: Suppose signals are objectively independent ($p=p_0)$, but agents believe that they are positively correlated according to the subjective model $q=p_{\rho}$, for a known $\rho>0$. In this case, agents associate every streak of observed actions to correlation, so they \textit{underreact} to signals that are equally informative. Suppose agents are Bayesians with prior $\pi(\theta)=\text{Beta}(\alpha,\beta)$, so the inference procedure on $\theta$ is the posterior $F^{q}_{n,z}(\theta|\rho)=\eta_z(\rho)\hspace{0.02in}\text{\normalfont Beta}(\alpha+nz,\beta+n(1-z))
+
(1-\eta_z(\rho))\hspace{0.02in}\text{\normalfont Beta}(\alpha+z,\beta+1-z)$ (Observation \ref{thm:beta2}). Set $\alpha=\beta=1$ and $n=2$. Then, for every $\rho\in[0,1]$, a unique CoSESI solves
\begin{align*}
    1-\theta&=(1-\theta)^2C^{p_{\rho}}_{2,0}+2(1-\theta)\theta C^{p_{\rho}}_{2,1/2}+\theta^2C^{p_{\rho}}_{2,1}\\
    &=(1-\theta)^2\hspace{0.02in}\frac{\rho+2/5}{7(\rho+2)}
    +
    2(1-\theta)\theta\frac{1}{7}
    +
    \theta^2\hspace{0.02in}\frac{\rho+6}{7(\rho+2)}.
\end{align*}
where $C^{p_{\rho}}_{2,z}=\int_0^1\theta^4dF^{q}_{2,z}(\theta|\rho)$, for $z\in\{0,1/2,1\}$. 
Recall that NE is $0.72$. If $\rho=0$, then CoSESI coincides with SESI, which is about $0.72$, so even with just two signals, Bayesian agents' behavior in SESI nearly matches NE. However, if $\rho=1/2$, then CoSESI is about $0.74$. If $\rho=1$, then CoSESI becomes roughly $0.75$. This ranking suggests  that when agents suffer from  hot-hand fallacy, they underreact to signals and therefore underestimate the cost of taking action $A$, so too many agents end up taking $A$. Notice that this mechanism is the opposite of the one induced by correlation neglect.

\subsection{Informativeness}

This section introduces what it means for a joint distribution of actions to be ``informative'' for the aggregate state $\theta$. This notion will play a central role in our analysis.

\par For any two probability mass functions $P$ and $Q$, the total variation (TV) distance between $P$ and $Q$, denoted $d_{TV}(P,Q)$, is defined as
\begin{align*}
    d_{TV}(P,Q):=\frac{1}{2}\sum_{\forall k}|P(k)-Q(k)|\in[0,1],
\end{align*}
 which is a popular measure of discrepancy between probability distributions.
\begin{definition}[Informativeness]\normalfont\label{def:infor}
For fixed $n$ and $\theta$, a joint distribution $p\in\Delta_n$ is
\begin{itemize}
    \item \textit{fully informative} for $\theta$ if $d_{TV}\big(\mathscr{L}_n(p),\mathscr{B}_n(\theta)\big)=0$;
    \item \textit{partially informative} for $\theta$ if $d_{TV}\big(\mathscr{L}_n(p),\mathscr{B}_n(\theta)\big)\in(0,1)$;
    \item \textit{fully uninformative} for $\theta$ if $d_{TV}\big(\mathscr{L}_n(p),\mathscr{B}_n(\theta)\big)=1$.
\end{itemize}
\end{definition}
This notion of \textit{informativeness} aims to capture the ``amount'' of information about $\theta$ that agents can learn based on signals $(x_i)_{i=1}^n\sim p$. The binomial distribution $\mathscr{B}_n(\theta)$ is the benchmark for informativeness because the sum $\sum_{i=1}^nx_i\sim\mathscr{B}_n(\theta)$ is a sufficient statistic for $\theta$. On one extreme, if $p$ is \textit{fully informative}, then agents can infer $\theta$ based on signals from $p$. On the other extreme, if $p$ is \textit{fully uninformative}, then agents cannot infer $\theta\in(0,1)$ based on signals from $p$.\footnote{A distribution is fully uninformative if its support and that of the binomial distribution are disjoint.} In fact, the next observation shows that $p_{\rho}$ in eq. (\ref{eq:prho}) is the ``worst'' joint distribution for inferring $\theta$ because it can be fully uninformative.
\begin{obs}\label{thm:nongen}
  As $n\rightarrow\infty$,  $p_{\rho}$ in eq. (\ref{eq:prho}) becomes fully uninformative for $\theta$ when $\rho=1$.
\end{obs}
Our notion of informativeness is useful in part because it characterizes SESI.
\begin{proposition}[Characterization of SESI]\label{thm:sesi2}
    Fix $n\geq1$, $q=p_0$, $F^q=G$, and $\upsilon=\mathcal{U}[0,1]$. Then, given any joint distribution $p\in\bar{\Delta}_n$ and  $(x_i)_{i=1}^n\sim p$, the following are equivalent:
    \begin{align*}
   (1) & \quad     \text{\normalfont CoSESI } \theta_{n,G}(p)\text{ \normalfont coincides with SESI } \theta_{n,G}(p_{0}).\\
(2) &\quad d_{TV}\big(\mathscr{L}_n(p),\mathscr{B}_n(\theta_{n,G}(p_{0}))\big)=0. \\
 (3) &\quad\sum_{1\leq i_1<i_2<\dots<i_j\leq n}\E_p\Bigg[\prod_{k=1}^j x_{i_k}\Bigg]=\binom{n}{j}\theta_{n,G}^j(p) \quad \forall j\in\{1,\dots,n\}.
\end{align*}

\end{proposition}
Proposition \ref{thm:sesi2} leverages informativeness to characterize SESI without independence.
 Proposition \ref{thm:sesi2} is useful  
because it highlights the fact that there exist  many joint distributions of nonindependent Bernoulli variables whose sum is binomial. This happens precisely when Proposition \ref{thm:sesi2}.(3) holds,\footnote{Proposition \ref{thm:sesi2}.(3) is particularly useful in practice because it can be tested in finite sample using (or building on) the simple procedures in \citet[][Remarks 2.3--2.4]{binom01}. } i.e., when the correlation structure is \textit{balanced} in such a way that summing the Bernoulli random variables cancels out their dependence.

\subsection{Discussion: CoSESI}
\label{sec:disc}
 CoSESI captures three procedural constraints on agents’ decision-making process.  1) \textit{Informational} constraint: agents obtain data on the behavior of a subset of their peers due to limited time or accessibility. 2) \textit{Cognitive} constraint: agents use only their data and statistical inference to estimate the distribution of actions. These constraints are also present in existing sampling equilibria. If either 1) or 2) is dropped, then agents would fully learn the correct equilibrium action profile, and hence CoSESI would coincide with NE.  Our key novelty is the third constraint 3), which is \textit{behavioral}: agents' subjective models may be misspecified---they may incorrectly perceive the correlation structure.

\section{Aggregate Efficiency}\label{sec:gen}
Recall that CoSESI, $\theta_{n,F^q}(p)$, depends on all the primitives of the subjective game $\langle u_{\gamma},\upsilon,p,F^{q}\rangle_n$. This poses a challenge in practice because a social planner needs to know $\langle u_{\gamma},\upsilon,p,F^{q}\rangle_n$ to analyze equilibrium outcomes such as social welfare. 
Thus, we take an approach that is agnostic about $\langle u_{\gamma},\upsilon,p,F^{q}\rangle_n$ by answering the question: \textit{absent specific knowledge of} $\langle u_{\gamma},\upsilon,p,F^{q}\rangle_n$, \textit{which CoSESIs will most likely be played}? 
\par As our baseline, we endow  $\Delta_n$ with the uniform probability measure, denoted $\lambda_n$,\footnote{Formally, since $\Delta_n$ is the simplex, the uniform probability measure on $\Delta_n$ is the normalized Lebesgue measure. Thus, when we say a subset of $\Delta_n$ has ``measure zero,'' we mean zero Lebesgue measure.} to quantify the likelihood of joint distributions of actions in $\Delta_n$.  The intuition is the standard game of incomplete information logic of \citet{harsanyi67}: Nature moves first and draws  uniformly at random  from $\Delta_n$ a joint distribution of actions $p$ (viewed as a ``state of nature''), then agents acquire their signals from $p$ and play the game in Section \ref{sec:cosesi}.\footnote{As \citet[][p. 159]{harsanyi67} suggests: ``The original game can be replaced by a game where nature first conducts a lottery in accordance with the basic probability distribution, and the outcome of this lottery will decide which particular subgame will be played.'' Uniformity therefore captures  the idea that Nature may not be a ``strategic agent'' in the sense that she may not favor any particular correlation structure ex ante, so by Laplace's principle of insufficient reason, she treats them all as equally probable.} Section \ref{sec:marketmicro} uses a competitive market to microfound this intuition. 
\par We use $\lambda_n$ to define what it means for a distribution of actions to be ``generic.'' 
\begin{definition}\label{def:genericity}\normalfont
    An arbitrary distribution  $\mathscr{L}^*_n\in\Delta(\{0,\dots,n\})$ is $\lambda_n$-\textit{generic} if, $\forall\epsilon>0$, 
    \begin{align*}
        \lambda_n\Big(\Big\{p'\in\Delta_n:d_{TV}\big(\mathscr{L}_n(p'),\mathscr{L}^*_n\big)\leq\epsilon\Big\}\Big)\longrightarrow1 \quad \text{as }n\rightarrow\infty.
    \end{align*}
\end{definition}
In words, a distribution of actions $\mathscr{L}^*_n$ is $\lambda_n$-generic (or generic) if Nature will almost surely select a joint distribution  $p'\in\Delta_n$ whose induced distribution of the sum $\mathscr{L}_n(p')$ is arbitrarily close to $\mathscr{L}^*_n$ in large sample. Conversely, a distribution of actions is \textit{nongeneric} if the set of joint distributions that induce it has measure zero in large sample. 

Given $n$, let $\mathcal{C}_n$ contain all equilibrium distributions that are induced by \textit{all} CoSESIs: 
\begin{align*}
    \mathcal{C}_n:=\Big\{\mathscr{L}_n(p)\in\Delta(\{0,\dots,n\}):\exists\langle u_{\gamma},\upsilon,p,F^{q}\rangle_n\text{ s.t. }\theta_{n,F^q}(p)\text{ is a CoSESI}\Big\}.
\end{align*}
Given $n$, let $\mathcal{S}_n$  contain all equilibrium distributions that are induced by all SESIs:
\begin{align*}
    \mathcal{S}_n:=\Big\{\mathscr{B}_n(\theta_{n,G}(p_0)):\exists\langle u_{\gamma},\upsilon,p,G\rangle_n\text{ s.t. }\theta_{n,G}(p_0)\text{ is a SESI}\Big\},
\end{align*}
so $\mathcal{S}_n\subset \mathcal{C}_n$ (SESIs are a subset of CoSESIs). Note that $\mathcal{C}_n$ and
$\mathcal{S}_n$ collect equilibrium distributions across \textit{all} possible subjective games, so they are properties of the equilibrium concept itself, not of any particular game. Given $n$ and any $\epsilon>0,$ let $\mathcal{S}^{\epsilon}_n$ denote the set of distributions of actions that are within an $\epsilon$-ball of those induced by all SESIs:
\begin{align*}
    \mathcal{S}^{\epsilon}_n:=\Big\{P\in\Delta(\{0,\dots,n\}):\exists \hspace{0.02in}Q\in \mathcal{S}_n \text{ s.t. }  d_{TV}(P,Q)\leq\epsilon\Big\}.
\end{align*}

\subsection{Main Result: Aggregate Efficiency Benchmark}
We now present our main result, which establishes the \textit{aggregate efficiency benchmark}. 

\begin{theorem}[Aggregate Efficiency]\label{thm:generic}\hfill
\begin{enumerate}
    \item For every $\epsilon>0$, $\lambda_n\big(\big\{p\in\Delta_n:\mathscr{L}_n(p)\in\mathcal{C}_n\backslash\mathcal{S}^{\epsilon}_n\big\}\big)\longrightarrow0$ as $n\rightarrow\infty$. 
    \item Suppose an equilibrium distribution $\mathscr{L}^*_n\in\mathcal{C}_n$ is $\lambda_n$-generic. Then, there exists a SESI, $\theta_{n,G}(p_0)\in[0,1]$, such that $d_{TV}\big(\mathscr{L}^*_n,\mathscr{B}_{n}(\theta_{n,G}(p_0))\big)\longrightarrow0$ as $n\rightarrow\infty$.
\end{enumerate}
\end{theorem}
\noindent It is perhaps easier to understand Theorem \ref{thm:generic} in terms of the questions that it answers: 
\begin{enumerate}
    \item ``Suppose Nature draws uniformly at random an arbitrary joint distribution of actions. What equilibrium distribution would typically be played?'' Theorem \ref{thm:generic}.1 answers this by showing that the equilibrium distribution of actions will almost surely be induced by \citeauthor{stat}'s (\citeyear{stat}) SESI. Thus, when agents have access to a lot of data, equilibria in which correlations create aggregate inefficiencies are exceptionally rare in the sense that they have measure zero in large sample. 
    \item ``Suppose the social planner observes an arbitrary equilibrium distribution that is $\lambda_n$-generic. What structure does this distribution have?''  
Theorem \ref{thm:generic}.2 answers this by showing that any such distribution must be induced by a SESI in large sample. Thus, SESIs are the \textit{unique} equilibrium distributions of actions that are $\lambda_n$-generic.
\end{enumerate}

 Notably, since we have shown that statistical inference with correlated signals is very challenging, Theorem \ref{thm:generic} is a positive benchmark because it suggests that if agents could acquire a lot of data in generic environments, they could get away with behaving \textit{as if} their signals were independent. This is because they would almost surely play a SESI in equilibrium, which is a solution concept where signals are objectively independent (Corollary \ref{thm:sesi1}).  
A sketch of Theorem \ref{thm:generic} and more intuitions are provided in Section \ref{sec:sketch}.

\subsubsection{Discussion}
Suppose a social planner suspects that agents suffer from behavioral biases and is concerned that these biases may lead to aggregate inefficiencies. She is therefore considering whether to intervene in the game to correct for these biases and restore efficiency.
\par Theorem \ref{thm:generic} provides guidance by establishing an aggregate efficiency benchmark: absent structural manipulation of the information structure, the social planner should not intervene to correct for agents' biases because these biases will almost surely cancel out in aggregate. Specifically, in this benchmark, the equilibrium that is most likely to be played is a SESI, where agents effectively behave as if they are correctly specified. This correction of  individual-level biases can also be interpreted as a ``statistical invisible hand'' because it ensures that aggregate behavior of boundedly rational agents mimics that of fully rational agents. Notably, \citet[][p. 834]{crockett21} find experimental evidence of a form of our aggregate efficiency benchmark in large markets by reporting that diverse individual-level biases do not prevent ``efficient aggregate equilibrium.'' In our framework, Theorem \ref{thm:generic} then implies  that if aggregate inefficiencies are observed in practice, they are likely due to the presence of structural economic forces---such as strategic networks and platform algorithms---that systematically select those pathological equilibria where aggregate efficiency fails. Thus, just as the First Welfare Theorem is a benchmark that helps identify market power, Theorem \ref{thm:generic} is a benchmark that may help identify ``information power'' in large population games. The next section shows that aggregate efficiency stems from a robust concentration-of-measure result (Lemma \ref{thm:lebesgue}). 

\subsubsection{Proof Sketch of Theorem \ref{thm:generic}, Robustness, and an old Conjecture}\label{sec:sketch}
To prove Theorem \ref{thm:generic}, we start by proving in Lemma \ref{thm:lebesgue} some general concentration inequalities regarding arbitrary sequences of numbers that are elements of the set
\begin{align}\label{eq:Theta}
    \Theta:=\Big\{(\theta_{n})_{n\geq1}:\big|\theta_{n} - 1/2\big|=o(1/\sqrt{n}),\theta_n\in[0,1]\Big\}.
\end{align}
This set contains sequences of numbers that are close to $1/2$ in large sample. Define also
\begin{align}\label{eq:delta}
    \Delta^{\epsilon}_{n}:=\Big\{ p \in \Delta_n : \underset{0\leq k\leq n}{\text{max}}\big|\mathscr{L}_n(p)(k)- \mathscr{B}_n(\theta_n)(k)\big| \le \epsilon\Big\},
\end{align}
which is the set of all joint distributions $p\in\Delta_n$ whose induced distribution of the sum $\mathscr{L}_n(p)$ is in an $\epsilon$-ball of the binomial distribution $\mathscr{B}_n(\theta_n)$, where $\theta_n\in[0,1]$. The next result is the main technical result that allows us to prove Theorem \ref{thm:generic}.
\begin{lemma}[Genericity]\label{thm:lebesgue}
Let $\theta_n\in[0,1]$ be an arbitrary number satisfying
$(\theta_n)_{n\geq1}\in \Theta$. Then, $\forall\epsilon>0$, there exists an integer $n_{\epsilon}\geq1$ such that, for all $n\geq n_{\epsilon}$, $$\lambda_{n}(\Delta^{\epsilon}_{n})\geq 1 - \frac{ 4}{\epsilon^2 (2^{n}+1)} \hspace{0.1in} \text{\normalfont and}\hspace{0.1in} \lambda_{n} \Big( \Big\{ p \in \Delta_n : d_{TV}\big(\mathscr{L}_n(p), \mathscr{B}_n(\theta_n)\big) \le \epsilon \Big\} \Big) \geq 1 - \frac{ 2(n+1)^2}{\epsilon^2 (2^{n}+1)}.$$
\end{lemma}
This result shows that the distribution of the sum of arbitrary Bernoulli variables \textit{concentrates} near the binomial distribution. Note that it does \textit{not} make any assumption regarding their correlation structure.  
Since SESIs induce binomial distributions, Lemma \ref{thm:lebesgue} establishes that any SESI, $\theta_{n,G}(p_0)$, satisfying $\big(\theta_{n,G}(p_0)\big)_{n\geq1}\in\Theta$ is $\lambda_n$-generic.  Lemma \ref{thm:sesigen} shows that such SESIs exist. Intuitively, extreme SESIs---those near the boundary $\{0,1\}$---are highly \textit{biased} toward one action, so they fall outside the set $\Theta$ and hence have measure zero. Thus, $\Theta$ captures the fact that, in the generic benchmark, no action should have an inherent structural advantage as $n\rightarrow\infty$. Observe also that if $\epsilon=0.01$, then the lower bounds in Lemma \ref{thm:lebesgue} are $>0.98$ even when $n$ is as small as $30$.

\begin{remark}[Robustness and a long-standing \textit{conjecture}]
    \normalfont Lemma \ref{thm:dir} (Section \ref{sec:robust}) shows that the conclusion of Lemma \ref{thm:lebesgue} is robust to the choice of the measure because it holds even when the uniform distribution $\lambda_n$ is replaced with any other symmetric Dirichlet distribution. Notably, Lemma \ref{thm:dir} confirms in the positive a conjecture made in \citet[][Remark 3.2]{binom01}:\footnote{\citet[][Theorem 2.1]{binom01} provide the first characterization of the binomial distribution. Specifically, it is determined uniquely by the condition in Proposition \ref{thm:sesi2}.(3).} 
    ``For a general $n$, the set $B_n(p)$ is determined by $(2^n - 1)$-dimensional vectors of probabilities and only $n$ coordinates have to satisfy $n$ conditions [...] and the remaining $2^n- n - 1$ coordinates could be arbitrary probabilities. Hence, for moderately large $n$, the distribution of $S_n$ is quite likely to follow or to be close to the binomial distribution.'' In this quote, $S_n$ denotes the sum of arbitrary Bernoulli random variables, $B_n(p)$ denotes the set of all joint distributions that are fully informative for the success probability, denoted $p\equiv\theta$ (Definition \ref{def:infor}). In summary, they conjectured that, in large sample, the distribution of the sum of arbitrary Bernoulli random variables is ``likely'' to concentrate on the binomial distribution. This is precisely the conclusion of  Lemma \ref{thm:dir}, where the family of Dirichlet distributions formalizes the notion of likelihood.\footnote{Online Appendix \hyperref[app:multi]{B.I} extends a version of Lemmas \ref{thm:lebesgue}--\ref{thm:dir} to a \textit{multinomial}-concentration result.} 
\end{remark}

\subsubsection{Microfoundation: Aggregate Efficiency Benchmark}\label{sec:marketmicro}
Is it reasonable to assume that Nature draws the true joint distribution of actions, $p$, uniformly at random? We provide a microfoundation for this intuition by modeling the selection of $p$ as the unique Walrasian equilibrium of a competitive market for \textit{exposure}.
\par Consider a ``meta-economy'' where coalitions---representing different information sources or signal structures---compete for the attention of agents in a population. Each coalition raises a budget from many decentralized sponsors (e.g., advertisers or donors) to purchase exposure share. We show in Online Appendix \hyperref[app:market]{A.II} that the unique Walrasian equilibrium of this market generates a distribution over exposure shares. Specifically, when budgets originate from independent sources, the equilibrium probability of realizing any specific joint distribution of actions $p$ follows a Dirichlet distribution.
\par The uniform distribution $\lambda_n$ constitutes the equilibrium outcome of a perfectly decentralized market for exposure. This microfounds our generic benchmark: just as perfect competition eliminates excess profits, decentralized funding washes out complex correlation structures. Thus, aggregate efficiency is the result of competitive arbitrage. Conversely, nongeneric equilibria arise only when this market is disrupted by strategic correlations in funding (e.g., coalitions form networked alliances). This is consistent with Online Appendix \hyperref[app:network]{A.I} which shows that strategic network formation induces some nongeneric distributions of actions such as the correlated binomial distribution in eq. (\ref{eq:distpositive}).

\subsection{Empirical Test of Genericity}\label{sec:test}
In practice, it would be useful if a social planner could determine whether an equilibrium is generic without knowing the correlation structure. This section shows that our aggregate efficiency benchmark is testable with finite data using simple hypothesis tests. It will matter whether the social planner is able to acquire multiple samples of agents' actions. 

\subsubsection{Case I: Single Sample}
Suppose the social planner has access to a single sample of $n$ agents' actions $(x_i)_{i=1}^n$. Recall the sets $\Theta$ and $\Delta^{\epsilon}_{n}$ in eqs. (\ref{eq:Theta}) and (\ref{eq:delta}), and let $\alpha\in(0,1)$. Let $\theta_n\in[0,1]$ be any number satisfying $(\theta_n)_{n\geq1}\in\Theta$ and fix any number $\epsilon>0$ such that $1-\frac{4}{\epsilon^2(2^n+1)}\leq 1-\alpha/2$. For example, if $\alpha=0.01$ and $\epsilon=0.05$, then the inequality holds for all $n\geq19$.
\par Suppose the social planner's null hypothesis is that the true joint distribution satisfies $p\in\Delta^{\epsilon}_{n}$, i.e., its induced distribution of the sum $\mathscr{L}_n(p)$ is in an $\epsilon$-ball of $\mathscr{B}_n(\theta_n)$, and therefore $\lambda_n(\Delta^{\epsilon}_{n})\geq1-\alpha/2$ by Lemma \ref{thm:lebesgue}. The goal of this hypothesis test is to test whether the realized value of the sum $y=\sum_{i=1}^nx_i$ is atypical compared to a random variable $\hat{y}\sim \mathscr{B}_n(\theta_n)$. Formally, a conservative test of level $\alpha$ can be performed, where the social planner would reject her null hypothesis if $y\notin[\underline{y},\overline{y}]$, where $\Proba\big(\hat{y}\notin [\underline{y},\overline{y}]\big)\leq\alpha/2$.

\subsubsection{Case II: Multiple Samples}
Suppose the social planner has access to $N>1$ iid (or at least mixing)\footnote{The proof of Proposition \ref{thm:test} extends directly to allow mixing, but this would deliver weaker bounds.} samples each of size $n$. For example, she might observe the actions of agents from different regions or across time.
For each $j=1,\dots,N$, the sum of the signals $(x_{ij})_{i=1}^n$ is $S_n^{(j)}:=\sum_{i=1}^nx_{ij}$. Then, for each $k=0,\dots,n$, define the empirical distribution $P_N(k):=\frac{1}{N}\sum_{j=1}^N\mathds{1}_{S_n^{(j)}=k}$. Given the multiplicity of samples, we propose a more reliable conservative level-$\alpha$ test.

\begin{proposition}\label{thm:test}
   Fix any $\epsilon>0$, $\theta_n\in[0,1]$ with $(\theta_n)_{n\geq1}\in\Theta$,  $\alpha\in[0,1]$, and consider  $$\text{\normalfont H}_0:\text{\normalfont the true joint distribution satisfies }p\in\Delta^{\epsilon}_{n}.$$ Given a realization $k\geq0$ of the sum $S_n^{(1)}$, suppose the null hypothesis $\text{\normalfont H}_0$ is rejected when  
    \begin{align*}
    \big|P_N(k)-\mathscr{B}_n(\theta_n)(k)\big|>\text{\normalfont max}\Bigg\{\sqrt{\frac{2\text{\normalfont log}(4/\alpha)}{N}},\sqrt{\frac{32}{\alpha(2^n+1)}}\Bigg\}.
\end{align*}
Then, for every $\epsilon>0$ and $\alpha\in(0,1)$, there exists $n_0(\epsilon,\alpha)\ge 1$ such that for all
$n\ge n_0(\epsilon,\alpha)$ the following holds: if $p$ is drawn from
$\lambda_n$ and, conditional on $p$, the sample
$S_n^{(1)},\dots,S_n^{(N)}$ is drawn iid according to $\mathscr{L}_n(p)$,
then the probability of rejecting $\text{\normalfont H}_0$ is at most $\alpha$.
\end{proposition}
Notice that even for moderate $n$ (e.g., $n\geq20$), the rejection bound is driven by the left term, which depends only on $N$. Thus, if WHO's director could observe vaccination data across many regions (periods) $N$, she could easily test whether the vaccination environment is generic or nongeneric. Then, Theorem \ref{thm:generic} recommends that if the test deems the environment generic, the director should not waste resources on an intervention.

\subsection{Discussion: Harmful Equilibria}
Recall that when agents know $\theta$ ex ante, the resulting equilibrium is NE. We then say that correlation neglect is \textit{harmful} if CoSESI does not converge to NE in large sample, i.e., agents fail to aggregate information. This section shows that the set of equilibria where correlation neglect is harmful is very small. We define a weaker notion of informativeness.
\begin{definition}[Asymptotic Informativeness]\normalfont\label{def:asinfo}
    A joint distribution $p\in\bar{\Delta}_n$ is \textit{asymptotically informative} for $\theta$ if, for $(x_i)_{i=1}^n\sim p$, $z=\sum_{i=1}^nx_i/n$ is a consistent estimator of $\theta$. 
\end{definition}

 \par A monotone inference procedure $G$ is said to be \textit{convergent} if it converges in probability to $\mathds{1}_{\theta\geq z}$ as $n\rightarrow\infty$. This says that agents' estimates become concentrated around the sample mean $z$ in large sample, which holds for many inference procedures such as MLE. The next result shows that, in games of strategic substitutes, CoSESI converges to the unique NE in large sample when $G$ is convergent and $p$ is asymptotically informative.
\begin{proposition}\label{thm:asinfo}
  Fix $\langle u_{-1},\mathcal{U}[0,1],p,G\rangle_n$, where $G$ is convergent, and $p\in\bar{\Delta}_n$ is asymptotically informative. Then, CoSESI, $\theta_{n,G}(p)$, converges to NE, $\normalfont\theta_{\text{\tiny NE}}$, as $n\rightarrow\infty$.
\end{proposition}
Since all generic equilibria are asymptotically informative, all harmful equilibria are necessarily nongeneric. For example, $p_{\rho}$ is a harmful distribution because it is not asymptotically informative when $\rho>0$ (Section \ref{sec:ex1}). Theorem \ref{thm:generic} showed that the set of nongeneric equilibria is very small. Proposition \ref{thm:asinfo} complements this by showing that the set of equilibria where correlation neglect is harmful is even smaller. It shows that there exist many nongeneric equilibria that are not harmful. Section \ref{sec:mono} shows that a subset of asymptotically informative distributions are those represented by \textit{ergodic} Markov processes. Below, we illustrate a simple example of a nongeneric joint distribution of actions that is asymptotically informative and can model memory-sampling processes.
\begin{ex}\normalfont
Let $p^{\kappa}\in\bar{\Delta}_n$ denote the joint distribution of the signals $(x_i)_{i=1}^n$ that satisfies
\begin{align*}
\Proba\big(x_{i+1}=1\big|\mathcal{F}_i\big)=(1-\kappa)\theta+\kappa z_i,
\end{align*}
where, for all $i\geq1$, $\Proba(x_{i}=1)=\theta$, $\kappa\in[0,1)$, $z_i:=\sum_{j=1}^ix_j/i$, and $\mathcal{F}_i$ is the sigma-field generated by $(x_j)_{j=1}^i$. When $\kappa>0$, observing someone taking action $A$ ($B$) is more likely in signal $i+1$ when the previous average $z_i$ is greater (less) than $\theta$. By \citet[][p. 54]{heyde04},  $z=\sum_{i=1}^nx_i/n$ is a consistent estimator of $\theta$ for all $\kappa$, so $p^{\kappa}$ is asymptotically informative for $\theta$. Here are two interpretations of $p^{\kappa}$. 
(1) \textit{Online echo chambers}: Consider settings where agents observe their peers' actions online. If an agent observes action $A$ in their first few search results, a search engine's algorithm may start prioritizing showing action $A$. (2) \textit{Associative memory}: Consider settings where agents estimate the prevalence of action $A$ by retrieving information from their own memory (e.g., past interactions).  
The idea is that agents may have been involved in similar interactions in the past and encoded in their memory the distribution of actions. However, the retrieval of one memory can prime access to similar memories. Such memory biases are called ``associative memory'' and are the focus of recent cognitive models \citep[][]{kos22,fuden24}.   
\end{ex}

 In summary, Theorem \ref{thm:generic} and Proposition \ref{thm:asinfo} show that the set of equilibria in which correlations are harmful is very small. However, as noted before, structural economic forces can systematically select equilibria from this set to exploit agents' biases. Section \ref{sec:simplecosesi} characterizes a class of equilibria that belongs to this set to demonstrate that policy intervention is necessary when aggregate efficiency fails to correct for agents' biases.  

\section{A Nongeneric Harmful Equilibrium}\label{sec:simplecosesi} 
  This section characterizes the effect of (complete) correlation neglect in nongeneric environments. We focus on games of strategic substitutes where $\gamma=-1$, $c$ is increasing, and $\upsilon=\mathcal{U}[0,1]$. Recalling Definition \ref{def:corrneglect}, correlation neglect means that $q=p_0$ and $F^q=G$, but $p\neq p_0$. Let $p=p_{\rho}$ in eq. (\ref{eq:prho}), so the distribution of the number of agents taking action $A$ in any sample $(x_i)_{i=1}^n$ is (eq. (\ref{eq:distpositive})): $\mu^{\rho}_n(y|\theta)=(1-\rho)\mu^{0}_n(y|\theta)+\rho\mu^{1}_n(y|\theta)$,
where $y=\sum_{i=1}^nx_i$, and common pairwise correlation $\text{corr}(x_i,x_j)=\rho\geq0$ for all $i\neq j$. This setting will permit very tractable comparisons between CoSESI, NE, and SESI.

\subsection{Definition of Simple CoSESI}\label{sec:def}

We introduce a tractable class of CoSESIs called \textit{simple} CoSESI, denoted $\theta^{(\rho)}_{n,G}$.  
\begin{definition}[Simple CoSESI]\label{def:sCoSESI}\normalfont
A \textit{simple} CoSESI, $\theta^{(\rho)}_{n,G}$, corresponds to the subjective game where $\gamma=-1$, $c$ is increasing, $\upsilon=\mathcal{U}[0,1]$, $p=p_{\rho}$, for $\rho\geq0$, $q=p_0$, and $F^{q}=G$. 
\end{definition}
 
\par In this notation, SESI is $\theta^{(0)}_{n,G}$. Notably, the distribution $\mu^{\rho}_n$ can be interpreted as an \textit{overdispersed} binomial distribution because it allows higher variance than the binomial distribution, i.e., $\text{var}_{\mu^{\rho}_n}(y|\theta)=\theta(1-\theta)\big(n+\rho n(n-1)\big)$, which grows quadratically in $n$. 

\begin{remark}[Asymptotic Overdispersion]\label{rem:overdis}
  \normalfont The overdispersion in simple CoSESI relative to SESI is \textit{persistent} because it does not vanish in large samples. To see this, the variance of the number of agents $y=\sum_{i=1}^nx_i$ taking action $A$ in sample $(x_i)_{i=1}^n$ of size $n$ under $\mu^{\rho}_n$ in eq. (\ref{eq:distpositive}) can be decomposed as the sum of the ``sampling error'' and ``sampling bias'':
  \begin{align*} 
      \text{var}_{\mu^{\rho}_n}(y|\theta)=\underbrace{\theta(1-\theta)n}_{\text{ sampling error}}+\quad\underbrace{\rho(n-1) \theta(1-\theta)n}_{\text{sampling bias }},
  \end{align*} so the limiting variance of the sample mean $z=y/n$ becomes $\text{var}_{\mu^{\rho}_{\infty}}(z|\theta)=\rho\theta(1-\theta)>0$, for all $\theta\in(0,1)$ and $\rho>0$. In contrast, the variance of $y$ when signals are independent is the sampling error $\theta(1-\theta)n$, which implies that the limiting variance of the sample mean $\theta(1-\theta)/n$ under SESI is mechanically normalized to zero, and hence no sampling bias. In this sense, simple CoSESI can be interpreted as an \textit{overdispersed} SESI. \hfill  
\end{remark} 
  Simple CoSESI predicts that,  in large population games with strategic substitutes, correlation neglect causes persistent \textit{overprecision} bias---agents perceive their signals to be more informative than they actually are. This prediction is consistent with economics papers which find that correlation neglect causes overprecision in different settings \citep[][]{pol15}.\footnote{\citet[][Sections III.C.1--III.C.2]{school23} survey experimental evidence of some forms of correlation neglect and overprecision in matching markets.} It is also consistent with psychology papers which classify overprecision bias as the \textit{most} persistent form of overconfidence \citep[][]{conf08}.

\subsection{Existence and Uniqueness of Simple CoSESI}\label{sec:exist}

\par Proposition \ref{thm:cosesi} establishes the existence and uniqueness of simple CoSESI $\theta^{(\rho)}_{n,G}$.
\begin{proposition}\label{thm:cosesi}
  There exists a unique simple CoSESI, $\theta^{(\rho)}_{n,G}\in[0,1]$, for any $n$, $G$, and $\rho$.
\end{proposition} 
Online Appendix \hyperref[app:learning]{A.III} provides a dynamic foundation for simple CoSESI, where $\rho$ changes over time and converges to the long-run distribution of agents' sampling behavior. 

\begin{remark}[Proof Sketch]\label{rem:equicond}\normalfont Fix any $n$, $G$, and $\rho\geq0$. Proposition \ref{thm:cosesi} is proved by showing that simple CoSESI $\theta^{(\rho)}_{n,G}$ is the unique value of $\theta$ that satisfies the equation 
\begin{align}\label{eq:equi}
1-\theta=(1-\rho)\mathcal{B}_n(\theta;C_{n})+\rho\mathbb{B}_n(\theta;C_{n}),
\end{align}
 where $\mathbb{B}_n(\theta;C_{n}):=\theta C_{n,1}+(1-\theta)C_{n,0}$ and  $C_{n,z}=\int_0^1c(\theta)dG_{n,z}(\theta)$. When $\rho=0$, CoSESI coincides with SESI, $\theta^{(0)}_{n,G}$, which solves $1-\theta=\mathcal{B}_n(\theta;C_{n}):=\sum_{y=0}^n\binom{n}{y}\theta^y(1-\theta)^{n-y}C_{n,y/n}$. The function $\mathcal{B}_n(.;C_{n})\in[0,1]$ is known as the $n^{\text{th}}$-order \textit{Bernstein polynomial} of $C_n$ \citep[see,][eq. (7.1)]{phil}. The convex combination on the right-hand side of eq. (\ref{eq:equi}), denoted $\varPsi_n(\theta,\rho;C_n):=(1-\rho)\mathcal{B}_n(\theta;C_{n})+\rho\mathbb{B}_n(\theta;C_{n})$, can be interpreted as the $n^{\text{th}}$-order $\rho$-\textit{weighted} Bernstein polynomial of $C_n$. In terms of Observation \ref{thm:cosesichar}, $\varPsi_n(\theta,\rho;C_n)=1-\sum_{y=0}^n \mathscr{L}_n(p_{\rho})(y|\theta)\hspace{0.01in} \sigma(y/n)$, where $\mathscr{L}_n(p_{\rho})(y|\theta)=\mu^{\rho}_n(y|\theta)$, and $\sigma(y/n)=C_{n,y/n}$ $\forall y\geq0$. 
 
\end{remark}

In SESI, the Bernstein polynomial $\mathcal{B}_n(\theta;C_{n})$ plays a central role because it captures the total measure of agents who choose action $B$. However, this is no longer the case  in the presence of correlations. In simple CoSESI, $\mathcal{B}_n(\theta;C_{n})$ is replaced by the $\rho$-weighted Bernstein polynomial $\varPsi_n(\theta,\rho;C_n)$ in eq. (\ref{eq:equi}), which captures the fact that correlation neglect can lead society to occasionally become \textit{polarized} with probability $\rho\in[0,1]$. At the extreme when $\rho=1$, the total measure of agents who choose action $B$ becomes $\mathbb{B}_n(\theta;C_n)=\theta C_{n,1}+(1-\theta)C_{n,0}$ in eq. (\ref{eq:equi}) because a $\theta$ fraction of these agents observe only action $A$ and the remaining $1-\theta$ fraction observe only action $B$ being taken in their respective samples. Section \ref{sec:samp} will illustrate these differences using comparative statics. 
\subsection{Failure of Information Aggregation}\label{sec:prop}

Next, we show that agents fail to aggregate information in simple CoSESI when $\rho>0$.

\begin{proposition}[Harmful Equilibrium]\label{thm:noconv}
Suppose $\rho>0$, $G$ is convergent, and $c$ is not linear. Then, simple CoSESI, $\theta^{(\rho)}_{n,G}$, fails to converge to NE, $\normalfont\theta_{\text{\tiny NE}}$, as $n\rightarrow\infty$.
\end{proposition}

\par  Unlike SESI and some CoSESIs which converge to NE in large sample (Proposition \ref{thm:asinfo}), Proposition \ref{thm:noconv} shows that simple CoSESI does not converge to NE when $\rho>0$. Information aggregation fails here due to persistent overprecision bias: agents na{\"i}vely expect their beliefs to become increasingly more accurate in large sample, but the variance of their estimates $G_{n,z}$ never vanishes when $\rho>0$ because $\text{var}_{\mu^{\rho}_{\infty}}(z|\theta)=\rho\theta(1-\theta)>0$. 

\subsection{Comparative Statics: Inefficiencies}\label{sec:samp}
This section shows that the fraction of agents taking the objective action is smaller in CoSESI than in NE. 
To show this in small sample, a monotone inference procedure $G$ is said to be  \textit{sample-unbiased} if $\int^1_0\theta\hspace{0.03in}dG_{n,z}(\theta)=z$ for all $n$ and $z$. That is, an agent's estimate is concentrated on the sample mean for all $n$ and $z$, which, for example, holds for MLE but fails  for Bayesian inference because posteriors depend on the prior mean.
\begin{proposition}[Inefficiencies]\label{thm:compNE} Let $c$ be convex. Then, the following hold:
   \begin{enumerate}
       \item Suppose $G$ is sample-unbiased. Then, $\theta^{(\rho)}_{n,G}\leq\theta_{\text{\normalfont\tiny NE}}$, for any $n$ and $\rho\in[0,1]$.
       \item Suppose $n\rightarrow\infty$ and  $G$ is convergent. Then,  $\theta^{(\rho)}_{\infty,G}\leq \theta_{\text{\normalfont\tiny NE}}$ for all $\rho\in[0,1]$. Moreover, $\theta^{(\rho)}_{\infty,G}$ is a decreasing function of  $\rho\in[0,1]$. 
   \end{enumerate}  

\end{proposition} 
Proposition \ref{thm:compNE} demonstrates that correlation neglect leads many agents to select objectively inferior actions, so it amplifies the information friction created by correlations. 
\subsubsection{Example}
\begin{figure*}[hbt!]
\caption{Equilibrium Outcomes and Ranking of simple CoSESIs}
\label{fig:cosesi}
\centering
\includegraphics[width=0.85\textwidth,height=0.485\textwidth]{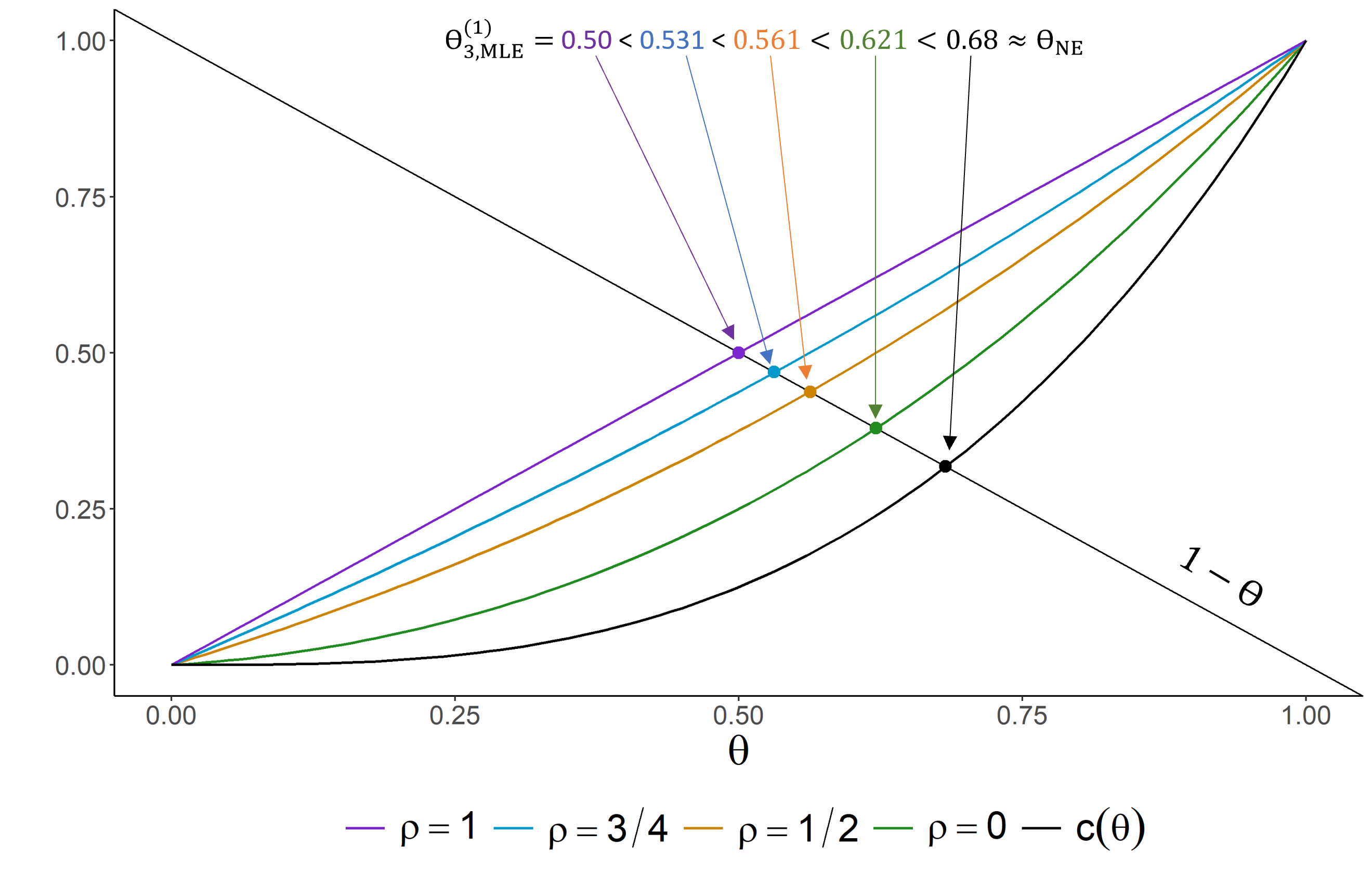}
\end{figure*}

The next example illustrates the negative effects of correlation neglect in simple CoSESI.
\begin{ex}[Comparative Statics]\label{ex:protest}\normalfont
Let $c(\theta)=\theta^3$ and $G=\text{MLE}$. To visualize the effect of correlation neglect on equilibrium outcomes, Figure \ref{fig:cosesi} plots 
both sides of eq. (\ref{eq:equi}): the line $1-\theta$ against the $\rho$-weighted Bernstein polynomial's curve, $\varPsi_n(\theta,\rho;C_n)$, for various values of $\rho\in\big\{0,1/2,3/4,1\big\}$ and $n=3$. When $\rho=0$ (green curve), simple CoSESI $\theta^{(\rho)}_{3,\text{\tiny MLE}}$ coincides with SESI $\theta^{(0)}_{3,\text{\tiny MLE}}=0.621$. The objective distribution of actions is $\theta_{\text{\tiny NE}}\approx0.68$ (intersection between $1-\theta$ and $c(\theta)$). Figure \ref{fig:cosesi} shows that the fraction of agents taking the objective action in simple CoSESI decreases as the correlation $\rho$ increases. 
\par As $n$ grows, SESI $\theta^{(0)}_{n,\text{\tiny MLE}}$ converges to NE $\theta_{\text{\tiny NE}}\approx0.68$. In contrast, Proposition \ref{thm:noconv} shows that,  when $\rho>0$, simple CoSESI $\theta^{(\rho)}_{n,\text{\tiny MLE}}$ will not converge to NE as $n$ grows. To see this, let $\rho=1$ then simple CoSESI is $\theta^{(1)}_{n,\text{\tiny MLE}}=1/2$ for all $n$ but $\theta_{\text{\tiny NE}}\approx0.68$. Thus, even when given access to infinitely many signals, 18\% of the population  in this simple CoSESI take the objectively inferior action ($B$)  in equilibrium due to correlation neglect.  
\end{ex}

 \subsubsection{Discussion: CoSESI vs. SESI and NE}

  \par Example \ref{ex:protest} leads to three observations about NE, SESI, and CoSESI. (i) In NE, all agents hold the same correct belief about the fraction $\theta$ of agents taking action $A$. This induces a positive sorting of types to actions: if an agent takes action $A$, then agents with higher types also take action $A$. (ii) In SESI ($\rho=0$), agents’ estimates differ from one another depending on their respective samples. Agents with a larger sample mean tend to take action $A$ less frequently than agents with a smaller sample mean, so there is a weaker positive sorting than in NE. (iii) Simple CoSESI ($\rho\geq0$) inherits SESI's properties and provides novel insights: when $\rho\approx1$, there is \textit{no} sorting from types to actions for any sample size $n$ because agents' actions depend only on their sample and not their type. This lack of sorting implies that correlation neglect causes not only a distortion in the level of participation but it also breaks down the idea of \textit{revealed preferences}: observing an agent take an action no longer indicates that they benefit from it. To prevent this, a social planner may focus on regulating the information structure because, as noted in the introduction, correcting individuals' behavior is known to be challenging in the literature.
  
\section{Applications}\label{sec:equiapp}
Theorem \ref{thm:generic} and Proposition \ref{thm:asinfo} established that harmful equilibria are mathematically rare. This raises a question: \textit{if these environments are so rare, why should we care about them?} This section demonstrates how some economic mechanisms can select these rare equilibria in markets. Section \ref{sec:mono} shows that a monopolist can maximize profit by engineering the correlation structure of $p_{\rho}$ in eq. (\ref{eq:prho}), where $\rho\rightarrow1$. Section \ref{sec:twosided} indicates that job-referrals networks in labor markets can generate positive correlation structures as captured by $p_{\rho}$.

\subsection{Monopoly Pricing}\label{sec:mono}

\citet{corr22} show that a third-party can manipulate  an agent who has correlation neglect. This application finds a similar result: a monopolist can gain from consumers having correlation neglect in nongeneric settings, but this gain vanishes in generic settings. 
\par Consider a market composed of consumers who have preference for uniqueness---their consumption utility from a good decreases in the number of agents consuming the good. 
Examples arise when a good conveys prestige (e.g., jewelry, artwork). Each individual has a consumption value of $\xi\sim\mathcal{U}[0,1]$ for the good, and a disutility of $-1$ if she meets one or more individuals who also consume the good in $t$ encounters. We allow these encounters to be correlated according to \citeauthor{klotz73}'s (\citeyear{klotz73}) Markov process. Formally, suppose that whether or not a consumer observes the good being consumed in a sequence of encounters is summarized by the following transition probability matrix $\Pi = (\pi_{jk})_{j, k \in \{0,1\}}$:
\begin{align}\label{eq:transition}
\Pi = \begin{pmatrix} \frac{1-2\theta+\phi\theta}{1-\theta} & \frac{(1-\phi)\theta}{1-\theta} \\1-\phi & \phi \end{pmatrix},
\end{align}
where the transition probabilities are explicitly defined as
\begin{align*}
\pi_{11} &= \mathbb{P}(x_i=1|x_{i-1}=1) = \phi \\
\pi_{10} &= \mathbb{P}(x_i=0|x_{i-1}=1) = 1-\phi \\
\pi_{01} &= \mathbb{P}(x_i=1|x_{i-1}=0) = (1-\phi)\theta/(1-\theta) \\
\pi_{00} &= \mathbb{P}(x_i=0|x_{i-1}=0) = (1-2\theta+\phi\theta)/(1-\theta),
\end{align*}
and $\mathbb{P}(x_i=1) = \theta\in(0,1)$, for all $i$. To ensure that all entries in $\Pi$ are between 0 and 1, set $\phi \in \big[\max\{0,(2\theta-1)/\theta\},1\big]$. Let $p^{\phi}$ denote the joint distribution representing eq. (\ref{eq:transition}), where the parameter $\phi$ captures the dependence structure \citep[][eq. (2.6)]{klotz73}. Conditional on having observed the good being consumed, the probability of observing another consumer who consumes the good in a consecutive encounter is $\pi_{11} = \phi$. If $\phi = \theta$, the encounters become random as in \citet[][Section 6]{stat}. However, if $\phi > \theta$, consumers tend to observe either clusters of others consuming the good or clusters of those not consuming the good, whereas $\phi < \theta$ represents the lack of any such clusters.\footnote{The Markov dependence should be viewed as a reduced-form representation of \textit{spatial} social interactions, as formalized in \citet{topa01}. Specifically, Topa models information exchange on a spatial lattice, where agents' Markov transition probabilities depend on the state of their neighbors, generating an equilibrium with spatial clustering. In such an environment, a sequential sampling of peers---traversing the links of a social network---naturally follows a Markov process. In our case, when $\phi>\theta$, observing a peer who consumes the good implies the agent is currently sampling from a ``high-density'' cluster, thereby increasing the conditional probability that the next sampled peer will also be consuming the good.} The intuition for allowing correlations is that if the good is very popular or trendy, then consumers may observe the good being consumed at a much higher rate than random.
\par If a $\theta$ fraction consumes the good and they suffer from correlation neglect, their perceived consumption utility is $u_{\gamma}(\xi,\theta) =\xi+\gamma\big(1-c_t(\theta)\big)$, where $\gamma=-1$. Notice that the true specification of the cost is $c_t(\theta|\phi)=(1-2\theta+\phi\theta)^{t-1}\big/(1-\theta)^{t-2}$, which depends on $\phi$, but since consumers have correlation neglect, they na{\"i}vely assume encounters are random by setting $\phi=\theta$, and hence fail to internalize $\phi$ by using $c_t(\theta):=c_t(\theta|\theta)=(1-\theta)^t$. 
\par A monopolist produces the good at zero marginal cost and sets a price $\psi$ to maximize profit. Consumers observe the price and estimate the expected consumer demand for the good at this price. Then, they decide whether to purchase the good. We follow \citet[][Section 6]{stat} and assume that the monopolist is rational in the sense that he knows how consumers make decisions and takes this knowledge into account when pricing. Here, a NE solves $1-\theta=\psi+c_t(\theta)$, which yields the equilibrium demand under rational expectations, i.e., when the demand is common knowledge to all consumers.  Let $\varLambda_{\psi,G}(y/n):=\text{min}\{1+\psi-C_{n,y/n},1\}$ denote the fraction that conditional on observing $y$ successes in $n$ signals do not purchase the good, where $C_{n,y/n}=\int_0^1c_t(\theta)\hspace{0.03in}dG_{n,y/n}(\theta)$. Thus, for each price $\psi$, CoSESI solves the equation $1-\theta=\sum_{y=0}^n\mathscr{L}_n(p^{\phi})(y)\varLambda_{\psi,G}(y/n)$.
\begin{obs}\label{thm:monoCoSESI}
    A CoSESI, $\theta_{n,G}(p^\phi|\psi)\in[0,1]$, exists for every $n$, $\psi$, $\phi$, and $G$.
\end{obs}
\par The next result analyzes how the monopolist's profit depends on  $\phi$. We first consider the case where $\phi=\theta+\frac{\rho^{1/t}}{1-\theta}$, for all $t\geq1$ and $\rho\in[0,1)$, so that $p^{\phi}=p_{\rho}$ in eq. (\ref{eq:prho}).

\begin{proposition}\label{thm:monocomp}
Let $\phi=\theta+\frac{\rho^{1/t}}{1-\theta}$, for all $t\geq1$, where $\rho\in[0,1)$. The following hold:
    \begin{enumerate}
    \item The CoSESI, $\theta_{n,G}(p_{\rho}|\psi)\in[0,1]$, is unique for all $n$, $\psi$, $\rho$, and $G$.
        \item Suppose the monotone inference procedure $G$ is sample-unbiased. For all $\rho$, the monopolist's profit in CoSESI is larger than in NE. When $\rho\rightarrow1$, the monopolist's profit in CoSESI is also larger than in SESI, for all $n>1$.
        \item Suppose the monotone inference procedure $G$ is convergent. Then, as $n\rightarrow\infty$, the monopolist's profit is larger in CoSESI than in NE and SESI, and it increases in $\rho$.
    \end{enumerate} 
\end{proposition}
Proposition \ref{thm:monocomp} shows that the monopolist can gain from consumers having correlation neglect when encounters are highly clustered. The next result demonstrates that this gain vanishes in large sample under more general Markov dependence structures.

\begin{obs}\label{thm:assinfomono}
    Let $\phi\neq\theta+\frac{\rho^{1/t}}{1-\theta}$, for all $t\geq1$ and $\rho\in[0,1)$, and fix any convergent $G$. Then, the monopolist's profit in CoSESI converges to the profit in NE as $n\rightarrow\infty$.
\end{obs}
This result exploits the cases where $p^{\phi}$ is asymptotically informative (Definition \ref{def:asinfo}), i.e., when the Markov process in eq. (\ref{eq:transition}) is \textit{ergodic}. This happens because when a two-state (time-homogeneous) Markov process is ergodic, the range of dependence will decay (geometrically fast) the farther apart encounters become \citep[][eq. (5.4)]{klotz73}.
\par To summarize, the monopolist can gain from consumers having correlation neglect, but this gain may vanish if their encounters are not very clustered. Thus, the monopolist has an incentive to design nongeneric environments where consumers' encounters are so clustered to the point that observed encounters become \textit{uninformative} for the market demand (i.e., $\rho\rightarrow1$). In practice, the monopolist can achieve this using the so-called ``rarity principle''---making a good highly \textit{exclusive} by segmenting the clientele into groups of consumers with varying purchasing privileges. \citet{lux95} report that this is consistent with the practice of luxury brands such as Rolex and De Beers.

\subsection{Two-Sided Markets}\label{sec:twosided}
\citet[][p. 391]{petro} remark that: ``Frictions are likely to be more important in the labor market than in other markets.'' We therefore introduce simple CoSESI in a two-sided labor market to show that correlations amplify matching frictions and reduce participation and matches when participants have correlation neglect.
\par  We build on \citeauthor{stat}'s (\citeyear[][Section 7]{stat}) two-side labor market.  Consider a unit mass of workers, where each worker decides whether to search for jobs at an idiosyncratic cost $\xi\sim\mathcal{U}[0,1]$, and a unit mass of firms, where each firm decides whether to post a job opening at an idiosyncratic cost $\omega\sim\mathcal{U}[0,1]$. If $\alpha$ workers search for jobs and  $\beta$ firms search for workers, the number of jobs created is given by the matching function $$m(\alpha,\beta)=\zeta\alpha^v\beta^{1-v},$$ where $\zeta\in(0,1)$ is a matching friction that prevents full employment even when all workers and all firms participate in the market and $v\in(0,1)$ is a constant. \citet{petro} justify this Cobb-Douglas functional form in their  empirical survey. A worker who searches for a job finds one with probability $\frac{m(\alpha,\beta)}{\alpha}$ when $\alpha$ workers and $\beta$ firms participate in the market. Assuming that each match creates a surplus of 2 split equally between the worker and the firm, the expected utility of a worker who searches for a job becomes $\frac{m(\alpha,\beta)}{\alpha} -\xi$. The expected utility of a firm that posts a job opening becomes $\frac{m(\alpha,\beta)}{\beta} -\omega$. Firms and workers who do not participate obtain zero utility. 
\par Market thickness will play a central role because it affects the likelihood that a job search is successful. If rational expectations is assumed, i.e., workers and firms have the same correct belief about the market thickness on both sides of the market, then the objective market participation is a NE pair $(\alpha_{\text{\tiny NE}},\beta_{\text{\tiny NE}})$---a solution to the system of equations $\alpha=\zeta\alpha^{v-1}\beta^{1-v}$ and $\beta=\zeta\alpha^v\beta^{-v}$ \citep[][eqs. (3)-(4)]{stat}. 
\par Following \citet{stat}, we first depart from the rational expectations assumption that all workers and firms form the same correct belief about the market thickness on either side of the market. Instead, workers only have a good sense of their side of the market, but to understand the firms' side, each worker obtains data on a few firms and uses an inference procedure $G^w$ to estimate firms’ participation. Similarly, each firm accurately estimates the market thickness of the firms’ side, so each one obtains data on a few workers and uses an inference procedure $G^f$ to estimate workers’ participation. 
\par We then also depart from \citeauthor{stat}'s (\citeyear{stat}) independent-sampling assumption by allowing the signals obtained by workers to be correlated with pairwise correlation $\rho\in[0,1]$, and similarly for firms' signals $\varphi\in[0,1]$.\footnote{Our positive correlation structure is a simple reduced-form representation of  \textit{job-referral} networks in labor markets. This is consistent with \citet[][p. 426]{matt04} who report: ``We develop a model where agents obtain information about job opportunities
through an explicitly modeled network of social contacts. We show that employment
is positively correlated across time and agents.''}  
 We now apply simple CoSESI, where workers' and firms' joint distribution of actions are, respectively,  $p_{\rho}\in\bar{\Delta}_k$ and $p_{\varphi}\in\bar{\Delta}_n$. 
\begin{definition}\label{def:cosesim}\normalfont
    Simple CoSESI in two-sided markets is  $\big(\alpha^{(\rho)}_{k,G^w},\beta^{(\varphi)}_{n,G^f}\big)\in[0,1]^2$, where:
    \begin{enumerate}
    \item[(i)] \textit{Workers}. 
 Fraction $\alpha^{(\rho)}_{k,G^w}$ of workers search for jobs when each worker (1) obtains $k$ Bernoulli signals about firms’ behavior from the joint distribution $p_{\rho}\in\bar{\Delta}_k$ with success probability $\beta^{(\varphi)}_{n,G^f}$ and common pairwise correlation $\rho\in[0,1]$; (2) best responds to the na{\"i}ve estimate formed based on the inference procedure $G^w$.
         \item[(ii)] \textit{Firms}. Fraction $\beta^{(\varphi)}_{n,G^f}$ of firms search for workers when each firm (1) obtains $n$ Bernoulli signals about workers’ behavior the joint distribution $p_{\varphi}\in\bar{\Delta}_n$ with success probability $\alpha^{(\rho)}_{k,G^w}$ and common pairwise correlation $\varphi\in[0,1]$; (2) best responds to the na{\"i}ve estimate formed  based on the inference procedure $G^f$. 
    \end{enumerate}
\end{definition}
\par Specifically, simple CoSESI in this market consists of a pair $\big(\alpha^{(\rho)}_{k,G^w},\beta^{(\varphi)}_{n,G^f}\big)$ that satisfies the following system of simultaneous equations characterizing each side of the market
\begin{align}
    \alpha&=\sum_{r=0}^{k}\Bigg[(1-\rho)\binom{k}{r}\beta^r(1-\beta)^{k-r}+\rho\Big( \beta\mathds{1}_{r=k}+(1-\beta)\mathds{1}_{r=0}\Big)\Bigg] M_{\alpha,\zeta}(k,r/k),\nonumber\\
\beta&=\sum_{s=0}^{n}\Bigg[(1-\varphi)\binom{n}{s}\alpha^s(1-\alpha)^{n-s}+\varphi\Big( \alpha\mathds{1}_{s=n}+(1-\alpha)\mathds{1}_{s=0}\Big)\Bigg] M_{\beta,\zeta}(n,s/n),\nonumber
\end{align}
for all $(\rho,\varphi)$, where each equilibrium equation in this system is equivalent to eq. (\ref{eq:equi}) applied to each side of the market. The fraction of workers who participate conditional on observing $k$ signals with mean $z=r/k$ is $M_{\alpha,\zeta}(k,z)=\zeta\alpha^{v-1}M^w_{k,z}$, where $M^w_{k,z}=\int_{0}^1\beta^{1-v}dG^w_{k,z}(\beta)$, and similarly, the corresponding fraction of firms with $n$ signals and mean $z'=s/n$ who participate is $M_{\beta,\zeta}(n,z')=\zeta\beta^{-v} M^f_{n,z'}$, where $M^f_{n,z'}=\int_0^1\alpha^vdG^f_{n,z'}(\alpha)$.   If $\rho=\varphi=0$, i.e., workers and firms acquire independent signals, then the above system reduces to the equations for SESI $\big(\alpha^{(0)}_{k,G^w},\beta^{(0)}_{n,G^f}\big)$ \citep[][eqs. (5)-(6)]{stat}.
\par Proposition \ref{thm:2cosesi} describes the effect of correlation neglect in two-side markets. It focuses on inference procedures that \textit{preserve shape} (e.g., MLE): $G$ is said to preserve shape if the expected cost $C_{n,z}$ is convex (concave) whenever the cost $c$ is convex (concave).
\begin{proposition}\label{thm:2cosesi}
   Fix any $(k,n)$, $(\rho,\varphi)\in[0,1]^2$, and  $(G^w,G^f)$ preserves shape.
   \begin{enumerate}
       \item There exists a unique CoSESI, $\big(\alpha^{(\rho)}_{k,G^w},\beta^{(\varphi)}_{n,G^f}\big)$, with positive employment.
       \item Let $(G^w,G^f)$ be sample-unbiased. Then, market thickness and employment are smaller in CoSESI than in the unique NE with positive employment.
       \item Let $(G^w,G^f)$ be convergent and $k,n\rightarrow\infty$. Then, market thickness and employment in CoSESI decrease as $\rho$ or $\varphi$ increase, and are smaller than in NE and SESI.
   \end{enumerate}
\end{proposition}
Proposition \ref{thm:2cosesi} reports the adverse effects of correlation neglect on employer-employee matching and participation. Proposition \ref{thm:2cosesi}.3 shows that these adverse effects get worse as the signals' correlations increase. 
 Example \ref{ex:labor} provides some simple comparative statics. 
\begin{ex}\normalfont\label{ex:labor}
Let $v=1/2$, $G^w=G^f=\text{MLE}$, so Proposition \ref{thm:2cosesi} applies. NE solves $\alpha=\zeta(\beta/\alpha)^{1/2}$ and $\beta=\zeta(\alpha/\beta)^{1/2}$, so participation in NE is $\alpha_{\text{\tiny NE}}=\beta_{\text{\tiny NE}}=\zeta$ and employment $m(\alpha_{\text{\tiny NE}},\beta_{\text{\tiny NE}})=\zeta^2$. Now, let $k=n=2$, $\rho=\varphi=0$, then by \citet[Theorem 7 and Example 6]{stat}, SESI's participation $\big(\alpha^{(0)}_{2,\text{\tiny MLE}},\beta^{(0)}_{2,\text{\tiny MLE}}\big)$ satisfies $\zeta^2<\alpha^{(0)}_{2,\text{\tiny MLE}}\leq\alpha_{\text{\tiny NE}}$ and $\zeta^2<\beta^{(0)}_{2,\text{\tiny MLE}}\leq\beta_{\text{\tiny NE}}$, so SESI's employment satisfies $\zeta^3< m(\alpha^{(0)}_{2,\text{\tiny MLE}},\beta^{(0)}_{2,\text{\tiny MLE}})\leq m(\alpha_{\text{\tiny NE}},\beta_{\text{\tiny NE}})=\zeta^2$. Now, if $\rho=\varphi=1$, then simple CoSESI solves $\alpha=(\zeta\beta)^{2/3}$ and $\beta=(\zeta\alpha)^{2/3}$, so participation $\alpha^{(1)}_{2,\text{\tiny MLE}}=\beta^{(1)}_{2,\text{\tiny MLE}}=\zeta^2$ and employment $m\big(\alpha^{(1)}_{2,\text{\tiny MLE}},\beta^{(1)}_{2,\text{\tiny MLE}}\big)=\zeta^{3}$ are both smaller than in NE by a factor of $\zeta$ and are also smaller than in SESI.  
\end{ex}
Example \ref{ex:labor} shows that the gap between CoSESI and NE is proportional to $\zeta$, which suggests the following mechanism: in labor markets where workers' and firms' signals originate predominantly from job-referral networks, correlation neglect negatively affects participation and employment by amplifying the fluctuations of the matching friction.

\section{Extensions and Concluding Remarks}\label{sec:conc}

\subsection{Robustness of Aggregate Efficiency}\label{sec:robust}
In this section, we show that aggregate efficiency---Theorem \ref{thm:generic}---is robust to the choice of the measure on the simplex $\Delta_n$. To this end, consider the symmetric Dirichlet distribution on $\Delta_n$, denoted $\nu_{\kappa,n}:=\text{Dir}(\underbrace{\kappa,\dots,\kappa}_{2^n \text{ times}})$, for any constant $\kappa>0$. For example, when $\kappa=1$, $\nu_{1,n}$ coincides with the uniform distribution $\lambda_n$ used in Section \ref{sec:gen}.
\begin{definition}\normalfont 
    An arbitrary distribution  $\mathscr{L}^*_n\in\Delta(\{0,\dots,n\})$ is $\nu_{\kappa,n}$-\textit{generic} if, $\forall\epsilon>0$, 
    \begin{align*}
        \nu_{\kappa,n}\Big(\Big\{p'\in\Delta_n:d_{TV}\big(\mathscr{L}_n(p'),\mathscr{L}^*_n\big)\leq\epsilon\Big\}\Big)\longrightarrow1 \quad \text{as }n\rightarrow\infty.
    \end{align*}
\end{definition}
 The next result extends Lemma \ref{thm:lebesgue} by showing that the concentration around the binomial distribution persists even when we consider any symmetric Dirichlet distribution on the simplex $\Delta_n$. Recall the sets $\Theta$ and $\Delta^{\epsilon}_{n}$ in eqs. (\ref{eq:Theta}) and (\ref{eq:delta}), respectively.
 
\begin{lemma}[Genericity, extended]\label{thm:dir}
Let $\theta_n\in[0,1]$ be an arbitrary number satisfying
$(\theta_n)_{n\geq1}\in \Theta$ and fix any $\kappa>0$. Then, $\forall\epsilon>0$, there exists $n_{\epsilon}\geq1$ such that, $\forall n\geq n_{\epsilon}$, $$\nu_{\kappa,n}(\Delta^{\epsilon}_{n})\geq 1 - \frac{ 4}{\epsilon^2 (2^{n}\kappa+1)} \hspace{0.04in} \text{\normalfont and}\hspace{0.04in} \nu_{\kappa,n} \Big( \Big\{ p \in \Delta_n : d_{TV}\big(\mathscr{L}_n(p), \mathscr{B}_n(\theta_n)\big) \le \epsilon \Big\} \Big) \geq 1 - \frac{ 2(n+1)^2}{\epsilon^2 (2^{n}\kappa+1)}.$$
\end{lemma}
Lemma \ref{thm:dir} also reveals that extending to the more general case of Dirichlet distributions remains very tractable because the lower bounds are explicit functions of the new parameter $\kappa$. Notice that when $\kappa=1$, these bounds are identical to those in Lemma \ref{thm:lebesgue}. 
\begin{theorem}[Aggregate Efficiency, extended]\label{thm:generic2} Fix any $\kappa>0.$ Then,\hfill
\begin{enumerate}
    \item For every $\epsilon>0$, $\nu_{\kappa,n}\big(\big\{p\in\Delta_n:\mathscr{L}_n(p)\in\mathcal{C}_n\backslash\mathcal{S}^{\epsilon}_n\big\}\big)\longrightarrow0$ as $n\rightarrow\infty$. 
    \item Suppose an equilibrium distribution $\mathscr{L}^*_n\in\mathcal{C}_n$ is $\nu_{\kappa,n}$-generic. Then, there exists a SESI, $\theta_{n,G}(p_0)\in[0,1]$, such that $d_{TV}\big(\mathscr{L}^*_n,\mathscr{B}_{n}(\theta_{n,G}(p_0))\big)\longrightarrow0$ as $n\rightarrow\infty$.
\end{enumerate}
\end{theorem}
Theorem \ref{thm:generic2} shows that the conclusion of Theorem \ref{thm:generic} is not an artifact of the uniform distribution. Online Appendix \hyperref[app:multi]{B.I} extends a version of Lemma \ref{thm:lebesgue} to \textit{multinomial} settings where signals take finitely many values. This multi-dimensional extension highlights that our concentration-of-measure result is also robust to the dimension in the sense that it extends to a multinomial case. This extension indicates that SESIs remain generic in the sense of Theorem \ref{thm:generic} even in multi-action games, where there are more than two actions. 

\subsection{Some Extensions}
In Online Appendix \hyperref[app:loc]{B.II}, we characterize the location and multiplicity of CoSESIs in games with strategic complements. In Online Appendix \hyperref[app:assort]{B.III}, we introduce \textit{assortativity} in CoSESI by allowing agents' types to depend on the correlation structure. In Online Appendix \hyperref[app:hetero]{B.IV}, we extend our framework to settings where agents are heterogeneous.

\subsection{Conclusion}\label{sec:conclusion}
This paper establishes a new testable aggregate efficiency benchmark in large population games: decentralized aggregation generically corrects for individual-level behavioral biases. Thus, aggregate inefficiencies in these settings are not driven solely by bounded rationality, but by structural economic forces such as strategic network formation and profit-maximizing platforms that systematically exploit individuals' biases and cause aggregate inefficiencies. We characterize these inefficiencies in monopoly and two-sided labor markets. Our results therefore suggest that policy should shift focus from correcting individuals' psychology to regulating the sources of correlation structures.

\phantomsection\label{sec:proof} 
\section*{Appendix A: Proofs for the Main Text}

\subsection*{Proof of Observations \ref{thm:beta2}--\ref{thm:beta}}
The proof of Observations \ref{thm:beta2}--\ref{thm:beta} are simple (but long) calculations of Bayesian posteriors, so they are relegated to Online Appendix \hyperref[app:proof]{C}.

\subsection*{Proof of Observation \ref{thm:cosesichar}}
Fix a sample size $n\geq 1$ and a subjective game $\langle u_{\gamma},\upsilon,p,F^{q}\rangle_n$.
Throughout the proof, we keep $n$, $p$, $q$, $F^{q}$, and $\upsilon$ fixed and suppress this dependence to ease notation.

For $\mathscr{L}_n(p)$, $\sigma(\cdot)$, $\theta\in[0,1]$, and each $k\in\{0,\dots,n\}$, 
\[
\mathscr{L}_n(p)(k\mid\theta)
= \Proba_p\Big(\sum_{i=1}^n x_i = k  \Big| \theta\Big)
\]
is the probability that an agent's sample has $k$ successes when signals are drawn from $p(\cdot\mid\theta)$.
Conditional on observing the sample mean $z=k/n$, an agent of type $\xi$ chooses $A$ if and only if, given her posterior belief induced by $F^{q}$ and her payoff $u_{\gamma}$, action $A$ is a best reply.
By the definition, $\sigma(k/n)$ is the fraction of types who choose $A$ after observing $z=k/n$.
Hence, by the law of total probability, the function
\begin{align}\label{eq:aggt}
    T(\theta)
:= \sum_{k=0}^n \mathscr{L}_n(p)(k|\theta) \sigma(k/n)
\end{align}
is the fraction of agents who take action $A$ when the true success probability is $\theta\in[0,1]$.

We now prove the two implications in the statement of Observation \ref{thm:cosesichar}.
\begin{proof}[Proof of Observation \ref{thm:cosesichar}]
\medskip
\noindent\emph{($\Rightarrow$)}
Suppose $\theta_{n,F^{q}}(p)\in[0,1]$ is a CoSESI.
By Definition \ref{def:CoSESI}, when the true success probability is $\theta_{n,F^{q}}(p)$, a $\theta_{n,F^{q}}(p)$ fraction of agents take action $A$ in the subjective game $\langle u_{\gamma},\upsilon,p,F^{q}\rangle_n$.
From the definition of $T$ in eq. (\ref{eq:aggt}), this fraction is exactly $T\big(\theta_{n,F^{q}}(p)\big)$.
Thus, $T\big(\theta_{n,F^{q}}(p)\big) = \theta_{n,F^{q}}(p),$
so $\theta_{n,F^{q}}(p)$ is a fixed point of eq. (\ref{eq:aggt}).

\medskip
\noindent\emph{($\Leftarrow$)}
Conversely, suppose $\theta^{*}\in[0,1]$ satisfies $T(\theta^{*}) = \theta^{*}$ in eq. (\ref{eq:aggt}). By definition of the map $T$, when the true success probability is $\theta^{*}$ and agents form beliefs using $F^{q}$ and best respond according to $u_{\gamma}$, the fraction of agents who choose action $A$ in the subjective game $\langle u_{\gamma},\upsilon,p,F^{q}\rangle_n$ is $T(\theta^{*})$.
The fixed-point condition $T(\theta^{*})=\theta^{*}$ therefore implies that, under these primitives, a $\theta^{*}$ fraction of agents take action $A$ when signals are drawn from $p(\cdot\mid\theta^{*})$ and each agent infers and best responds as in Definition \ref{def:CoSESI}.
This means precisely that $\theta^{*}$ is a CoSESI, i.e., $\theta^{*}=\theta_{n,F^{q}}(p)$.
\end{proof}

\subsection*{Proof of  Proposition \ref{thm:coscont} and Corollary \ref{thm:sesi1}}

 \begin{proof}[Proof of Proposition \ref{thm:coscont}]
Since the type distribution $\upsilon$ is absolutely continuous, the fraction of indifferent agents has measure zero. Thus, the map $\sigma(.)$ is single-valued. For every $k\in\{0,\dots,n\}$, let $S_k:=\big\{(x_i)_{i=1}^n\in\{0,1\}^n:\sum_{i=1}^nx_i=k\big\}$. By definition of the distribution of the sum of Bernoulli random variables, $\mathscr{L}_n(p)(k|\theta):=\sum_{(x_i)_{i=1}^n\in S_k}p\big((x_i)_{i=1}^n\big|\theta\big)$. Since the map $\theta\mapsto p(.|\theta)$ is assumed to be continuous, the map  $\theta\mapsto\mathscr{L}_n(p)(k|\theta)$ is continuous for all $k$. Thus, the map $\theta\mapsto T(\theta):=\sum_{k=0}^n \mathscr{L}_n(p)(k|\theta)\hspace{0.03in} \sigma(k/n)$ is also continuous. Since the map $T:[0,1]\rightarrow[0,1]$ is continuous, it has at least one fixed point by Brouwer's fixed-point theorem. Every such fixed point is a CoSESI by Observation \ref{thm:cosesichar}. 
 \end{proof}
 
 \begin{proof}[Proof of Corollary \ref{thm:sesi1}]
    By definition, when $q=p=p_0$, the distribution of actions $\mathscr{L}_n(p_0)$ is a binomial distribution $\mathscr{B}_n(\theta)$, so the aggregate best-response map in Observation \ref{thm:cosesichar} (see, eq. (\ref{eq:aggt})) is a Bernstein polynomial. By \citet[][Observation 1]{stat}, SESIs are the only solutions to Bernstein polynomials based on a monotone inference procedure $F^q=G$. Existence and uniqueness in every objective game $\langle u_{-1},\mathcal{U}[0,1],p_0,G\rangle_n$ where $c$ is increasing follows by \citet[][Theorem 1]{stat}.
 \end{proof}

\subsection*{Proof of Observation \ref{thm:nongen}}
\begin{proof}
 Recall that, for all $\rho\in[0,1]$, $p_{\rho}$ induces the distribution of the sum $\mathscr{L}_n(p_{\rho})=\mu^{\rho}_n(y|\theta)$ in eq. (\ref{eq:distpositive}). When $\rho=1$, $\mu^{\rho}_n(y|\theta)$  becomes $\mu^{1}_n(y|\theta)=\theta \mathds{1}_{y=n}+(1-\theta)\mathds{1}_{y=0}.$
    Then, 
    \begin{align*}
        &d_{TV}(\mu^{1}_n,\mathscr{B}_n(\theta))=\frac{1}{2}\sum_{y=0}^n\Bigg|\binom{n}{y}\theta^y(1-\theta)^{n-y}-\Big[ \theta\mathds{1}_{y=n}+(1-\theta)\mathds{1}_{y=0}\Big]\Bigg|\\
        &=\frac{1}{2}\Bigg[\sum_{y=1}^{n-1}\binom{n}{y}\theta^y(1-\theta)^{n-y}+(1-\theta)-(1-\theta)^n+\theta-\theta^n\Bigg]\underset{n\rightarrow\infty}{\longrightarrow}\frac{1}{2}\Big[1+(1-\theta)+\theta\Big]=1\end{align*} \end{proof}
        
\subsection*{Proof of Proposition \ref{thm:sesi2}}
\begin{proof}
  --- $(1)\iff(2)$: This follows from \citet[][Observation 1]{stat} because when CoSESI coincides with SESI, it means that CoSESI induces a binomial distribution of actions. Then, apply the following identification property of the TV norm: for all distributions $P,Q$, $d_{TV}(P,Q)=0$ if and only if $P=Q$.\\
  --- $(2)\iff(3)$: This follows from \citet[][Theorem 2.1]{binom01} who prove that the distribution of the sum of arbitrary Bernoulli random variables coincides with a binomial distribution if and only if (3) holds. 
\end{proof}

\subsection*{Proof of Theorem \ref{thm:generic2}}

\noindent\textit{Outline.} The proof connects the concentration properties of the Dirichlet distribution to the equilibrium set via three steps. To apply Lemma \ref{thm:dir}, we first use Lemma \ref{thm:sesigen} to identify a reference sequence of SESIs, $\theta_n$, that satisfies the convergence rate required for the set $\Theta$ in eq. (\ref{eq:Theta}). Then, for \textit{Part 1}, Lemma \ref{thm:dir} establishes that under any symmetric Dirichlet distribution, the true distribution of actions $\mathscr{L}_n(p)$ concentrates around the binomial distribution $\mathscr{B}_n(\theta_n)$ as $n\rightarrow\infty$. Since this limit is a SESI outcome, the measure of joint distributions inducing non-SESI outcomes must vanish. For \textit{Part 2}, Lemma \ref{lem:unique_generic_center} shows that any generic equilibrium distribution $\mathscr{L}^*_n$ must be asymptotically close to $\mathscr{B}_n(\theta_n)$, as both are arbitrarily close to the random realization $\mathscr{L}_n(p)$ with high probability.

\begin{proof}[Proof of Lemma \ref{thm:dir}]
Fix some constants $\kappa>0$ and $\epsilon>0$. 
\par\noindent\textit{Step 1: The induced law of $\mathscr{L}_n(p)$ under $\nu_{\kappa,n}$ is Dirichlet.}
Index $\{0,1\}^n$ as $\{x^1,\dots,x^{2^n}\}$ and write $p=(p_1,\dots,p_{2^n})$ where $p_i:=p_{x^i}$.
Under $\nu_{\kappa,n}$ we have $p\sim \text{\normalfont Dir}(\kappa,\dots,\kappa)$ on $\Delta_n$.

We use the standard Gamma representation of the Dirichlet law: let $(Y_i)_{i=1}^{2^n}$ be independent with
$Y_i\sim \Gamma(\kappa,1)$, let $T:=\sum_{i=1}^{2^n}Y_i$, and set $p_i:=Y_i/T$. Then $p\sim\text{\normalfont Dir}(\kappa,\dots,\kappa)$.

For each $k\in\{0,\dots,n\}$ define the index set
\[
\mathcal{G}_k:=\big\{i\in\{1,\dots,2^n\}: \textstyle\sum_{j=1}^n x^i_j=k\big\},
\]
so $|\mathcal{G}_k|=\binom{n}{k}$ and by definition,
\[
\mathscr{L}_n(p)(k)=\sum_{i\in \mathcal{G}_k}p_i=\frac{\sum_{i\in \mathcal{G}_k}Y_i}{\sum_{i=1}^{2^n}Y_i}.
\]
Define $Z_k:=\sum_{i\in \mathcal{G}_k}Y_i$. Because the $\mathcal{G}_k$ are disjoint and the $Y_i$ are independent Gamma$(\kappa,1)$,
the random variables $(Z_k)_{k=0}^n$ are independent and $Z_k \sim \Gamma(\kappa|\mathcal{G}_k|,1)=\Gamma(\kappa\binom{n}{k},1).$
Moreover, $\sum_{k=0}^n Z_k = \sum_{i=1}^{2^n} Y_i = T$, hence
\[
\big(\mathscr{L}_n(p)(0),\dots,\mathscr{L}_n(p)(n)\big)
=\Big(\frac{Z_0}{\sum_{\ell=0}^n Z_\ell},\dots,\frac{Z_n}{\sum_{\ell=0}^n Z_\ell}\Big).
\]
By the Gamma representation again, this implies
\[
\big(\mathscr{L}_n(p)(0),\dots,\mathscr{L}_n(p)(n)\big)\sim \text{\normalfont Dir}\big(\alpha_0(\kappa),\dots,\alpha_n(\kappa)\big),
\qquad
\alpha_k(\kappa):=\kappa\binom{n}{k}.
\]
Let $\alpha_\bullet(\kappa):=\sum_{k=0}^n \alpha_k(\kappa)=\kappa\sum_{k=0}^n\binom{n}{k}=\kappa 2^n$.

\par\noindent\textit{Step 2: Dirichlet concentration around the mean.}
Let $P=(P_0,\dots,P_n)\sim \text{\normalfont Dir}(\alpha_0,\dots,\alpha_n)$ on the simplex
$D_n:=\{q\in[0,1]^{n+1}:\sum_{k=0}^n q_k=1\}$, where $\alpha_k>0$ and $\alpha_\bullet:=\sum_{k=0}^n\alpha_k$.
Let $q_k:=\mathbb{E}[P_k]$ and $\|P-q\|_\infty:=\max_{0\le k\le n}|P_k-q_k|$.

\begin{claim}\label{thm:claim1}
For each $k$,
\[
\mathbb{E}[P_k]=\frac{\alpha_k}{\alpha_\bullet},
\qquad
\mathrm{Var}(P_k)=\frac{\alpha_k(\alpha_\bullet-\alpha_k)}{\alpha_\bullet^2(\alpha_\bullet+1)}.
\] 
\end{claim}
\begin{proof}[Proof of Claim \ref{thm:claim1}]
    By the Gamma representation, take independent $Y_j\sim \Gamma(\alpha_j,1)$ and set $P_j:=Y_j/T$ where $T:=\sum_{j=0}^n Y_j$.
Fix $k$ and let $Z:=\sum_{j\ne k}Y_j$. Then $Y_k\sim\Gamma(\alpha_k,1)$, $Z\sim\Gamma(\alpha_\bullet-\alpha_k,1)$, and $Y_k$ and $Z$ are independent.
Hence, $P_k=\frac{Y_k}{Y_k+Z}$
has a $\mathrm{Beta}(\alpha_k,\alpha_\bullet-\alpha_k)$ distribution. For $X\sim\mathrm{Beta}(a,b)$, $\mathbb{E}[X]=\frac{a}{a+b}$ and $\mathrm{Var}(X)=\frac{ab}{(a+b)^2(a+b+1)}$. Substituting $a=\alpha_k$ and $b=\alpha_\bullet-\alpha_k$ yields the claim.
\end{proof}

\begin{claim}\label{thm:claim2}
    \[
\sum_{k=0}^n \mathrm{Var}(P_k)\le \frac{1}{\alpha_\bullet+1}.
\]
\end{claim}
\begin{proof}[Proof of Claim \ref{thm:claim2}]
Using Claim \ref{thm:claim1},
\[
\sum_{k=0}^n \mathrm{Var}(P_k)
=\frac{1}{\alpha_\bullet^2(\alpha_\bullet+1)}\sum_{k=0}^n \alpha_k(\alpha_\bullet-\alpha_k)
=\frac{1}{\alpha_\bullet^2(\alpha_\bullet+1)}\Big(\alpha_\bullet\sum_{k=0}^n\alpha_k-\sum_{k=0}^n\alpha_k^2\Big).
\]
Since $\sum_{k=0}^n\alpha_k=\alpha_\bullet$ and $\sum_{k=0}^n\alpha_k^2\ge 0$,
\[
\sum_{k=0}^n \mathrm{Var}(P_k)\le \frac{1}{\alpha_\bullet^2(\alpha_\bullet+1)} \alpha_\bullet^2=\frac{1}{\alpha_\bullet+1}.
\]
\end{proof}

\begin{claim}\label{thm:claim3}
    For every $\eta>0$,
\begin{align}
\mathbb{P}\big(\|P-q\|_\infty\ge \eta\big) &\le \frac{1}{\eta^2(\alpha_\bullet+1)}, \label{eq:dir_bound_infty}\\
\mathbb{P}\Big(\sup_{I\subseteq\{0,\dots,n\}}\big|P(I)-q(I)\big|\ge \eta\Big)
&\le \frac{(n+1)^2}{\eta^2(\alpha_\bullet+1)}. \label{eq:dir_bound_sets}
\end{align}
\end{claim} 
\begin{proof}[Proof of Claim \ref{thm:claim3}]
For each $k$, Chebyshev's inequality yields $\mathbb{P}(|P_k-q_k|\ge \eta)\le \mathrm{Var}(P_k)/\eta^2$.
Thus, by a union bound and Claim \ref{thm:claim2},
\[
\mathbb{P}(\|P-q\|_\infty\ge \eta)
\le \sum_{k=0}^n \mathbb{P}(|P_k-q_k|\ge \eta)
\le \frac{1}{\eta^2}\sum_{k=0}^n \mathrm{Var}(P_k)
\le \frac{1}{\eta^2(\alpha_\bullet+1)},
\]
which is \eqref{eq:dir_bound_infty}. For \eqref{eq:dir_bound_sets}, note that for any $I\subseteq\{0,\dots,n\}$,
\[
|P(I)-q(I)|=\Big|\sum_{k\in I}(P_k-q_k)\Big|
\le \sum_{k\in I}|P_k-q_k|
\le |I|\hspace{0.02in}\|P-q\|_\infty
\le (n+1)\|P-q\|_\infty.
\]
Hence $\sup_I |P(I)-q(I)| \le (n+1)\|P-q\|_\infty$, so
\[
\Big\{\sup_I |P(I)-q(I)|\ge \eta\Big\}\subseteq \Big\{\|P-q\|_\infty\ge \eta/(n+1)\Big\}.
\]
Applying \eqref{eq:dir_bound_infty} with $\eta/(n+1)$ yields \eqref{eq:dir_bound_sets}. \end{proof}

\par\noindent\textit{Step 3: Apply Step 2 to $\mathscr{L}_n(p)$ and identify the mean.}
From Step 1, under $\nu_{\kappa,n}$ the vector $\big(\mathscr{L}_n(p)(0),\dots,\mathscr{L}_n(p)(n)\big)$ is Dirichlet with parameters
$\alpha_k(\kappa)=\kappa\binom{n}{k}$ and total mass $\alpha_\bullet(\kappa)=\kappa 2^n$.
Therefore, by Claim \ref{thm:claim1}, for every $k\in\{0,\dots,n\}$,
\[
\mathbb{E}_{\nu_{\kappa,n}}\big[\mathscr{L}_n(p)(k)\big]
=\frac{\alpha_k(\kappa)}{\alpha_\bullet(\kappa)}
=\frac{\kappa\binom{n}{k}}{\kappa 2^n}
=\binom{n}{k}2^{-n}
=\mathscr{B}_n(1/2)(k).
\]
Applying \eqref{eq:dir_bound_infty}--\eqref{eq:dir_bound_sets} with $\alpha_\bullet=\kappa 2^n$ gives: for all $\eta>0$,
\begin{align}
\nu_{\kappa,n}\Big(\max_{0\le k\le n}\big|\mathscr{L}_n(p)(k)-\mathscr{B}_n(1/2)(k)\big|\ge \eta\Big)
&\le \frac{1}{\eta^2(\kappa 2^n+1)}, \label{eq:L_vs_B12_point}\\
\nu_{\kappa,n}\Big(\sup_{I\subseteq\{0,\dots,n\}}\big|\mathscr{L}_n(p)(I)-\mathscr{B}_n(1/2)(I)\big|\ge \eta\Big)
&\le \frac{(n+1)^2}{\eta^2(\kappa 2^n+1)}. \label{eq:L_vs_B12_set}
\end{align}

\par\noindent\textit{Step 4: Control the distance between $\mathscr{B}_n(1/2)$ and $\mathscr{B}_n(\theta_n)$.}
Let $\theta\in(0,1)$ and write $\delta:=\theta-1/2$. Also, let $D_{KL}(.\lVert.)$ denote the relative entropy. Simple calculation yields
\[
D_{KL}\big(\mathrm{Bern}(1/2)\hspace{0.02in}\|\hspace{0.02in}\mathrm{Bern}(\theta)\big)
=\frac12\log\Big(\frac{1}{4\theta(1-\theta)}\Big)
=\frac12\log\Big(\frac{1}{1-4\delta^2}\Big).
\]
If $|\delta|\le 1/4$, then $x:=4\delta^2\in[0,1/2]$ and $\log\big(\frac{1}{1-x}\big)\le 2x$, hence
\[
D_{KL}\big(\mathrm{Bern}(1/2)\hspace{0.02in}\|\hspace{0.02in}\mathrm{Bern}(\theta)\big)\le \frac12\cdot 2\cdot 4\delta^2 = 4\delta^2.
\]
By additivity of relative entropy for iid products,
\[
D_{KL}\big(\mathscr{B}_n(1/2)\hspace{0.02in}\|\hspace{0.02in}\mathscr{B}_n(\theta)\big)\le 4n(\theta-1/2)^2.
\]
Pinsker's inequality yields
\[
d_{TV}\big(\mathscr{B}_n(1/2),\mathscr{B}_n(\theta)\big)
\le \sqrt{\frac12\hspace{0.02in}D_{KL}\big(\mathscr{B}_n(1/2)\hspace{0.02in}\|\hspace{0.02in}\mathscr{B}_n(\theta)\big)}
\le \sqrt{2n}\hspace{0.02in}|\theta-1/2|.
\]
Let $\delta_n:=\sqrt{2n}\hspace{0.02in}|\theta_n-1/2|.$
Since $(\theta_n)_{n\ge 1}\in\Theta$, $\sqrt{n}\hspace{0.02in}|\theta_n-1/2|\to 0$, hence $\delta_n\to 0$.
Moreover,
\begin{align}
\sup_{I\subseteq\{0,\dots,n\}}\big|\mathscr{B}_n(1/2)(I)-\mathscr{B}_n(\theta_n)(I)\big|
&= d_{TV}\big(\mathscr{B}_n(1/2),\mathscr{B}_n(\theta_n)\big)\le \delta_n, \label{eq:Bn_sets}\\
\max_{0\le k\le n}\big|\mathscr{B}_n(1/2)(k)-\mathscr{B}_n(\theta_n)(k)\big|
&\le \sum_{k=0}^n \big|\mathscr{B}_n(1/2)(k)-\mathscr{B}_n(\theta_n)(k)\big|\nonumber\\
&=2\hspace{0.02in}d_{TV}\big(\mathscr{B}_n(1/2),\mathscr{B}_n(\theta_n)\big)\le 2\delta_n. \label{eq:Bn_points}
\end{align}

Fix $\epsilon>0$. Since $\delta_n\to 0$ and $\theta_n\to 1/2$, there exists an integer $n_\epsilon$ such that for all $n\ge n_\epsilon$, $\delta_n\le \epsilon/4$
 and
$|\theta_n-1/2|\le 1/4.$ We fix $n\ge n_\epsilon$ hereafter.

\par\noindent\textit{Step 5: Proof of the first inequality.}
For any $p\in\Delta_n$ and $k\in\{0,\dots,n\}$,
\[
\big|\mathscr{L}_n(p)(k)-\mathscr{B}_n(\theta_n)(k)\big|
\le \big|\mathscr{L}_n(p)(k)-\mathscr{B}_n(1/2)(k)\big|
+\big|\mathscr{B}_n(1/2)(k)-\mathscr{B}_n(\theta_n)(k)\big|.
\]
Taking the maximum over $k$ and using \eqref{eq:Bn_points} with $\delta_n\le \epsilon/4$,
\begin{align*}
\max_k \big|\mathscr{L}_n(p)(k)-\mathscr{B}_n(\theta_n)(k)\big|
&\le \max_k \big|\mathscr{L}_n(p)(k)-\mathscr{B}_n(1/2)(k)\big| + 2\delta_n\\
&\le \max_k \big|\mathscr{L}_n(p)(k)-\mathscr{B}_n(1/2)(k)\big| + \epsilon/2.
\end{align*}
Therefore,
\[
\Big\{\max_k \big|\mathscr{L}_n(p)(k)-\mathscr{B}_n(\theta_n)(k)\big|\ge \epsilon\Big\}
\subseteq
\Big\{\max_k \big|\mathscr{L}_n(p)(k)-\mathscr{B}_n(1/2)(k)\big|\ge \epsilon/2\Big\}.
\]
Applying \eqref{eq:L_vs_B12_point} with $\eta=\epsilon/2$ yields
\[
\nu_{\kappa,n}\Big(\max_k \big|\mathscr{L}_n(p)(k)-\mathscr{B}_n(\theta_n)(k)\big|\ge \epsilon\Big)
\le \frac{1}{(\epsilon/2)^2(\kappa 2^n+1)}
=\frac{4}{\epsilon^2(\kappa 2^n+1)}.
\]

\par\noindent\textit{Step 6: Proof of the TV inequality.}
For any $p\in\Delta_n$ and any $I\subseteq\{0,\dots,n\}$,
\begin{align*}
\big|\mathscr{L}_n(p)(I)-\mathscr{B}_n(\theta_n)(I)\big|
&\le \big|\mathscr{L}_n(p)(I)-\mathscr{B}_n(1/2)(I)\big|
+\big|\mathscr{B}_n(1/2)(I)-\mathscr{B}_n(\theta_n)(I)\big|
\\
&\le \sup_J \big|\mathscr{L}_n(p)(J)-\mathscr{B}_n(1/2)(J)\big|+\delta_n,
\end{align*}
where we used \eqref{eq:Bn_sets}. Taking the supremum over $I$ and using $\delta_n\le \epsilon/4$ gives
\[
\sup_I \big|\mathscr{L}_n(p)(I)-\mathscr{B}_n(\theta_n)(I)\big|
\le \sup_I \big|\mathscr{L}_n(p)(I)-\mathscr{B}_n(1/2)(I)\big|+\epsilon/4,
\]
hence
\[
\Big\{\sup_I \big|\mathscr{L}_n(p)(I)-\mathscr{B}_n(\theta_n)(I)\big|\ge \epsilon\Big\}
\subseteq
\Big\{\sup_I \big|\mathscr{L}_n(p)(I)-\mathscr{B}_n(1/2)(I)\big|\ge 3\epsilon/4\Big\}.
\]
Applying \eqref{eq:L_vs_B12_set} with $\eta=3\epsilon/4$ yields
\[
\nu_{\kappa,n}\Big(\sup_I \big|\mathscr{L}_n(p)(I)-\mathscr{B}_n(\theta_n)(I)\big|\ge \epsilon\Big)
\le \frac{(n+1)^2}{(3\epsilon/4)^2(\kappa 2^n+1)}
=\frac{16(n+1)^2}{9\epsilon^2(\kappa 2^n+1)}
\le \frac{2(n+1)^2}{\epsilon^2(\kappa 2^n+1)}.
\]
Since $n\ge n_\epsilon$ was arbitrary, both inequalities hold for all $n\ge n_\epsilon$.
\end{proof}

The next result proves the existence of some SESIs that are elements of $\Theta$ in eq. (\ref{eq:Theta}).
\begin{lemma}\label{thm:sesigen}
    Consider an objective game $\langle u_{-1},\mathcal{U}[0,1],p_0,G^{\mathrm{MLE}}\rangle_n$, where $c$ is increasing and twice continuously differentiable on $[0,1]$.
    Suppose the unique NE is $\theta_{\text{\normalfont\tiny NE}} = 1/2$ (e.g., $c(\theta)=\theta^2/2+3/8$).
    Then, for each $n$, the SESI $\theta_{n,G^{\mathrm{MLE}}}(p_0)$ satisfies
    \[
    \Big|\theta_{n,G^{\mathrm{MLE}}}(p_0) - \frac{1}{2}\Big| = O\Big(\frac{1}{n}\Big).
    \]
    In particular, this sequence of SESIs, $(\theta_{n,G^{\mathrm{MLE}}}(p_0))_{n\geq 1}$, satisfies $(\theta_{n,G^{\mathrm{MLE}}}(p_0))_{n\geq 1} \in \Theta$.
\end{lemma}

The proof of Lemma \ref{thm:sesigen} is straightforward, so it is in Online Appendix \hyperref[app:proof]{C}.

\begin{lemma}\label{lem:unique_generic_center}
Fix any sequence of probability measures $(\gamma_n)_{n\ge1}$ on $\Delta_n$.
Let $(B_n)_{n\ge1}$ and $(F_n)_{n\ge1}$ be sequences in $\Delta(\{0,\dots,n\})$.
Suppose that for every $\epsilon>0$,
\[
\gamma_n\Big(\big\{p\in\Delta_n: d_{TV}(\mathscr{L}_n(p),B_n)\ge \epsilon\big\}\Big)\to 0
\text{ and }
\gamma_n\Big(\big\{p\in\Delta_n: d_{TV}(\mathscr{L}_n(p),F_n)\ge \epsilon\big\}\Big)\to 0
\]
as $n\to\infty$. Then, $d_{TV}(B_n,F_n)\to 0$ as $n\to\infty$.
\end{lemma}
The proof of Lemma \ref{lem:unique_generic_center} is also  straightforward, so it in Online Appendix \hyperref[app:proof]{C}.

\begin{proof}[Proof of Theorem \ref{thm:generic2}]
Fix $\kappa>0$. Choose a SESI $\theta_n:=\theta_{n,G^{\mathrm{MLE}}}(p_0)$ in Lemma \ref{thm:sesigen}, so we have $(\theta_n)_{n\ge1}\in\Theta$. By definition of the set of all possible SESIs $\mathcal{S}_n$,  
 $\mathscr{B}_n(\theta_n)\in\mathcal{S}_n$ $\forall n$.

\medskip
\noindent\textit{(1)} Fix $\epsilon>0$. Since $(\theta_n)_{n\ge1}\in\Theta$, Lemma \ref{thm:dir} implies
\[
\nu_{\kappa,n}\Big(\Big\{p\in\Delta_n: d_{TV}\big(\mathscr{L}_n(p),\mathscr{B}_n(\theta_n)\big)\le \epsilon\Big\}\Big)\to 1
\qquad\text{as }n\to\infty.
\]
For any $p\in\Delta_n$, if $d_{TV}\big(\mathscr{L}_n(p),\mathscr{B}_n(\theta_n)\big)\le \epsilon$, then (since
$\mathscr{B}_n(\theta_n)\in\mathcal{S}_n$) we have $\mathscr{L}_n(p)\in\mathcal{S}_n^\epsilon$ by the definition of the set $\mathcal{S}_n^\epsilon$.
Thus,
\[
\Big\{p\in\Delta_n: d_{TV}\big(\mathscr{L}_n(p),\mathscr{B}_n(\theta_n)\big)\le \epsilon\Big\}
\subseteq
\Big\{p\in\Delta_n: \mathscr{L}_n(p)\in\mathcal{S}_n^\epsilon\Big\}.
\]
Taking complements and using the preceding convergence,
\[
\nu_{\kappa,n}\Big(\Big\{p\in\Delta_n: \mathscr{L}_n(p)\notin\mathcal{S}_n^\epsilon\Big\}\Big)\to 0.
\]
Finally, since $\{p: \mathscr{L}_n(p)\in\mathcal{C}_n\setminus\mathcal{S}_n^\epsilon\}\subseteq \{p: \mathscr{L}_n(p)\notin\mathcal{S}_n^\epsilon\}$,
we obtain
\[
\nu_{\kappa,n}\Big(\Big\{p\in\Delta_n: \mathscr{L}_n(p)\in\mathcal{C}_n\setminus\mathcal{S}_n^\epsilon\Big\}\Big)\to 0.
\]

\medskip
\noindent\textit{(2)} Suppose $\mathscr{L}^*_n\in\mathcal{C}_n$ is $\nu_{\kappa,n}$-generic. By Definition \ref{def:genericity}, for every $\epsilon>0$,
\[
\nu_{\kappa,n}\Big(\Big\{p\in\Delta_n: d_{TV}\big(\mathscr{L}_n(p),\mathscr{L}^*_n\big)\ge \epsilon\Big\}\Big)\to 0.
\]
On the other hand, since $(\theta_n)_{n\ge1}\in\Theta$, Lemma \ref{thm:dir} implies that for every $\epsilon>0$,
\[
\nu_{\kappa,n}\Big(\Big\{p\in\Delta_n: d_{TV}\big(\mathscr{L}_n(p),\mathscr{B}_n(\theta_n)\big)\ge \epsilon\Big\}\Big)\to 0.
\]
Applying Lemma \ref{lem:unique_generic_center} yields $d_{TV}\big(\mathscr{L}^*_n,\mathscr{B}_n(\theta_n)\big)\to 0$ as $n\to\infty$.
\end{proof}

\subsection*{Proof of Proposition \ref{thm:test}}
\begin{proof}
We work under the multi-stage experiment in which
$p\sim\lambda_n$ is first drawn from the normalized Lebesgue measure
on $\Delta_n$, and then, conditional on $p$, we observe the independent
sample $S_n^{(1)},\dots,S_n^{(N)}$ from $\mathscr{L}_n(p)$.

Fix $\epsilon>0$, $\alpha\in(0,1)$, and $n,N\ge 1$. Let
$k\in\{0,\dots,n\}$ be fixed. We study the deviation
$\bigl|P_N(k)-\mathscr{B}_n(\theta_n)(k)\bigr|$.

\medskip\noindent
\emph{Step 1: Decomposition of the error event.}
For any $p\in\Delta_n$ we have
\[
\bigl|P_N(k)-\mathscr{B}_n(\theta_n)(k)\bigr|
\le \bigl|P_N(k)-\mathscr{L}_n(p)(k)\bigr|
   +\bigl|\mathscr{L}_n(p)(k)-\mathscr{B}_n(\theta_n)(k)\bigr|.
\]
Hence, for every $a>0$,
\[
\bigl\{|P_N(k)-\mathscr{B}_n(\theta_n)(k)|>a\bigr\}
\subset
\bigl\{|P_N(k)-\mathscr{L}_n(p)(k)|>\tfrac{a}{2}\bigr\}
 \cup 
\bigl\{|\mathscr{L}_n(p)(k)-\mathscr{B}_n(\theta_n)(k)|>\tfrac{a}{2}\bigr\}.
\]
For brevity, denote
\[
E_1(p) := \Bigl\{|P_N(k)-\mathscr{L}_n(p)(k)|>\tfrac{a}{2}\Bigr\},\qquad
E_2(p) := \Bigl\{|\mathscr{L}_n(p)(k)-\mathscr{B}_n(\theta_n)(k)|>\tfrac{a}{2}\Bigr\}.
\]

\medskip\noindent
\emph{Step 2: Sampling fluctuation bound.}
Conditional on $p$, the variables
\[
Z_i := \mathds{1}_{S_n^{(i)}=k},\qquad i=1,\dots,N,
\]
are independent Bernoulli variables with mean $\mathscr{L}_n(p)(k)$. Then,
$P_N(k)=(1/N)\sum_{i=1}^N Z_i$, and Hoeffding's inequality implies that
for any $t>0$,
\[
\mathbb{P}_p\bigl(|P_N(k)-\mathscr{L}_n(p)(k)|>t\bigr)
\le 2\exp(-2Nt^2),
\]
where $\mathbb{P}_p$ denotes probability conditional on $p$. Applying this with $t=a/2$a:
\[
\mathbb{P}_p(E_1(p))
= \mathbb{P}_p\Bigl(|P_N(k)-\mathscr{L}_n(p)(k)|>\tfrac{a}{2}\Bigr)
\le 2\exp\Bigl(-\frac{Na^2}{2}\Bigr).
\]
If we choose $a$ so that $a^2  \ge  \frac{2}{N} \log\big(\frac{4}{\alpha}\big),$
then
\[
2\exp\Bigl(-\frac{Na^2}{2}\Bigr)
\le 2\exp\Bigl(-\log\Bigl(\frac{4}{\alpha}\Bigr)\Bigr)
= \frac{\alpha}{2}.
\]
Thus, for such $a$, we have the uniform bound
\begin{equation}\label{eq:sample-bound}
\mathbb{P}_p(E_1(p))\le \frac{\alpha}{2}
\quad\text{for all }p\in\Delta_n.
\end{equation}

\medskip\noindent
\emph{Step 3: Model deviation bound.}
We now control the second term $E_2(p)$ under the Lebesgue prior
$\lambda_n$. By Lemma \ref{thm:lebesgue}, for every
$\epsilon>0$, there exists $n_0(\epsilon)\ge 1$ such that for all
$n\ge n_0(\epsilon)$,
\[
\lambda_n\Bigl(
  \bigl\{p\in\Delta_n :
    \max_{0\le j\le n}\big|\mathscr{L}_n(p)(j)-\mathscr{B}_n(\theta_n)(j)\big|>\epsilon
  \bigr\}
  \Bigr)
 \le  \frac{4}{\epsilon^2 (2^n+1)}.
\]
In particular, for any fixed coordinate $k$ and any $\epsilon>0$,
\[
\lambda_n\Bigl(
  \bigl\{p\in\Delta_n :
    |\mathscr{L}_n(p)(k)-\mathscr{B}_n(\theta_n)(k)|>\epsilon
  \bigr\}
  \Bigr)
\le \frac{4}{\epsilon^2 (2^n+1)}
\]
for all $n\ge n_0(\epsilon)$, since the event inside is contained in the
event that the sup–norm exceeds $\epsilon$. Now set $\epsilon=a/2$. For all integers $n\ge n_0(a/2)$, we get
\[
\lambda_n\Bigl(
  \bigl\{p\in\Delta_n :
    |\mathscr{L}_n(p)(k)-\mathscr{B}_n(\theta_n)(k)|>\tfrac{a}{2}
  \bigr\}
  \Bigr)
\le \frac{4}{(\tfrac{a}{2})^2 (2^n+1)}
= \frac{16}{a^2 (2^n+1)}.
\]
If we choose $a$ so that $a^2  \ge  \frac{32}{\alpha (2^n+1)},$
then
\[
\frac{16}{a^2 (2^n+1)}
\le \frac{16}{\frac{32}{\alpha (2^n+1)} (2^n+1)}
= \frac{\alpha}{2},
\]
and hence, for all $n\ge n_0(a/2)$,
\begin{equation}\label{eq:model-bound}
\lambda_n\bigl(\{p\in\Delta_n : E_2(p)\text{ occurs}\}\bigr)
\le \frac{\alpha}{2}.
\end{equation}

\medskip\noindent
\emph{Step 4: Joint error probability and choice of $a(n,N,\alpha)$.}
We now combine the bounds \eqref{eq:sample-bound} and
\eqref{eq:model-bound}. Consider the joint probability of rejection in
the multi-stage experiment, i.e., with $p\sim\lambda_n$ and, given $p$,
$S_n^{(1)},\dots,S_n^{(N)}$ drawn from $\mathscr{L}_n(p)$. Define the test
\[
\phi_{n,N} := \mathds{1}\Bigl\{
  \bigl|P_N(k)-\mathscr{B}_n(\theta_n)(k)\bigr|>a
\Bigr\}.
\]
Then,
\[
\mathbb{P}(\phi_{n,N}=1)
= \int_{\Delta_n}\mathbb{P}_p\bigl(|P_N(k)-\mathscr{B}_n(\theta_n)(k)|>a\bigr) 
  d\lambda_n(p).
\]
Since
$\{|P_N(k)-\mathscr{B}_n(\theta_n)(k)|>a\}\subset E_1(p)\cup E_2(p)$ and the union
bound, we get
\[
\mathbb{P}(\phi_{n,N}=1)
\le
\int_{\Delta_n}\mathbb{P}_p(E_1(p)) d\lambda_n(p)
+\lambda_n\bigl(\{p\in\Delta_n:E_2(p)\text{ occurs}\}\bigr).
\]
By \eqref{eq:sample-bound}, the integrand in the first term is bounded
by $\alpha/2$ uniformly in $p$, so
\[
\int_{\Delta_n}\mathbb{P}_p(E_1(p)) d\lambda_n(p)
\le \frac{\alpha}{2}.
\]
By \eqref{eq:model-bound}, the second term is at most $\alpha/2$ for all
$n\ge n_0(a/2)$. Thus, if we choose $a$ so that
\[
a^2 \ge \frac{2}{N} \log\Bigl(\frac{4}{\alpha}\Bigr)
\quad\text{and}\quad
a^2 \ge \frac{32}{\alpha (2^n+1)},
\]
then, for all $n\ge n_0(a/2)$, $\mathbb{P}(\phi_{n,N}=1)
\le \frac{\alpha}{2}+\frac{\alpha}{2}
= \alpha.$ By definition, rejection happens at
\[
a(n,N,\alpha)
:= \max\!\left\{
  \sqrt{\frac{2}{N} \log\Bigl(\frac{4}{\alpha}\Bigr)},
   \sqrt{\frac{32}{\alpha (2^n+1)}}
\right\},
\]
so this choice satisfies both inequalities, and the test;
reject if $|P_N(k)-\mathscr{B}_n(\theta_n)(k)|>a(n,N,\alpha)$ has joint
rejection probability at most $\alpha$ for all
$n\ge n_0\bigl(a(n,N,\alpha)/2\bigr)$.
\end{proof}

\subsection*{Proof of Proposition \ref{thm:asinfo}}

\begin{proof}

We must show that $\varPhi_n(\theta) := 1 - \E_p\big[C_{n,Z}\big]$ converges to $1 - c(\theta)$ in probability, where $Z$ denotes the sample mean (as a random variable) of signals drawn from $p$ given $\theta$. Since $\theta_{\text{\tiny NE}}$ is the unique fixed point of $1-c(\theta)$ and $\theta_{n,G}(p)$ is a fixed point of $\varPhi_n$, this would imply that $\theta_{n,G}(p)$ converges to  $\theta_{\text{\tiny NE}}$ as $n\rightarrow\infty$.

\medskip\noindent\textit{Step 1: Decomposition of the Error.}
Consider $\big|\E_p[C_{n, Z}] - c(\theta)\big|.$ By triangle inequality, 
\[
\big|\E_p[C_{n, Z}] - c(\theta)\big| \le \underbrace{\E_p\big[|C_{n, Z} - c(Z)|\big]}_{\text{Inference Error}} + \underbrace{\E_p\big[|c(Z) - c(\theta)|\big]}_{\text{Sampling Error}}.
\]

\medskip\noindent\textit{Step 2: Bounding the Inference Error.}
Since $G$ is a convergent monotone inference procedure, the estimate $G_{n,z}$ converges in probability to a unit mass on the sample mean $z$ as $n \to \infty$. Thus, $C_{n, Z}=\int_0^1c(\vartheta)\hspace{0.02in}dG_{n,Z}(\vartheta)$ converges in probability to $c(Z)$ as $n\rightarrow\infty$, so $\E_p\big[|C_{n, Z} - c(Z)|\big] \to 0$ as $n\rightarrow\infty$ because $|C_{n, Z}|\leq1$ a.s. for all $n$.

\medskip\noindent\textit{Step 3: Bounding the Sampling Error.}
We now bound the second term, $\E_p\big[|c(Z) - c(\theta)|\big]$.
\begin{enumerate}
    \item \textit{Uniform Continuity}: Since $c(\cdot)$ is continuous on the compact set $[0, 1]$, it is uniformly continuous. Therefore, for any $\epsilon > 0$, there exists $\delta > 0$ such that whenever $|z - \theta| < \delta$, we have $|c(z) - c(\theta)| < \epsilon$.
    \item \textit{Consistency}: Since $p$ is \textit{asymptotically informative} (Definition \ref{def:asinfo}), the sample mean $Z$ is a consistent estimator for $\theta$. That is, for any $\delta > 0$:
    \[
    \Proba_n(|Z - \theta| \ge \delta) \to 0 \quad \text{as } n \to \infty.
    \]
    \item \textit{Bounding the Expectation}: Let $M := \sup_{x \in [0,1]} |c(x)|$. We decompose the expectation based on the event $\{|Z - \theta| < \delta\}$:
    \begin{align*}
    \E_p\big[|c(Z) - c(\theta)|\big] &= \E_p\Big[|c(Z) - c(\theta)| \cdot \mathds{1}_{\{|Z - \theta| < \delta\}}\Big] + \E_p\Big[|c(Z) - c(\theta)| \cdot \mathds{1}_{\{|Z - \theta| \ge \delta\}}\Big].
    \end{align*}
    For the first term, when $|Z - \theta| < \delta$, we have $|c(Z) - c(\theta)| < \epsilon$. For the second term, we use the upper bound $|c(Z) - c(\theta)| \le 2M$. Therefore:
    \[
    \E_p\big[|c(Z) - c(\theta)|\big] \le \epsilon \cdot \Proba_n(|Z - \theta| < \delta) + 2M \cdot \Proba_n(|Z - \theta| \ge \delta) \le \epsilon + 2M \cdot \Proba_n(|Z - \theta| \ge \delta).
    \]
\end{enumerate}

\medskip\noindent\textit{Step 4: Convergence to NE.}
Taking the limit as $n \to \infty$, since $\Proba_n(|Z - \theta| \ge \delta) \to 0$, the sampling error is bounded by $\epsilon$. Since $\epsilon$ was arbitrary, $\E_p[c(Z)] \to c(\theta)$.

Combining Step 2 and Step 3, we have: $\varPhi_n(\theta) = 1 - \E_p[C_{n, Z}] \overset{}{\longrightarrow} 1 - c(\theta)$ as $n\rightarrow\infty$.
Let $\varPhi_{\text{\tiny NE}}(\theta) = 1-c(\theta)$. Since $c(\theta)$ is increasing, $\varPhi_{\text{\tiny NE}}(\theta)$ is strictly decreasing, ensuring $\theta_{\text{\tiny NE}}$ is the unique fixed point of $\theta = \varPhi_{\text{\tiny NE}}(\theta)$.
Since $\varPhi_n(\theta)$ converges uniformly to $\varPhi_{\text{\tiny NE}}(\theta)$ as $n\rightarrow\infty$, $\theta_{n,G}(p)$ must converge to the unique fixed point $\theta_{\text{\tiny NE}}$ as $n\rightarrow\infty$.
\end{proof}

\subsection*{Proof of Proposition \ref{thm:cosesi}}
 We begin by analyzing some properties of the $\rho$-weighted Bernstein polynomial of $C_n$, $\varPsi_n(\theta,\rho;C_n)$, which we will use in later proofs. 
The idea is that $\varPsi_n(\theta,\rho;C_n)$ inherits and preserves the key properties of the Bernstein polynomial $\mathcal{B}_n(\theta;C_{n})$, for any $\rho\in[0,1]$. 
\begin{lemma}[Properties of $\rho$-weighted Bernstein Polynomials]\label{thm:cosesilemma} The following hold:
\begin{enumerate}
    \item $\varPsi_n(\theta,\rho;C_n)$ increases in $\theta$ when $C_n$ is increasing;
    \item $\varPsi_n(0,\rho;C_n)=C_{n,0}$ and $\varPsi_n(1,\rho;C_n)=C_{n,1}$, for all $\rho\in[0,1]$;
    \item $\varPsi_n(\theta,\rho;C_n)$ is convex (concave) in $\theta$ when $C_n$ is convex (concave);
    \item $\varPsi_n(\theta,\rho;C_n)$ is Lipschitz continuous in $\theta$;
    \item $\varPsi_n(\theta,\rho;C_n)=c(\theta)$ when $G$ is sample-unbiased and $c$ is linear, for all $\rho\in[0,1]$.
\end{enumerate}

\end{lemma}
\begin{proof}[Proof of Lemma \ref{thm:cosesilemma}] We prove this result one part at a time.
    \par---\textit{Lemma \ref{thm:cosesilemma}.1}:
 \citet[][Property 1]{stat} show that $\mathcal{B}_n(\theta;C_{n})$ is increasing when  $C_n$ is increasing. It remains to show that the other component $\mathbb{B}_n(\theta;C_{n})$ is increasing. This is easily seen by taking the first derivative: $\frac{\partial}{\partial\theta}\mathbb{B}_n(\theta;C_{n})=C_{n,1}-C_{n,0}\geq0$ since $C_{n}$ is increasing. Thus, when $C_n$ is increasing, $\varPsi_n(\theta,\rho;C_n)$ is a convex combination of two increasing functions in $\theta$, so it is itself increasing in $\theta$.
    \par---\textit{Lemma \ref{thm:cosesilemma}.2}: This follows immediately by noting that \begin{align*}
        \varPsi_n(0,\rho;C_n)&=(1-\rho)\mathcal{B}_n(0;C_{n})+\rho\mathbb{B}_n(0;C_{n})=(1-\rho)C_{n,0}+\rho C_{n,0}=C_{n,0}\\
        \varPsi_n(1,\rho;C_n)&=(1-\rho)\mathcal{B}_n(1;C_{n})+\rho\mathbb{B}_n(1;C_{n})=(1-\rho)C_{n,1}+\rho C_{n,1}=C_{n,1},
    \end{align*}
    because by definition $\mathcal{B}_n(0;C_{n})=C_{n,0}$ and $\mathcal{B}_n(1;C_{n})=C_{n,1}$.
 \par---\textit{Lemma \ref{thm:cosesilemma}.3}: When $C_n$ is convex (concave), $\varPsi_n(\theta,\rho;C_n)$ becomes a convex combination of two convex (concave) functions: $\mathcal{B}_n(\theta;C_{n})$ is convex by \citet[][Property 2]{stat} and a linear function $\mathbb{B}_n(\theta;C_{n})$, so $\varPsi_n(\theta,\rho;C_n)$ is convex (concave).
  \par---\textit{Lemma \ref{thm:cosesilemma}.4}: Because  $\mathcal{B}_n(\theta;C_{n})$ is a polynomial of degree $n$ in $\theta$ bounded on $[0,1]$, for $C_n\in[0,1]$,  it is Lipschitz in $\theta$. Thus, $\varPsi_n(\theta,\rho;C_n)$ is Lipschitz because it is a convex combination between two Lipschitz continuous functions: $\mathcal{B}_n(\theta;C_{n})$ and 
  $\mathbb{B}_n(\theta;C_{n})$.
  \par---\textit{Lemma \ref{thm:cosesilemma}.5}: When $G$ is sample-unbiased and $c$ is linear, we have $C_{n,z}=c$. Thus, $\mathcal{B}_{n}(\theta;C_{n})=\mathcal{B}_{n}(\theta;c)=c$ because the Bernstein polynomial of a linear function coincides with the function, so the $\rho$-weighted Bernstein polynomial becomes $$\varPsi_n(\theta,\rho;C_{n})=(1-\rho)\mathcal{B}_{n}(\theta;C_{n})+\rho\Big[\theta C_{n,1}+(1-\theta)C_{n,0}\Big]=(1-\rho)c(\theta)+\rho\Big[\theta c(1)+(1-\theta)c(0)\Big].$$ Then, $\varPsi_n(\theta,\rho;C_{n})=c(\theta)$ 
because when $c$ is linear, $c(\theta)=\theta c(1)+(1-\theta)c(0)$.
\end{proof}

\begin{proof}[Proof of Proposition \ref{thm:cosesi}]
  The probability of observing $y$ successes in $n$ samples with common correlation $\rho$ is $\mu^{\rho}_n(y|\theta)$ in eq. (\ref{eq:distpositive}). Conditional on observing $y$ successes, all agents with $\xi\leq C_{n,y/n}$ take action $B$, where $\xi\sim\mathcal{U}[0,1]$. Thus, the fraction of agents who observe $y$ successes and take the action $B$ is $\mu^{\rho}_n(y|\theta)C_{n,y/n}$, so summing over all possible successes $y$ yields the total measure of agents taking the action $B$, which is $\sum_{y=0}^n\mu^{\rho}_n(y|\theta)C_{n,y/n}$. Putting all this together, the equilibrium equation that any CoSESI must satisfy becomes
\begin{align}
1-\theta&=\sum_{y=0}^n\mu^{\rho}_n(y|\theta)C_{n,y/n}\nonumber\\
&=\sum_{y=0}^n\Bigg\{(1-\rho)\binom{n}{y}\theta^y(1-\theta)^{n-y}+\rho\Big[ \theta\mathds{1}_{y=n}+(1-\theta)\mathds{1}_{y=0}\Big]\Bigg\}C_{n,y/n}\nonumber\\
&=(1-\rho)\underbrace{\sum_{y=0}^n\binom{n}{y}\theta^y(1-\theta)^{n-y}C_{n,y/n}}_{\mathcal{B}_n(\theta;C_n)}+\rho\sum_{y=0}^n\Big[ \theta\mathds{1}_{y=n}+(1-\theta)\mathds{1}_{y=0}\Big]C_{n,y/n}\nonumber\\
&=(1-\rho)\mathcal{B}_n(\theta;C_n)+\rho\underbrace{\Big[\theta C_{n,1}+(1-\theta)C_{n,0}\Big]}_{\mathbb{B}_n(\theta;C_n)}\nonumber\\
&=(1-\rho)\mathcal{B}_n(\theta;C_{n})+\rho\mathbb{B}_n(\theta;C_{n})\nonumber\\
&=\varPsi_n(\theta,\rho;C_{n}).\label{eq:equi2}
\end{align}
Since the cost $c$ is increasing, then so is $C_n$, and hence we can apply Lemma \ref{thm:cosesilemma}.1 to conclude that $\varPsi_n(\theta,\rho;C_{n})$ is increasing in $\theta$. Further, $0\leq \varPsi_n(0,\rho;C_{n})<1$ because $\varPsi_n(0,\rho;C_{n})=(1-\rho)\mathcal{B}_n(0;C_n)+\rho C_{n,0}=C_{n,0}<C_{n,1}\leq1$, where the second equality follows by Lemma \ref{thm:cosesilemma}.2, and similarly, $0<\varPsi_n(1,\rho;C_{n})\leq1$. Lastly, since $1-\theta$ is strictly decreasing in $\theta$, there is a unique fixed point to eq. (\ref{eq:equi2}) on $[0,1]$, for any $\rho\in[0,1]$.
\end{proof}

\subsection*{Proof of Proposition \ref{thm:noconv}}
\begin{proof}
We want to show that $\varPsi_n(\theta,\rho;C_n)$ in eq. (\ref{eq:equi}) does not converge to $c(\theta)$ for convergent inference procedures. This follows by observing that as $n\rightarrow\infty$, $\mathcal{B}_n(\theta;C_n)\overset{}{\rightarrow} c(\theta)$ \citep[][Observation 2]{stat} and $\mathbb{B}_n(\theta;C_n)\overset{}{\rightarrow}\theta c(1)+(1-\theta)c(0)\neq c(\theta)$ when $c$ is not linear. Thus, when $\rho>0$ and $c$ is not linear:$$\varPsi_n(\theta,\rho;C_n)\overset{}{\longrightarrow}(1-\rho)c(\theta)+\rho\Big[\theta c(1)+(1-\theta)c(0)\Big]\neq c(\theta).$$

\end{proof}

\subsection*{Proof of Proposition \ref{thm:compNE}}
\begin{proof}
   ---Proposition \ref{thm:compNE}.1: Fix a sample-unbiased $G$.  The inequality is proved in two steps. (1) \citet[][Theorem 2]{stat} show that $\mathcal{B}_n(\theta;C_n)\geq c(\theta)$, for any sample-unbiased $G$. (2) By sample-unbiasedness and convexity, $\mathbb{B}_n(\theta;C_n)\geq c(\theta)$ because any estimate of $G$ is a mean-preserving spread of the corresponding MLE estimate for any sample $(n,z)$ \citep[][Property 3]{stat}, and the expected cost function with respect to MLE is $c$ itself. Since eq. (\ref{eq:equi}) and $1-\theta=c(\theta)$ have the same l.h.s, comparing their r.h.s by putting (1) and (2) together gives $\varPsi_n(\theta,\rho;C_n)\geq c(\theta)$, so $\theta^{(\rho)}_{n,G}\leq\theta_{\text{\normalfont\tiny NE}}$ for all $n$ and $\rho\in[0,1]$.
 \par\noindent---Proposition \ref{thm:compNE}.2: When $n\rightarrow\infty$, the Bernstein polynomial  $\mathcal{B}_n(\theta;C_n)$ converges to $C_\infty$ \citep[][]{phil}, which equals $c(\theta)$ because $G$ is convergent. Thus, $\theta^{(\rho)}_{\infty,G}\leq \theta_{\text{\tiny NE}}$ because $\mathbb{B}_\infty(\theta;C_\infty)=\theta c(1)+(1-\theta)c(0)\geq c(\theta)$ by convexity.   Let's show that $\varPsi_\infty(\theta,\rho;C_\infty)$ is an increasing function of $\rho$, for all $\theta$. This function is linear in $\rho$, so its first derivative is $\frac{\partial}{\partial \rho}\varPsi_\infty(\theta,\rho;C_\infty)=\mathbb{B}_\infty(\theta;C_\infty)-\mathcal{B}_\infty(\theta;C_\infty)$. Notice that this difference is nonnegative: 
    \begin{align*}
        \frac{\partial}{\partial \rho}\varPsi_\infty(\theta,\rho;C_\infty)= \theta c(1)+(1-\theta)c(0)-c(\theta)\geq0,
    \end{align*}
    where we use the fact that $G$ is convergent and $c$ is convex for the inequality.
\end{proof}

\subsection*{Proof of Observation \ref{thm:monoCoSESI}}
\begin{proof}
    The joint distribution $p^\phi(.|\theta)$ is continuous in $\theta\in(0,1)$ for all $n$ and $\phi$ \citep[see,][eq. (2.6)]{klotz73}. By Proposition \ref{thm:coscont}, a CoSESI exists for all $\psi$, $\phi$, $n$, and $G$.
\end{proof}
\subsection*{Proof of Proposition \ref{thm:monocomp}}

\begin{proof}
  \par --- Proposition \ref{thm:monocomp}.1: By \citeauthor{klotz73}'s (\citeyear[][eq. (5.4)]{klotz73}), when $\phi=\theta+\frac{\rho^{1/t}}{1-\theta}$ for all $t\geq1$ and $\rho\in[0,1)$, $p^{\phi}=p_{\rho}$ in eq. (\ref{eq:prho}). This means that CoSESI becomes simple CoSESI.   For each price $\psi$ and correlation $\rho$, a unique simple CoSESI exists and solves the equation 
\begin{align*}
\theta=\sum_{y=0}^n\mu^{\rho}_n(y|\theta)\hspace{0.03in}\text{max}\big\{C_{n,y/n}-\psi,0\big\}\iff
    1-\theta=\sum_{y=0}^n\mu^{\rho}_n(y|\theta)\varLambda_{\psi,G}(y/n),
\end{align*}
where we recall that $\varLambda_{\psi,G}(y/n):=\text{min}\{1+\psi-C_{n,y/n},1\}$ (Proposition \ref{thm:cosesi} and eq. (\ref{eq:equi})). 
\par --- Proposition \ref{thm:monocomp}.2: Since $c_t(\theta)=(1-\theta)^t$ is convex and decreasing for all $t\geq0$, and $G$ preserves shape, then $C_{n}(.):=C_{n,.}$ is convex and decreasing, so $\varLambda_{\psi,G}(y/n)$ is concave and increasing. We can therefore apply Proposition \ref{thm:compNE}.1 because the inequality in its statement reverses under concavity, which holds here. For each $\psi$ and $\rho$, CoSESI demand is therefore larger than NE demand, where the latter solves $1-\theta=c_t(\theta)+\psi$. This then implies the ranking of the monopolist's profit. When $\rho=0$ and $n=1$, SESI coincides with CoSESI when $\rho=1$ for any $n$. Thus, when $\rho=1$, CoSESI is larger than SESI for all $n>1$ by \citet[][Theorem 3]{stat}.
\par --- Proposition \ref{thm:monocomp}.3: apply Proposition \ref{thm:compNE}.2 using the same reasoning above. \end{proof}

\subsection*{Proof of Observation \ref{thm:assinfomono}}
\begin{proof}
    When $\phi\neq\theta+\frac{\rho^{1/t}}{1-\theta}$ for all $t\geq1$ and $\rho\in[0,1)$, $p^{\phi}\neq p_{\rho}$, so we are in \citeauthor{klotz73}'s (\citeyear[][eq. (5.4)]{klotz73}) Markov dependence structure. By \citet[][Theorem]{klotz73}, the sample mean $z$ is a consistent estimator for $\theta$, so by Definition \ref{def:asinfo}, $p^{\phi}:=p^{\phi}(.|\theta)$ is asymptotically informative for $\theta$.  By Proposition \ref{thm:asinfo}, CoSESI converges to NE in large sample, so monopoly profit in CoSESI also converges to monopoly profit in NE.
\end{proof}

\subsection*{Proof of Proposition \ref{thm:2cosesi}}
The system of equations for simple CoSESI in Section \ref{sec:twosided} can be simplified to
\begin{align}
    \alpha^{(\rho)}_{k,G^w}(\beta)&=\Bigg(\zeta\sum_{r=0}^{k}\Bigg[(1-\rho)\binom{k}{r}\beta^r(1-\beta)^{k-r}+\rho\Big( \beta\mathds{1}_{r=k}+(1-\beta)\mathds{1}_{r=0}\Big)\Bigg] M^w_{k,r/k}\Bigg)^{\frac{1}{2-v}}\nonumber\\\label{eq:emp2}&=\varPsi_k\big(\beta,\rho;M^w_{k}\big)^{\frac{1}{2-v}}\\
    \beta^{(\varphi)}_{n,G^f}(\alpha)&=\Bigg(\zeta\sum_{s=0}^{n}\Bigg[(1-\varphi)\binom{n}{s}\alpha^s(1-\alpha)^{n-s}+\varphi\Big( \alpha\mathds{1}_{s=n}+(1-\alpha)\mathds{1}_{s=0}\Big)\Bigg]M^f_{n,s/n}\Bigg)^{\frac{1}{1+v}}\nonumber\\\label{eq:firm2}  &=\varPsi_n\big(\beta,\varphi;M^f_{n}\big)^{\frac{1}{1+v}}.
\end{align}

\begin{proof}
    We aim to show that there exists a unique pair $\big(\alpha^{(\rho)}_{k,G^w},\beta^{(\varphi)}_{n,G^f}\big)\in[0,1]^2$ that solves eqs. (\ref{eq:emp2})-(\ref{eq:firm2}) with positive employment, for all pairs of correlation $(\rho,\varphi)\in[0,1]^2$. We first note that $\alpha^{(\rho)}_{k,G^w}(\beta)$ in eq. (\ref{eq:emp2}) is strictly concave and strictly increasing in $\beta$. This is because (1) $\beta^{1-v}$ is concave and strictly increasing in $\beta$, so (2) $M^w_{k,z}$ is concave and strictly increasing in $z$ because $G_w$ preserves shape, and therefore (3) the $\rho$-weighted Bernstein polynomial $\varPsi_k(\beta,\rho;M^w_{k})$ is strictly increasing and concave by Lemma \ref{thm:cosesilemma}.1 and \ref{thm:cosesilemma}.3, respectively. Thus, $\varPsi_k(\beta,\rho;M^w_{k})$ raised to the power of $\frac{1}{2-v}<1$ is strictly increasing and strictly concave in $\beta$. A similar argument establishes that $\beta^{(\varphi)}_{n,G^f}(\alpha)$ in eq. (\ref{eq:firm2}) is strictly concave and strictly increasing in $\alpha$, which implies that its inverse function $\hat{\alpha}^{(\varphi)}_{n,G^f}(\beta)$ is strictly convex and strictly increasing in $\beta$.
    \par Given the above, $\beta$ is part of a simple CoSESI if and only if it is a point in which the functions $\alpha^{(\rho)}_{k,G^w}(\beta)$ and $\hat{\alpha}^{(\varphi)}_{n,G^f}(\beta)$ intersect. Notice that $\alpha^{(\rho)}_{k,G^w}(\beta)$
    is strictly concave and $\hat{\alpha}^{(\varphi)}_{n,G^f}(\beta)$ is strictly convex and both are increasing, so they intersect in at most one positive point. Thus, such a point exists for all $(\rho,\varphi)\in[0,1]^2$ because: when $\beta=0$, we use Lemma \ref{thm:cosesilemma}.2 to find that $\alpha^{(\rho)}_{k,G^w}(0)=\big(\zeta M^w_{k,0}\big)^{\frac{1}{2-v}}=0=\hat{\alpha}^{(\rho)}_{n,G^f}(0)$, and when $\beta=1$, $\alpha^{(\rho)}_{k,G^w}(1)=\zeta^{\frac{1}{2-v}}<\hat{\alpha}^{(\rho)}_{n,G^f}(1)$. We have therefore proved Proposition \ref{thm:2cosesi}.1.
    \par A unique NE with positive employment exists \citep[][Theorem 7]{stat}. By sample-unbiasedness, Proposition \ref{thm:compNE}.1 can be used to show that the
function $\alpha^{(\rho)}_{k,G^w}(\beta)$ lies below the function $\alpha_{\text{\tiny NE}}(\beta)$, for any $\beta\in(0,1)$. Similarly,  $\hat{\alpha}^{(\varphi)}_{n,G^f}(\beta)$ lies above $\hat{\alpha}_{\text{\tiny NE}}(\beta)$. This implies the
ranking of simple CoSESI and NE, which proves Proposition \ref{thm:2cosesi}.2. 
\par Lastly, Proposition \ref{thm:compNE}.2 shows that if $\rho$ increases to $\rho'$, then the
function $\alpha^{(\rho')}_{\infty,G^w}(\beta)$ lies below the function $\alpha^{(\rho)}_{\infty,G^w}(\beta)$, for any $\beta\in(0,1)$. Similarly, if $\varphi$ increases to $\varphi'$, then $\hat{\alpha}^{(\varphi')}_{\infty,G^f}(\beta)$ lies above $\hat{\alpha}^{(\varphi)}_{\infty,G^f}(\beta)$. Thus, market thickness and employment decrease.
\end{proof}

\begin{singlespace}
\bibliography{ref}
\bibliographystyle{apalike}
\end{singlespace}

\clearpage

\phantomsection\label{app:online}
\begin{center}
{\LARGE\bf Online Appendix: \par ``Aggregate Efficiency in Games''}
\end{center}

\par This online appendix contains the microfoundations, extensions, and proofs.
Online Appendix \hyperref[app:micro]{A} provides the microfoundations: \hyperref[app:network]{A.I} derives the correlated binomial distribution from strategic network formation; \hyperref[app:market]{A.II} uses a competitive market to interpret aggregate efficiency; and \hyperref[app:learning]{A.III} establishes CoSESI as the steady state of learning dynamics.
Online Appendix \hyperref[app:ext]{B} explores various extensions: \hyperref[app:multi]{B.I} extends CoSESI to multi-action games; \hyperref[app:loc]{B.II} characterizes multiplicity of CoSESI in games of strategic complements; \hyperref[app:assort]{B.III} introduces assortative sampling; \hyperref[app:hetero]{B.IV} allows for agent heterogeneity; and \hyperref[app:ABEE]{B.V} formulates ABEE as a CoSESI.
Online Appendix \hyperref[app:proof]{C} contains all omitted proofs.

\phantomsection\label{app:micro}
\section*{Online Appendix A: Microfoundations}

\phantomsection\label{app:network}
\subsection*{Online Appendix A.I: Network Microfoundation}

We provide a strategic network microfoundation for the correlated binomial sampling distribution $\mu_n^{\rho}(\cdot|\theta)$ in eq. (\ref{eq:distpositive}) to demonstrate how  correlation structures in our framework can arise from strategic network formations and communications. This microfoundation has four  key steps: (i) agents play a \textit{sociability game} by forming links strategically, which generates a random network with connected components; (ii) inside each component, agents play a simple \emph{source--selection game} that determines the probability of relying on a common information source; (iii) conditional on this choice, signals are generated according to the \emph{process with correlated signals} of \citet{stra20}; and (iv) under the truth-telling conditions in their Propositions 1 and 4, the resulting sampling distribution of the number of observed actions coincides with $\mu_n^{\rho}(\cdot|\theta)$ in eq. (\ref{eq:distpositive}).

Recall that $\theta\in[0,1]$ is the fraction of agents taking action $A$ in the population, and for each sample size $n\ge2$ the joint distribution of local actions $(x_i)_{i=1}^n$ conditional on $\theta$ is given by $p_\rho(\cdot|\theta)$ in eq. (\ref{eq:prho}), for $\rho\in[0,1]$. The induced sampling distribution of the sum $Y=\sum_{i=1}^n x_i$ is the correlated binomial distribution $\mu_n^{\rho}(\cdot|\theta)$ in eq. (\ref{eq:distpositive}). Our goal is to show how the correlation parameter $\rho$ can be interpreted as an equilibrium object determined by network formation and strategic communication.

\subsection*{A.I.1 Network formation and component sizes.}

Fix $N\ge2$ and let $V_N=\{1,\dots,N\}$ be the finite set of agents. A \emph{network} on $V_N$ is an undirected simple graph $\mathbb{G}_N=(V_N,E_N)$, where $E_N\subseteq\{\{i,j\}:i,j\in V_N, i\neq j\}$ is the set of undirected links. For distinct agents $i,j\in V_N$ we write $i\sim j$ if $\{i,j\}\in E_N$.

\begin{definition}\label{def:sociability}
Each agent $i\in V_N$ chooses a \emph{sociability} level $\sigma_i\in[0,1]$. Given a profile $\sigma=(\sigma_i)_{i\in V_N}$, links are formed independently across unordered pairs, with
\[
\Proba\big(\{i,j\}\in E_N\mid\sigma\big)=\sigma_i\sigma_j\quad\text{for all }i\neq j.
\]
\end{definition}
This is a probabilistic bilateral-consent reduced form: the probability of a link is the product of the two agents' intensities \citep[e.g.,][]{parkes05}. In a symmetric profile, this reduces to the Erd\H{o}s--R\'enyi random graph $K_{N,p}$, where each edge is present independently with the same probability $p$. Given $\sigma$, the \emph{expected degree} of agent $i$ is
\[
d_i(\sigma):=\sum_{j\neq i}\sigma_i\sigma_j
=\sigma_i\sum_{j\neq i}\sigma_j.
\]
Agent $i$'s payoff from sociability is given by
\[
U_i(\sigma):=B\big(d_i(\sigma)\big)-D\big(d_i(\sigma)\big),
\]
where $B:[0,\infty)\to\mathbb{R}$ is increasing and strictly concave (benefits from links), and $D:[0,\infty)\to\mathbb{R}$ is increasing and strictly convex (costs of maintaining links).

\begin{assumption}\label{ass:BC}
The functions $B(.)$ and $D(.)$ are twice continuously differentiable, $B'(0)>D'(0)$ and $B'(d)<D'(d)$ for all sufficiently large $d$.
\end{assumption}

Assumption \ref{ass:BC} ensures that there is a unique interior level of expected degree where marginal benefits equal marginal costs.

\begin{lemma}[Symmetric equilibrium network]\label{lem:symmetric_network}
Under Assumption \ref{ass:BC}, for each $N\ge2$ there exists a unique symmetric Nash equilibrium of the sociability game in which all agents choose the same $\sigma^*_N\in(0,1)$. In this equilibrium,
\[
d^*_N:=d_i(\sigma^*_N)
=(N-1)(\sigma^*_N)^2
\]
is the unique solution in $(0,\infty)$ to the first-order condition
\begin{equation}\label{eq:FOC_network}
B'(d)=D'(d).
\end{equation}
Moreover, conditional on $\sigma^*_N$, the resulting network $\mathbb{G}_N$ is an Erd\H{o}s--R\'enyi graph
\[
\mathbb{G}_N\sim K_{N,p_N},\qquad p_N:=\frac{d^*_N}{N-1}\in(0,1).
\]
\end{lemma}

The Erd\H{o}s--R\'enyi network $K_{N,p_N}$ decomposes into connected components. For $i\in V_N$, let $C(i)$ denote the connected component of $\mathbb{G}_N$ containing $i$, and let
\[
S(i):=|C(i)|
\]
be its size. If $I_N$ is drawn uniformly from $V_N$, we define the random component size
\[
S_N:=S(I_N)=|C(I_N)|.
\]
Economically, $S_N$ is the number of peers with whom a typical agent is indirectly connected through the equilibrium network.

\subsection*{A.I.2 Source selection inside components.}

Fix a realization of $\mathbb{G}_N$ and a connected component $C\subseteq V_N$ of size $s:=|C|\ge2$. We now define a simple strategic game in which agents in $C$ choose whether to rely on independent or common information sources.

\begin{definition}\label{def:source_game}
Given a component $C$ of size $s\ge2$, each agent $j\in C$ chooses an action $a_j\in\{\mathrm{I},\mathrm{K}\}$, where:
\begin{itemize}
\item $a_j=\mathrm{I}$ means that $j$ will later rely on an \emph{independent} information source;
\item $a_j=\mathrm{K}$ means that $j$ will later rely on a \emph{common} information source shared with all other agents in $C$ who also choose $\mathrm{K}$.
\end{itemize}
Let $m(a):=\sum_{j\in C}\mathds{1}_{\{a_j=\mathrm{K}\}}$ denote the number of agents choosing the common source in profile $a=(a_j)_{j\in C}$.
\end{definition}

We specify payoffs so that there is a tension between coordination (benefits from belonging to a large common-source group) and differentiation (benefits from having an independent source when others coordinate).

Fix parameters $\beta>0$ and $c>0$. In a profile $a$ with $m=m(a)$, the payoff of agent $j\in C$ is
\begin{equation}\label{eq:source_payoffs}
u_j(a):=
\begin{cases}
\beta \dfrac{m}{s}-c & \text{if }a_j=\mathrm{K},\\[0.5em]
\beta \dfrac{s-m}{s} & \text{if }a_j=\mathrm{I}.
\end{cases}
\end{equation}
If many agents choose $\mathrm{K}$, then belonging to the common-source group yields a high benefit $\beta m/s$, but at the cost $c$. If few agents choose $\mathrm{K}$, an agent who chooses $\mathrm{I}$ enjoys the benefit $\beta(s-m)/s$ of being relatively unique. This payoff is a reduced-form representation of strategic complementarities and induces a well-defined mixed equilibrium.

\begin{lemma}[Symmetric mixed equilibrium in a component]\label{lem:qs}
Fix $s\ge2$ and consider the source-selection game in $C$ with payoffs \eqref{eq:source_payoffs}. There exists a unique symmetric mixed Nash equilibrium in which each agent $j\in C$ chooses $a_j=\mathrm{K}$ with probability $q_s\in(0,1)$ and $a_j=\mathrm{I}$ with probability $1-q_s$. The equilibrium mixing probability $q_s$ is given by
\begin{equation}\label{eq:qs}
q_s
=\frac{s-1+\frac{cs}{\beta}}{2(s-1)}.
\end{equation}
For any given $s$, $\beta>0$ and $c>0$ can be chosen so that $q_s\in(0,1)$.
\end{lemma}

Given $q_s$, the probability that all $s$ agents in $C$ choose the common source is
\begin{equation}\label{eq:ks}
k(s):=\Proba\big(a_j=\mathrm{K} \forall j\in C\big)=q_s^s\in[0,1).
\end{equation}
We interpret $k(s)$ as the \emph{equilibrium probability} that information in $C$ is generated from a single common source. This creates ``informational homophily.'' The restriction $k(s)\ge0$ ensures that the correlated-signal process of \citet{stra20} is well-defined.

\subsection*{A.I.3 Information structure and truth-telling.}

We now embed, inside each component $C$, the process with correlated signals and the cheap-talk game analyzed by \citet{stra20}. Let $C$ be a component of size $s\ge2$, and conditional on $s$, fix a sample size $n\leq s$.\footnote{Fixing $n\leq s$ may be interpreted as agents acquiring small samples of their neighbors. However, this assumption is not necessary for this analysis because Online Appendix \hyperref[app:hetero]{B.IV} extends the CoSESI framework to allow agents to have heterogeneous sample sizes $n_i$ that depend on their local environment. Under that extension, agents in small components ($s < n$) would simply acquire a smaller sample $n_i = s$.} We interpret the $n$ agents whose actions will be observed by a focal agent as the $n$ \emph{senders} in the \citet{stra20} model, and a distinguished agent (possibly the focal agent herself) as the \emph{receiver}.

Conditional on $\theta\in(0,1)$ and the component size $s$, we adopt the signal generation process of \citet[][Section 2.2]{stra20} to propose a two-step generation process:

\begin{enumerate}
\item \emph{Regime selection.} Draw a Bernoulli random variable $W_C\in\{0,1\}$ such that
\[
\Proba(W_C=1\mid |C|=s)=k(s),\qquad \Proba(W_C=0\mid |C|=s)=1-k(s),
\]
with $k(s)$ as in \eqref{eq:ks}.
\item \emph{Signal generation.} Conditional on $W_C$ and $\theta$:\footnote{\citet[][p. 2191]{stra20} describe this signal generation as follows: ``The process generating $\text{Pr}(\m{s}|\theta)$ is then one in which senders either collect information from $n$ independent sources with probability $1-k$ or, with probability $k$, they collect information from perfectly correlated sources.'' In our notation, $\m{s}=(s_i)_{i=1}^n$.}
\begin{itemize}
\item If $W_C=0$, draw $n$ independent Bernoulli$(\theta)$ signals $(s_1,\dots,s_n)$.
\item If $W_C=1$, draw a single Bernoulli$(\theta)$ variable $Z_C$ and set $s_i:=Z_C$ for all $i=1,\dots,n$.
\end{itemize}
\end{enumerate}
The resulting joint distribution of $(s_i)_{i=1}^n$ conditional on $\theta$ and $|C|=s$ is exactly the distribution $p_{\rho}(\cdot|\theta)$ in eq. (\ref{eq:prho}) with correlation parameter
\[
\rho=k(s)\in[0,1).
\]
This is formalized in the following result.

\begin{lemma}\label{lem:equiv_prho}
Fix $\theta\in(0,1)$ and $s\ge n\geq 2$. Let $(s_i)_{i=1}^n$ be generated by the two-step process above. Then, conditional on $\theta$ and $|C|=s$, the joint distribution of $(s_i)_{i=1}^n$ is $p_{k(s)}(\cdot|\theta)$, for $k(s)\in[0,1]$, as in eq. (\ref{eq:prho}).
\end{lemma}

The restriction $k(s)\in[0,1)$ ensures that this process corresponds to the case of \emph{nonnegative} correlation studied by \citet{stra20}. In particular, negative values of $k$ would violate the nonnegativity of probabilities in eq. (10) of their paper and are thus excluded in our microfoundation.

We now place the \citet{stra20} cheap-talk game on top of this information structure. For each component $C$ and a given sample size $n\ge2$, consider $n$ senders $i=1,\dots,n$ and one receiver $R$. Sender $i$ has ideal bias $b_i\in\mathbb{R}$, while the receiver has ideal bias $b_R\in\mathbb{R}$. Payoffs are quadratic in the distance between the receiver's action and the preferred action, as in \citet{stra20}. Let
\[
d_{iR}:=|b_i-b_R|
\]
denote the \emph{distance in preferences} between sender $i$ and the receiver.

\medskip

\noindent\emph{Truth-telling conditions.} In the \citet{stra20} cheap-talk game with joint distribution of signals $p_{k}(\cdot|\theta)$, two key results hold:

\begin{itemize}
\item For $n=2$ senders and $k\in[0,1)$, Proposition 1 of \citet{stra20} shows that there exists a truth-telling equilibrium if and only if
\begin{equation}\label{eq:CUC2}
d_{iR}\le \frac{1}{8+4k}\quad\text{for }i=1,2.
\end{equation}
\item For $n>2$ senders and $k\in[0,1)$, Proposition 4 of \citet[][Section 2.2.1]{stra20} shows that there exists a unique $\bar k(n)\in(0,1)$ and a continuous function $d^*(n,k)>0$ defined for all $k\in[0,1)$ such that:
\begin{enumerate}
\item if $k\neq\bar k(n)$, a truth-telling equilibrium exists if and only if $d_{iR}\le d^*(n,k)$ for all senders $i$;
\item if $k=\bar k(n)$, a truth-telling equilibrium exists for \emph{any} distances in preferences $(d_{iR})_i$.
\end{enumerate}
\end{itemize}

In our context, for each component $C$ of size $s$ we have $k=k(s)\in[0,1)$ and, by Lemma \ref{lem:equiv_prho}, the joint distribution of signals coincides with $p_{k(s)}(\cdot|\theta)$. Hence, for any given sample size $n\ge2$ and any component $C$, the \citet{stra20} truth-telling conditions above apply with $k=k(s)$.

\subsection*{A.I.4 From components to the correlated binomial distribution.}

We now connect the above structure to the correlated binomial distribution $\mu_n^{\rho}(\cdot|\theta)$.

Fix $n\ge2$ and consider a \emph{typical} agent $I_N$ in $V_N$, drawn uniformly at random. Conditional on the network $\mathbb{G}_N$, let $C(I_N)$ be her connected component and $S_N:=|C(I_N)|$ be its size. Inside this component, the source-selection game yields the equilibrium probability $k(S_N)$ of using a common information source, and the correlated-signal process with parameter $k(S_N)$ generates signals $(s_i)_{i=1}^n$ for the $n$ neighbors whose actions are observed by $I_N$. In a truth-telling equilibrium, we can identify $x_i:=s_i$, so that the joint law of $(x_i)_{i=1}^n$ conditional on $(\theta,S_N)$ is $p_{k(S_N)}(\cdot|\theta)$. Define the sum $Y:=\sum_{i=1}^n x_i$
of the $n$ observed actions. The next lemma identifies the distribution of $Y$ conditional on $(\theta,S_N)$.

\begin{lemma}[Component-level correlated binomial]\label{lem:Y_condSN}
Fix $\theta\in(0,1)$. Conditional on $S_N=s\ge n\geq 2$, the distribution of $Y$ is
\[
\Proba(Y=y\mid \theta,S_N=s)
=(1-k(s)) \mu^0_n(y|\theta)+k(s) \mu^1_n(y|\theta),
\]
for all $y\in\{0,1,\dots,n\}$, where $\mu^0_n(\cdot|\theta)$ and $\mu^1_n(\cdot|\theta)$ are the two components in eq. (\ref{eq:distpositive}). In particular,
\[
\Proba(Y=\cdot\mid\theta,S_N=s)=\mu_n^{\rho}(\cdot|\theta)\quad\text{with }\rho=k(s).
\]
\end{lemma}

To obtain the unconditional sampling distribution, we average over the random component size $S_N$.

\begin{proposition}[Network microfoundation of eq. (\ref{eq:distpositive})]\label{prop:network_micro}
Fix $n\ge2$ and $\theta\in(0,1)$. Define
\begin{equation}\label{eq:rhoN}
\rho_N:=\mathbb{E}\big[k(S_N)|S_N\geq n\big]
=\sum_{s\ge n}k(s)\frac{\Proba(S_N=s)}{\Proba(S_N\geq n)}.
\end{equation}
If, for each component size $s\ge2$, the cheap-talk game with information structure $p_{k(s)}(\cdot|\theta)$ satisfies the truth-telling conditions of \citet[][Propositions 1 and 4]{stra20}, then the unconditional distribution of $Y$ is
\[
\Proba(Y=y\mid\theta, S_N\geq n)
=(1-\rho_N) \mu^0_n(y|\theta)+\rho_N \mu^1_n(y|\theta)
=\mu_n^{\rho_N}(y|\theta)
\]
for all $y\in\{0,1,\dots,n\}$. In particular, the correlated binomial sampling distribution $\mu^{\rho}_n(.|\theta)$ in eq. (\ref{eq:distpositive}) arises from the equilibrium network, with correlation parameter $\rho=\rho_N\in[0,1)$ determined by the network and communication structures.
\end{proposition}

\medskip

\noindent\textit{Summary}: The network formation game in Lemma \ref{lem:symmetric_network} determines a random graph with connected components of random size $S_N$. Inside each component, the source-selection game in Lemma \ref{lem:qs} generates an equilibrium probability $k(s)$ that all agents rely on a common information source. The correlated-signal process of \citet{stra20} then transforms $k(s)$ into a correlation parameter governing the joint distribution $p_{k(s)}(\cdot|\theta)$ of local signals, and their Propositions 1 and 4 provide conditions on the preference distances $(d_{iR})_i$ under which truth-telling in this environment is an equilibrium. From the perspective of CoSESI, the resulting sampling distribution of the number of observed actions is the correlated binomial $\mu_n^{\rho_N}(\cdot|\theta)$ with $\rho_N=\mathbb{E}[k(S_N)|S_N\geq n]$.\footnote{We define the correlation $\rho_N$ as the expected probability of using a common source, conditional on agents belonging to a component large enough to support the sample size $n$. As noted earlier, this assumption can be relaxed using Online Appendix \hyperref[app:hetero]{B.IV} to allow agents to have different sample sizes.} Thus, the correlation parameter $\rho$ in eq. (\ref{eq:distpositive}) is an equilibrium measure of the commonality of information sources induced jointly by strategic network formation and strategic communication.

\phantomsection\label{app:market}
\subsection*{Online Appendix A.II: Market Microfoundation for Genericity}

This appendix provides a microfoundation for the measure $\lambda_n$ (and its generalization $\nu_{\kappa,n}$) used to define genericity in Section \ref{sec:gen}. We model the selection of the true joint distribution $p \in \Delta_n$ as the unique equilibrium of a competitive market for ``exposure.''

\paragraph{The Meta-Economy.}
Let $\Omega = \{0,1\}^n$ be the set of all possible signal realizations (outcomes) of size $n$. The cardinality of this space is $R = 2^n$. By definition, $\Delta_n:=\Delta(\Omega)$, the simplex of dimension $R-1$. Recall that the true data-generating process (DGP) is a vector $p = (p_\omega)_{\omega \in \Omega} \in \Delta_n$, where $p_\omega$ is the probability of observing signal sequence $\omega$. 

We view $p$ as an allocation of \textit{exposure} in a competitive market with exogenous budgets. There is a single platform selling a divisible good: audience attention (normalized to 1). There are $R$ distinct \textit{media coalitions}, indexed by $\omega \in \Omega$. Coalition $\omega$ consists of decentralized information sources (e.g., news media, advertisers, content creators) whose objective is to secure exposure for the specific data pattern (or narrative) $\omega$.\footnote{In this context, setting $R=2^n$ captures a perfectly competitive market for information where coalitions represent content creators or ``micro-narratives'' rather than broad interest groups.}
\paragraph{Random Budgets and Decentralization.}
Before trading, each coalition $\omega$ draws a budget $B_\omega > 0$ to spend on exposure. We microfound these budgets as the aggregation of contributions from many decentralized sources. Specifically, assume coalition $\omega$ is supported by $a_\omega$ independent sponsors. Each sponsor $j$ contributes a budget $\xi_{j,\omega} \sim \text{Exponential}(1)$. The total budget is the sum of these independent contributions:
\[
B_\omega = \sum_{j=1}^{a_\omega} \xi_{j,\omega}.
\]
It is a standard result that the sum of $a_\omega$ independent unit-exponential variables follows a Gamma distribution with shape $a_\omega$ and scale 1. Thus, 
\[
B_\omega \sim \mathrm{Gamma}(a_\omega, 1).
\]
This captures the idea of \textit{financial decentralization}: no single entity coordinates the funding across different signal patterns or narratives. Instead, funding arises from the independent decisions of many small donors/sponsors. If funding were highly correlated, the resulting environment would be nongeneric. More formally, our modeling assumptions are consistent with the finance literature where the magnitude of positive external shocks (i.e., capital injections) is modeled using iid exponential random variables. \citet[][Section 2]{kou02} justifies iid exponential random variables because they induce return distributions that are fat-tailed (or leptokurtic), which is consistent with stylized facts, and they also possess the so-called ``overshoot'' property (due to their memoryless property). More generally, \citet[][Corollary 1, eq. (17)]{kou02} proposes a rational expectations model to show that  financial shocks should be modeled using the exponential family.

\paragraph{Market Equilibrium.}
The platform sells exposure at a competitive price $P > 0$. Let $x_\omega \ge 0$ be coalition $\omega$'s purchase of exposure share. The coalition maximizes logarithmic utility subject to budget constraint:
\[
\max_{x_\omega \ge 0}\hspace{0.03in} \log \hspace{0.03in}x_\omega \quad \text{ s.t. } \quad P x_\omega \le B_\omega.
\]
\begin{proposition}[Walrasian Equilibrium]\label{prop:market_eq}
Given realized budgets $B=(B_\omega)_{\omega \in \Omega}$, there exists a unique Walrasian equilibrium where the price is $P^* = \sum_{\omega} B_\omega$ and the exposure shares are proportional to budgets:
\[
x_\omega^* = \frac{B_\omega}{\sum_{\omega' \in \Omega} B_{\omega'}}.
\]
\end{proposition}

\paragraph{The Emergence of the Dirichlet Distribution.}
We define the random DGP $p\in\Delta_n$ as the equilibrium allocation $x^*$. Since the budgets $B_\omega$ are independent Gamma variables, the resulting shares follow a Dirichlet distribution.

\begin{corollary}[Dirichlet Distribution]\label{cor:market_dirichlet}
    If $B_\omega \stackrel{\text{\normalfont indep}}{\sim} \mathrm{Gamma}(a_\omega, 1)$, then the equilibrium DGP $p = x^*$ follows a Dirichlet distribution on $\Delta_n$:
    \[
    p \sim \mathrm{Dir}\big( (a_\omega)_{\omega \in \Omega} \big).
    \]
    In the symmetric case where $a_\omega = \kappa>0$ for all $\omega$, $p$ follows the symmetric Dirichlet distribution $\nu_{\kappa,n}$. If $a_\omega = 1$, $p$ follows the uniform distribution $\lambda_n$ on $\Delta_n$.
\end{corollary}
This result follows directly from the standard property that a vector of normalized independent Gamma random variables is Dirichlet distributed.

\par Corollary \ref{cor:market_dirichlet} provides an economic intuition for Theorem \ref{thm:generic}. Aggregate efficiency relies on Nature's probability measure being ``spread out'' over the simplex $\Delta_n$, which is consistent with the Principle of Maximum Entropy used in information theory. Corollary \ref{cor:market_dirichlet} shows that such probability measures arise naturally from financial decentralization in the market for agents' attention. In contrast, nongeneric distributions---which concentrate mass on correlated subgroups---require coordinated, non-independent funding shocks.

\paragraph{Endogenizing the Parameters via Investment.}
We now microfound the symmetric parameters ($a_\omega = \kappa>0$) required for the robustness analysis (Theorem \ref{thm:generic2}). Suppose that prior to the budget realization, each coalition $\omega$ chooses an investment level $\alpha_\omega \ge 0$ at cost $C(\alpha_\omega) = \frac{\kappa}{2}\alpha_\omega^2$, which determines the expected number of sponsors they attract. Specifically, let $a_\omega = a + \alpha_\omega$, where $a > 0$ is a baseline parameter common to all outcomes. Each coalition is remunerated at rate $\pi > 0$ per unit of expected market share.

Coalition $\omega$'s expected payoff is:
\[
\Pi_\omega(\alpha) = \pi  \mathbb{E}[p_\omega] - \frac{\kappa}{2}\alpha_\omega^2 
= \pi \frac{a + \alpha_\omega}{\sum_{\omega' \in \Omega} (a + \alpha_{\omega'})} - \frac{\kappa}{2}\alpha_\omega^2.
\]

\begin{lemma}\label{lem:concavity}
    Fix the investment profiles of other coalitions $\alpha_{-\omega}$. Let $A_{-\omega} = \sum_{j \neq \omega} (a + \alpha_j)$. Coalition $\omega$'s objective function
    \[
    \Pi_\omega(\alpha_\omega) = \pi \frac{a + \alpha_\omega}{A_{-\omega} + a + \alpha_\omega} - \frac{\kappa}{2}\alpha_\omega^2
    \]
    is strictly concave in $\alpha_\omega$ on $[0, \infty)$.
\end{lemma}

\begin{proposition}[Symmetric Investment Equilibrium]\label{prop:symmetric_invest}
    In a symmetric investment game, there exists a unique symmetric NE where $\alpha_\omega^* = \alpha^*$ for all $\omega$. Consequently, the equilibrium DGP is governed by a symmetric Dirichlet distribution $\nu_{\kappa^*,n}$ with concentration parameter $\kappa^* = a + \alpha^*$.
\end{proposition}

This result confirms that the robustness of genericity to symmetric Dirichlet distribution (Theorem \ref{thm:generic2}) is consistent with a market where information sources face similar costs of entry. As competition increases ($R \to \infty$), individual investment $\alpha^*$ declines, but the aggregate equilibrium distribution remains symmetric. 
\phantomsection\label{app:learning}
\subsection*{Online Appendix A.III: Dynamic Foundation}

We provide a dynamic microfoundation for simple CoSESI (Section \ref{sec:simplecosesi}) where the correlation parameter $\rho$ arises as the steady state of an evolving information environment. We interpret the dynamics using the vaccination game.

\medskip
\noindent\textit{The Vaccination Game.}
Consider a population where agents decide whether to get vaccinated (action $A$) or not (action $B$). Let $\theta_t$ denote the vaccination rate at time $t$. Agents are born/enter the population, sample the vaccination status of $n$ peers, and make a decision based on their inference. 

We model the evolution of this system through two components: a \textit{Sampling Distribution} (how information is gathered) and an \textit{Action Distribution} (how behavior evolves).

\bigskip
\noindent\textit{1. Sampling Distribution.}
In every discrete period $t=0,1,2,\dots$, the population is subject to a shock $\varLambda_t \in \{0,1\}$ representing the state of the information environment.
\begin{itemize}
    \item $\varLambda_t=0$ (\textit{Normal Times}): Agents sample peers randomly. The actions observed in a sample of size $n$ are mutually independent. 
    \item $\varLambda_t=1$ (\textit{Viral/Echo-Chamber Times}): The society is polarized or subject to common informational shocks, e.g., a viral news story about side effects or a severe outbreak cluster. In this state, agents observe correlated signals: with probability $\rho_t$, the sample is correlated, and with probability $1-\rho_t$, it remains independent.
\end{itemize}
We model the environment $\varLambda_t$ as a two-state Markov chain with transition matrix:
\begin{align}\label{eq:transmat}
    \varSigma:=\begin{pmatrix}
        1-\phi_{\varLambda} & \phi_{\varLambda}\\
        \varphi_{\varLambda}& 1-\varphi_{\varLambda}
    \end{pmatrix},
\end{align}
where $\phi_{\varLambda} = \mathbb{P}(\varLambda_t=1 \mid \varLambda_{t-1}=0)$. Let $\rho_t := \mathbb{P}(\varLambda_t=1)$ denote the probability that the environment is in the ``viral'' state at time $t$. This parameter $\rho_t$ serves as the \textit{pairwise correlation} between any two signals observed by an agent at time $t$. Specifically, if $x_{i,t}$ and $x_{j,t}$ are the actions of two distinct peers sampled at time $t$, their correlation is:
\[ \text{corr}(x_{i,t}, x_{j,t}) = \rho_t. \]
The next result characterizes the distribution of the number of vaccinated peers observed in a sample at each time $t$.

\begin{obs}\label{thm:dyndist}
Under the regime $\varLambda_t$ governed by $\varSigma$, the number of agents $y_t$ taking action $A$ (vaccinated) observed by a sampling agent at time $t$ follows the distribution:
\begin{align}\label{eq:dynamicdist}
\mu_{\rho_t}(y_t|\theta_t)=
(1-\rho_t)\underbrace{\binom{n}{y_t}\theta_t^{y_t}(1-\theta_t)^{n-y_t}}_{\text{Independent Sampling}}+\rho_t\underbrace{\Bigg[ \theta_t\mathds{1}_{y_t=n}+(1-\theta_t)\mathds{1}_{y_t=0}\Bigg]}_{\text{Correlated Sampling}},
\end{align} 
where $\theta_t$ is the true vaccination rate and $\rho_t$ is the pairwise correlation at time $t$.
\end{obs}
\noindent Note that eq. (\ref{eq:dynamicdist}) is the dynamic analogue of the static distribution in eq. (\ref{eq:distpositive}).

\bigskip
\noindent\textit{2. Action Distribution.}
Let $\theta_0 \in [0,1]$ be the initial vaccination rate. In every subsequent period $t$, a fraction $1-\delta$ of the population survives (or retains their previous decision), while a fraction $\delta \in (0,1]$ of new agents (or revisers) enters the population. 

New agents suffer from correlation neglect. They observe the actions of the previous generation, $y_{t-1} \sim \mu_{\rho_{t-1}}(\cdot|\theta_{t-1})$, but na{\"i}vely assume signals are independent. They estimate the vaccination rate using a monotone inference procedure $G$ and best respond to their estimate. 
Recall from Section \ref{sec:simplecosesi} that the aggregate probability of a na{\"i}ve agent choosing $A$ when the true rate is $\theta$ and correlation is $\rho$ is given by $1 - \varPsi_n(\theta, \rho; C_n)$.

Thus, the law of motion for the vaccination rate $\theta_t$ is:
\begin{align}\label{eq:process}
    \theta_t = (1-\delta)\theta_{t-1} + \delta \left( 1 - \varPsi_n(\theta_{t-1}, \rho_{t-1}; C_n) \right).
\end{align}
We define the steady-state vaccination rate $\theta^{(\rho^*)}$ as the fixed point of the dynamics under the long-run correlation $\rho^*$:
\begin{align}\label{eq:steady}
    \theta^{(\rho^*)} = 1 - \varPsi_n(\theta^{(\rho^*)}, \rho^*; C_n).
\end{align}
Recall that this is the equilibrium equation of simple CoSESI in eq. (\ref{eq:equi}).

\begin{proposition}\label{thm:maindynamic}
   Consider any initial vaccination rate $\theta_0 \in [0,1]$ and initial correlation parameter $\rho_0 \in [0,1]$. If the transition matrix $\varSigma$ is irreducible and aperiodic, then:
   \begin{enumerate}
       \item The pairwise correlation $\rho_t$ converges exponentially fast to $\rho^* = \frac{\phi_{\varLambda}}{\phi_{\varLambda}+\varphi_{\varLambda}}$ as $t\rightarrow\infty$.
       \item For sufficiently small $\delta$, the vaccination rate $\theta_t$ converges globally to the unique simple CoSESI $\theta^{(\rho^*)}$ as $t\rightarrow\infty$.
   \end{enumerate}
\end{proposition}

\noindent\textit{Summary.} Proposition \ref{thm:maindynamic} suggests that even if the vaccination game starts with arbitrary beliefs and a volatile information structure, the dynamical system will eventually settles. Notably, it will settle on the CoSESI $\theta^{(\rho^*)}$, not the NE. The limit correlation $\rho^*$ reflects the long-run frequency of ``viral" information regimes. If $\rho^*$ is high (frequent echo chambers), the steady-state vaccination rate $\theta^{(\rho^*)}$ will be lower than NE due to the persistent overprecision bias generated by the environment.

\phantomsection\label{app:ext}
\section*{Online Appendix B: Extensions}
This online appendix considers some extensions of our baseline model. 
Online Appendix \hyperref[app:multi]{B.I} extends CoSESI to settings where there is a finite number of actions. Online Appendix \hyperref[app:loc]{B.II} studies the multiplicity and location of CoSESI in games of strategic complements. Online Appendix \hyperref[app:assort]{B.III} introduces assortativity in CoSESI by making the correlation depend on agents' types. Online Appendix \hyperref[app:hetero]{B.IV} allows for cases where agents have heterogeneous subjective models, sample sizes and inference procedures. Online Appendix \hyperref[app:ABEE]{B.V} shows how to formulate \citeauthor{jehiel}'s (\citeyear{jehiel}) ABEE as a CoSESI in large population games.

\phantomsection\label{app:multi}
\subsection*{Online Appendix B.I: Multi-Action Games and Genericity}
This appendix extends CoSESI (Definition \ref{def:CoSESI}) to games with a finite
number of actions. 

\subsection*{B.I.1 Multi-action primitives and inference}

\paragraph{Actions and success probability vector.}
Let the finite action set be
\[
\mathcal{A}:=\{0,1,\dots,N\},\qquad N\geq1,
\]
and interpret action \(0\) as an outside option and actions \(1,\dots,N\) as ``active'' actions. Let
$\boldsymbol{\theta}:=(\theta_a)_{a\in\mathcal{A}}\in\Delta(\mathcal{A})$
denote the \emph{success probability vector}, where \(\theta_a\) is the fraction of
agents taking action \(a\).

\paragraph{Payoffs.}
We modify the primitives in the main text to accommodate $N$ actions. Let the type space be $\Xi := [-1, 1]^N$. An agent's type is a vector $\xi = (\xi_1, ..., \xi_n) \in \Xi$, distributed according to a probability measure $\upsilon$ that is absolutely continuous with respect to the Lebesgue measure on $\Xi$. The utility of the outside option $0$ is normalized to $0$. To be able to nest the existing SESI in multi-action games, we specify the utility of each action as in \citet[][Supplementary Material]{stat} and \citet[][Section 6.1]{dynamic23}: The utility of taking action $a \in \{1, ..., N\}$ is $u_{\gamma,a}(\xi_a, \theta_a) = \xi_a + \gamma c_a(\theta_a)$, where $\theta_a$ is the fraction of agents taking action $a$, $\gamma\in\{-1,1\}$ is a constant, and for all $a\in\mathcal{A}$, $c_a$ is continuous and satisfies $0\leq c_a(.)\leq 1$. Then, define $u_{\gamma}:=(u_{\gamma,a})_{a\in\mathcal{A}}$.

\paragraph{Sampling.} As in Section \ref{sec:sampling}, each agent observes a sample of size \(n\geq1\).
A realization of a sample is a vector
$x=(x_1,\dots,x_n)\in\mathcal{A}^n,$
where each component \(x_i\) is the observed action of a randomly selected agent in
the population. For each \(x\in\mathcal{A}^n\) and each \(a\in\mathcal{A}\), define the count and
empirical frequency of action \(a\) by
$y_a(x):=\bigl|\{i\in\{1,\dots,n\}:x_i=a\}\bigr|,
z_a(x):=\frac{y_a(x)}{n},$
and let the empirical share vector be
$\boldsymbol{z}(x):=(z_a(x))_{a\in\mathcal{A}}\in\Delta(\mathcal{A}).$
Let
$Z_n:=\{\boldsymbol{z}(x):x\in\mathcal{A}^n\}$
 denote the set of all possible empirical share vectors generated by samples of
size \(n\).

\paragraph{True joint distribution and correlation neglect.}
For each \(\boldsymbol{\theta}\in\Delta(\mathcal{A})\), let
\(\bar{\Delta}_n(\mathcal{A},\boldsymbol{\theta})\) be the set of joint distributions
on \(\mathcal{A}^n\) whose one-dimensional marginals all coincide with
\(\boldsymbol{\theta}\). A \emph{true joint distribution with success probability
vector \(\boldsymbol{\theta}\)} is any element
\[
p_{\boldsymbol{\theta}}\in\bar{\Delta}_n(\mathcal{A},\boldsymbol{\theta}).
\]

Agents do \emph{not} know the true joint distribution and may face arbitrary
correlation in the sample, as in Section \ref{sec:sampling}. We maintain
complete correlation neglect (Definition \ref{def:corrneglect}) in the following
multi-action sense: conditional on \(\boldsymbol{\theta}\), agents treat the \(x_i\)'s as
iid draws from the Categorical distribution with parameter
\(\boldsymbol{\theta}\). That is, their subjective model is
\begin{align}\label{eq:indep}
    q_{\boldsymbol{\theta}}\big((x_i)_{i=1}^n\big)
=\prod_{i=1}^n\theta_{x_i},
\end{align}
i.e., the multi-action analogue of the independent Bernoulli model \(p_0\) used in
Definition \ref{def:corrneglect}. The inference procedure is then applied to this subjective model.

\paragraph{Inference procedures.}
An inference procedure assigns to each
sample size and empirical share vector an estimate of \(\boldsymbol{\theta}\). It is the multi-action extension of
Definition \ref{def:infproc}--\ref{def:ninf}.

\begin{definition}[Inference procedure]\label{def:multi_infproc}
An \emph{inference procedure} is a family
$F:=\{F_{n,\boldsymbol{z}}\}_{\boldsymbol{z}\in Z_n},$
where for each \(\boldsymbol{z}\in Z_n\), \(F_{n,\boldsymbol{z}}\) is a probability
measure on $\Delta(\mathcal{A})$. It is interpreted as the agent's estimate of the success probability vector \(\boldsymbol{\theta}\) after observing a sample whose empirical share vector is \(\boldsymbol{z}\).
\end{definition}

For each \(a\in\mathcal{A}\), let \(G^{(a)}_{n,\boldsymbol{z}}\) denote the marginal
distribution of the \(a\)-th component of \(\boldsymbol{\theta}\) under
\(G_{n,\boldsymbol{z}}\). Since agents have correlation neglect (subjective model $q_{}$ in eq. (\ref{eq:indep}) treats signals as independent), we restrict attention to monotone inference procedures.

\begin{definition}[Monotone inference procedure]\label{def:multi_monotone}
A multi-action inference procedure \(G=\{G_{n,\boldsymbol{z}}\}_{\boldsymbol{z}\in Z_n}\)
is \emph{monotone} if for every action \(a\in\mathcal{A}\) and every two empirical share
vectors \(\boldsymbol{z},\widetilde{\boldsymbol{z}}\in Z_n\) satisfying
$\widetilde{z}_a>z_a
\quad\text{and}\quad
\widetilde{z}_b=z_b \text{ for all }b\neq a,$
the marginal \(G^{(a)}_{n,\widetilde{\boldsymbol{z}}}\) strictly first-order
stochastically dominates \(G^{(a)}_{n,\boldsymbol{z}}\). 
\end{definition}

\paragraph{Examples of Inference Procedures.}
We illustrate the two standard examples of monotone inference procedures in the multi-action game.

\medskip\noindent
\emph{Example B.I.1: MLE.}
Fix a sample of size \(n\) with empirical share vector \(\boldsymbol{z}\in Z_n\), and
let \(y_a:=n z_a\) be the count of action \(a\) in the sample. Under the independent
subjective model \(q_{\boldsymbol{\theta}}\), the likelihood of \(\boldsymbol{\theta}\) is
\[
L(\boldsymbol{\theta};\boldsymbol{y})
=\prod_{a\in\mathcal{A}}\theta_a^{y_a}
\qquad\text{for }\boldsymbol{\theta}\in\Delta(\mathcal{A}).
\]
Maximizing \(L(\boldsymbol{\theta};\boldsymbol{y})\) over \(\Delta(\mathcal{A})\) yields
the familiar result $\widehat{\theta}_a(\boldsymbol{z})=\frac{y_a}{n}=z_a$ for all $a\in\mathcal{A}.$
Thus, MLE sets the success probability vector equal to the
empirical share vector: $G^{\mathrm{MLE}}_{n,\boldsymbol{z}}:=\delta_{\boldsymbol{z}},$
the degenerate probability measure assigning mass \(1\) to \(\boldsymbol{\theta}=\boldsymbol{z}\).

\medskip\noindent
\emph{Example B.I.2 (Dirichlet estimation).}
This inference procedure is the multi-action generalization of the Beta-estimation
procedure, originally introduced in \citet[][Example 3]{stat}. A player
has ``complete ignorance'' about \(\boldsymbol{\theta}\); formally, her prior is the
limit of the symmetric Dirichlet distribution with parameter
\((\varepsilon,\dots,\varepsilon)\) as \(\varepsilon\to0\). After observing a sample of
size \(n\) with counts \(\boldsymbol{y}=(y_a)_{a\in\mathcal{A}}\), the posterior
distribution over \(\boldsymbol{\theta}\) is the Dirichlet distribution with parameter
\(\boldsymbol{y}\). In our notation,
\[
G^{\mathrm{Dir}}_{n,\boldsymbol{z}}
:=\text{Dirichlet}(y_0,\dots,y_N)
=\text{Dirichlet}\bigl(n z_0,\dots,n z_N\bigr).
\]

\paragraph{Decision-making process.}
Given an inference procedure $G_{n,\boldsymbol{z}}$, each agent calculates the expected social incentive of action $a$ using the marginal distribution of $\theta_a$:$$C_{n,z}^a = \int_{\Delta(\mathcal{A})} c_a(\theta_a)   \hspace{0.03in}dG^{(a)}_{n,\boldsymbol{z}}(\boldsymbol{\theta}).$$
Then, each agent then chooses action $a\in\{1,\dots,N\}$ if $\xi_a + \gamma C_{n,\boldsymbol{z}}^a > \xi_b + \gamma C_{n,\boldsymbol{z}}^b$ for all $b\in\{1,\dots,N\}$, $b\neq a$, and $\xi_a + \gamma C_{n,\boldsymbol{z}}^a >0$ (to account for the outside option).

\subsection*{B.I.2 Multi-action CoSESI and existence}

We now define multi-action CoSESI and prove an existence result under continuity of
the true joint distribution in the success probability vector. The definition below
extends Definition \ref{def:CoSESI} from a scalar success probability \(\theta\in[0,1]\) to a
vector \(\boldsymbol{\theta}\in\Delta(\mathcal{A})\).

\begin{definition}[Multi-action CoSESI]\label{def:multi_CoSESI}
Fix a sample size \(n\), a family of true joint distributions
\(\{p_{\boldsymbol{\theta}}\}_{\boldsymbol{\theta}\in\Delta(\mathcal{A})}
\subseteq\bigcup_{\boldsymbol{\theta}}\bar{\Delta}_n(\mathcal{A},\boldsymbol{\theta})\),
and a multi-action inference procedure \(G=\{G_{n,\boldsymbol{z}}\}_{\boldsymbol{z}\in Z_n}\).
A \emph{multi-action CoSESI} is a vector
\(\boldsymbol{\theta}_{n,G}(p)\in\Delta(\mathcal{A})\) such that an
\(\theta_{n,G,a}(p)\)-fraction of agents takes action \(a\in\mathcal{A}\) when:
\begin{enumerate}
    \item \textnormal{(Sampling)} Each agent acquires \(n\) (possibly correlated)
    signals from the true joint distribution
    \(p_{\boldsymbol{\theta}_{n,G}(p)}\in\bar{\Delta}_n(\mathcal{A},\boldsymbol{\theta}_{n,G}(p))\).
    \item \textnormal{(Inference)} Given the independent subjective model
    \(q_{\boldsymbol{\theta}}\) and the inference procedure
    \(G=\{G_{n,\boldsymbol{z}}\}_{\boldsymbol{z}\in Z_n}\), each agent forms an
    estimate of \(\boldsymbol{\theta}\) from her sample.
    \item \textnormal{(Optimality)} Each agent best responds to her estimate of
    \(\boldsymbol{\theta}\) by choosing an action in \(\mathcal{A}\), using the
    same payoff and type primitives as in the subjective game
    \(\langle u_{\gamma},\upsilon,p,G\rangle_n\) in Section \ref{sec:cosesi}.
\end{enumerate}
\end{definition}

Thus, a multi-action CoSESI is a fixed point in which the success probability vector of
the true sampling distribution coincides with the aggregate distribution of actions
induced by statistical decision-making under correlation neglect.

\paragraph{Continuity in success probability vector.}
For each \(\boldsymbol{\theta}\in\Delta(\mathcal{A})\) and
\(\boldsymbol{z}\in Z_n\), define
\begin{align}\label{eq:distsum}
   \mu_n(\boldsymbol{z}\mid\boldsymbol{\theta})
:=\mathbb{P}_{p_{\boldsymbol{\theta}}}\bigl(\boldsymbol{z}(x)=\boldsymbol{z}\bigr) ,
\end{align}
i.e., the probability (under the true joint distribution \(p_{\boldsymbol{\theta}}\))
that a sample of size \(n\) generates empirical share vector \(\boldsymbol{z}\).

\begin{assumption}\label{ass:multi_cont}
For every \(\boldsymbol{z}\in Z_n\), the map
$\Delta(\mathcal{A})\ni\boldsymbol{\theta}
\mapsto
\mu_n(\boldsymbol{z}\mid\boldsymbol{\theta})$
is continuous.
\end{assumption}

Assumption \ref{ass:multi_cont} extends the continuity condition in Proposition \ref{thm:coscont}, which required continuity of the joint distribution in the scalar success probability \(\theta\). Here, continuity is required with respect to the probability of each possible empirical share vector.

\begin{proposition}[Existence of multi-action CoSESI]\label{prop:multi_exist}
Fix a sample size \(n\), a family
\(\{p_{\boldsymbol{\theta}}\}_{\boldsymbol{\theta}\in\Delta(\mathcal{A})}\)
of true joint distributions as above, and a monotone inference procedure
\(G=\{G_{n,\boldsymbol{z}}\}_{\boldsymbol{z}\in Z_n}\). Suppose
Assumption \ref{ass:multi_cont} holds. Then, there exists at least one multi-action
CoSESI \(\boldsymbol{\theta}_{n,G}(p)\in\Delta(\mathcal{A})\) for the subjective game
\(\langle u_{\gamma},\upsilon,p,G\rangle_n\).
\end{proposition}

 Assumption \ref{ass:multi_cont} is implied by continuity of the true
joint distribution.

\begin{corollary}
\label{cor:multi_coscont}
Suppose that for every action profile \(x\in\mathcal{A}^n\), the map
\[
\Delta(\mathcal{A})\ni\boldsymbol{\theta}
\mapsto p_{\boldsymbol{\theta}}(x)
\]
is continuous. Then, Assumption \ref{ass:multi_cont} holds, and for every sample size
\(n\) and every multi-action monotone inference procedure
\(G=\{G_{n,\boldsymbol{z}}\}_{\boldsymbol{z}\in Z_n}\), there exists a multi-action
CoSESI \(\boldsymbol{\theta}_{n,G}(p)\in\Delta(\mathcal{A})\). This is the multi-action
analogue of Proposition \ref{thm:coscont}.
\end{corollary}

Corollary \ref{cor:multi_coscont} shows that the existence result in
Proposition \ref{thm:coscont} extends to games with an arbitrary finite number
of actions: under correlation neglect (Definition \ref{def:corrneglect}), if the true
joint distribution over samples is continuous in the success probability vector
\(\boldsymbol{\theta}\), then a multi-action CoSESI exists. Moreover, when the true joint distribution coincides with the Categorical distribution in eq. (\ref{eq:indep}), this multi-action CoSESI coincides with multi-action SESI in \citet[][Supplementary Material, Observation S1]{stat}.

\subsection*{B.I.3 Multi-Action SESI and Genericity}

We now characterize the specific equilibrium that arises when agents have correct beliefs about the sampling process in the presence of correlation neglect. This extends \citeauthor{stat}'s (\citeyear{stat}) SESI to the multi-action setting and establishes its generic robustness using a concentration of measure argument on the high-dimensional simplex.

\paragraph{Multi-Action SESI.}
Recall from Corollary \ref{cor:multi_coscont} that when the true joint distribution coincides with the Categorical distribution $q_{\boldsymbol{\theta}}$ in eq. (\ref{eq:indep}), signals are truly independent, so the resulting equilibrium is a multi-action SESI.
Formally, let $p_{\boldsymbol{\theta}}^{\text{ind}}$ denote the joint distribution where signals are iid draws from $\boldsymbol{\theta}$.
A multi-action SESI is a vector $\boldsymbol{\theta}^* \in \Delta(\mathcal{A})$ such that $\boldsymbol{\theta}^* = \boldsymbol{\theta}_{n,G}(p_{\boldsymbol{\theta}^*}^{\text{ind}})$.

The distribution of the count vector $\boldsymbol{y} = (y_a)_{a\in\mathcal{A}}$ in a sample of size $n$ under SESI follows a \textit{multinomial} distribution with parameters $n$ and $\boldsymbol{\theta}^*$, which is the multi-dimensional extension of the binomial distribution.
A specific case of interest is the \textit{symmetric} SESI, where $\theta^*_a = 1/|\mathcal{A}|$ for all $a \in \mathcal{A}$.
Let $K := |\mathcal{A}| \ge 2$. The induced distribution of counts is the symmetric multinomial distribution, denoted $M_{n,K}$:
\begin{align}\label{eq:sym_multinom}
    M_{n,K}(\boldsymbol{y}) := \frac{n!}{\prod_{a\in\mathcal{A}} y_a!} \left(\frac{1}{K}\right)^n, \qquad \boldsymbol{y} \in \mathcal{C}_{n,K},
\end{align}
where $\mathcal{C}_{n,K}$ is the set of all count vectors summing to $n$, formally defined below.

\paragraph{Genericity in Multi-Action Games.}
We now show that a version of the binomial-approximation Lemma \ref{thm:lebesgue} extends to multi-actions games.
To this end, we define the measure $\lambda_{n,K}$ as the normalized Lebesgue measure on $\Delta_n^K$, the simplex of all possible joint distributions over the outcome space $\Omega_n = \mathcal{A}^n$. Note that the dimension of this simplex is $K^n - 1$.

Proposition \ref{thm:multinom_approx} below is the multi-dimensional analogue of Lemma \ref{thm:lebesgue}. It establishes that as the complexity of the correlation structure increases (via $K^n$), the probability of observing nongeneric distributions vanishes exponentially fast. Formally, it shows that, as $K^n$ grows, the distributions of arbitrarily correlated actions concentrate on the symmetric multinomial distribution $M_{n,K}$ in eq. (\ref{eq:sym_multinom}). For two such mass functions $\mu$ and $\nu$ on $\mathcal{C}_{n,K}$, their sup-nom is $\|\mu-\nu\|_\infty
:= \max_{n\in\mathcal{C}_{n,K}} |\mu(n)-\nu(n)|$ and $\|\mu-\nu\|_{TV}$ their TV distance.

\begin{proposition}[Genericity, multinomial extension]\label{thm:multinom_approx}
    Fix sample size $n \ge 1$, number of actions $K = |\mathcal{A}| \ge 2$, and $\epsilon > 0$. Let $p$ be drawn from $\Delta_n^K$ according to $\lambda_{n,K}$. Let $\mathscr{L}_n^K(p)$ denote the induced law of the count vector. Then,
    \begin{enumerate}
        \item \textnormal{(Sup-norm bound)}
        \[
        \lambda_{n,K}\Bigl(\Big\{p\in\Delta_n^K : \|\mathscr{L}_n^K(p)-M_{n,K}\|_\infty \le \epsilon\Big\}\Bigr) \ge 1 - \frac{1}{\epsilon^2 K^n}.
        \]
        \item \textnormal{(Total Variation bound)}
        \[
        \lambda_{n,K}\Bigl(\Big\{p\in\Delta_n^K : \|\mathscr{L}_n^K(p)-M_{n,K}\|_{TV} \le \epsilon\Big\}\Bigr) \ge 1 - \frac{1}{4\epsilon^2 K^n}\binom{n+K-1}{K-1}.
        \]
    \end{enumerate}
\end{proposition}

Proposition \ref{thm:multinom_approx} indicates that SESIs are also generic in multi-action games.
The denominator $K^n$ grows exponentially with  $n$, ensuring that the measure of distributions distorting aggregate behavior vanishes rapidly.
Thus, a social planner in a multi-action game can be confident that, absent strategic control over the information structure, aggregate behavior will resemble that of agents sampling from a multinomial distribution.

\phantomsection\label{app:loc}
\subsection*{Online Appendix B.II: Multiplicity of CoSESI}
We now study how the correlation structure of signals may affect the number of CoSESIs in games. To this end, we focus on simple CoSESI in games of strategic complements (e.g., coordination games), where agents suffer from correlation neglect. Thus, set $q=p_0$, $\gamma=1$, $\Xi=[-1,0]$, $p=p_{\rho}$, for $\rho\geq0$, and a monotone inference procedure $G$. 
\par Consider a unit mass of consumers where each one is deciding whether to adopt a product at an idiosyncratic cost $\xi\sim\mathcal{U}[-1,0]$. The adoption benefit $c$ is $S$-shaped: it is positive and increasing in the number of adopters, convex up to an inflection point, and then concave. As in  \citet[][Section 5]{stat}, let $c$ cross the $45^\circ$ line three times, $c(0)>0$, $c(1)<1$, and consumers use MLE to estimate the fraction of adopters.
\par A simple CoSESI has to solve $\theta=\varPsi_n(\theta,\rho;c)$, where $C_n=c$  because $G$ is the MLE. There are three NEs because an NE has to solve $\theta=c(\theta)$ and $c$ crosses the $45^\circ$ line three times. The next result shows that the exact number of simple CoSESIs depends on $\rho$.
\begin{proposition}\label{thm:select}
Fix $n$. There exists at least one and at most three simple CoSESIs, for all $\rho\in[0,1]$.   The total number of simple CoSESIs decreases as $\rho$ increases from 0 to 1. 
\end{proposition}
The sharpest predictions about simple CoSESIs and their locations occur when $\rho=1$.
\begin{corollary}\label{thm:rho1}
    Assume $\rho=1$. Then, for all $n$, simple CoSESI is unique and is located between the smallest and second-largest NE.
\end{corollary}
When $\rho<1$, the number of simple CoSESIs can be characterized in terms of the sample size and the \textit{variation} in $c$ close to the boundary points 0 and 1. The variation of $c$ at the point $j/n$ is defined as $\delta^n_j=\Delta_{j+1}^n-\Delta_{j}^n$, where $\Delta_{m}^n=c\big(\frac{m}{n}\big)-c\big(\frac{m-1}{n}\big)$.
\begin{corollary}\label{thm:var}
Assume $\rho<1$. If $\delta^n_1\leq0$ or $\delta^n_{n-1}\geq0$, then there is a unique simple CoSESI for all $n\geq2$. Otherwise, there are at most three simple CoSESIs. Let $\bar{n}$ be the largest sample size $n$ such that either $\delta^n_1\leq0$ or $\delta^n_{n-1}\geq0$. Then, there is a unique simple CoSESI for all $n\leq \bar{n}$. 
\end{corollary}
Intuitively, $\delta^n_{n-1}\geq0$ ensures that the concave part of $c$ does not affect the curvature of its $\rho$-weighted Bernstein polynomial, which is convex and therefore intersects the $45^\circ$ line once. Similarly, $\delta^n_1\leq0$ guarantees that the convex part of $c$ does not affect the curvature of its $\rho$-weighted Bernstein polynomial, which is concave and therefore intersects the $45^\circ$ line once. When $\delta^n_1>0$ and $\delta^n_{n-1}<0$, the $\rho$-weighted Bernstein polynomial becomes $S$-shaped and hence may cross the $45^\circ$ line up to three times.
\par The location of the CoSESIs can also be characterized. To see this, let $\text{conv}(c)$ denote the convex hull of $c$ and $m_k=\text{arg max}_{\theta\in[0,1]}\frac{|c(k)-c(\theta)|}{1-\theta}$, for $k\in \{0,1\}$. The lower envelope of $\text{conv}(c)$ is $c$ itself from 0 up to the point $m_0$ and is then the line segment connecting $c(m_0)$ and $c(1)$. If $c(m_0)<m_0$, the smallest NE proportion is smaller than $m_0$. Similarly, the upper envelope of $\text{conv}(c)$ is the line segment connecting $c(0)$ and $c(m_1)$ up to the point $m_1$ and is then $c$ itself. If  $c(m_1)>m_1$, the largest NE proportion is larger than $m_1$. Since $\varPsi_n(\theta,\rho;c)$ is the convex combination of the values of $c$, its graph lies inside $\text{conv}(c)$. It is above $c$ between 0 and $m_0$ and below $c$ between $m_1$ and 1. This is formalized below.
\begin{obs}
  Assume $\rho<1$.  If $f(m_0)<m_0$ and $f(m_1)>m_1$, then any simple CoSESI is larger than the smallest NE and is smaller than the largest NE.
\end{obs}

\phantomsection\label{app:assort}
\subsection*{Online Appendix B.III: Assortative CoSESI}
We incorporate \textit{assortativity} in simple CoSESI to study its effects in  games with social incentives.  
By assortativity, we mean that the correlation $\rho\in[0,1]$ in eq. (\ref{eq:distpositive}) depends on the agents' types $\xi\sim\mathcal{U}[0,1]$ in some potentially complex fashion. This could happen, for instance, if agents of higher types are more likely to observe correlated signals than lower types. 
A leading example of this happens when the interactions among agents take place on online platforms (e.g., LinkedIn or Facebook). The data engineers or designers of such platforms may create assortativity by making the following observation. High-type users may be more active on the platform relative to low types because their preference for action $A$ might demand more assiduous online searches to determine the value of this action. Thus, the engineers may detect this and design recommender systems that endogenously show high types the action that best fits their on-platform activity. 
 
\par Formally, let $\rho:[0,1]\rightarrow[0,1]$ denote a random matching technology that continuously maps agents' types to some correlation structure in these agents' signals. That is, $\rho(\xi)\in[0,1]$ is the pairwise correlation between any two signals obtained by agents of type $\xi$. All agents suffer from ``correlated sorting neglect'', i.e., both correlation neglect and assortativity neglect. The latter just means that agents are unaware that the signals they acquire from their peers depend on their type. It captures the fact that even if agents were able to understand that the observed actions are correlated, they may fail to understand the extent to which this correlation differs across types. Proposition \ref{thm:assortcoss} shows that a simple CoSESI exists; in this environment we call it an \textit{assortative} CoSESI.
\begin{proposition}\label{thm:assortcoss}
    There exists an \textit{assortative} CoSESI, for any $n$, $G$, and $\rho(.)$. It solves
    \begin{align}\label{eq:assort}
      \normalfont  1-\theta=\mathcal{B}_n(\theta;C_{n})+\underbrace{\mathcal{B}_n(\theta;\varLambda_{C_n})-\mathbb{B}_n(\theta;\varLambda_{C_n})}_{\text{correlated sorting neglect}},
    \end{align}
    where $\varLambda_{C_n}=\int_{C_n}^1\rho(\xi)\hspace{0.03in}d\xi$.\hfill
\end{proposition}
When $\rho(\xi)=0$ for all types $\xi$, the unique solution to eq. (\ref{eq:assort}) is SESI. In contrast,  
an assortative CoSESI captures the effect of correlated sorting neglect when $\rho(\xi)>0$.

\begin{ex}[Correlated Sorting Neglect]\label{thm:assortcosesi}\normalfont
    When $c(\theta)=\theta$, $G=\text{MLE}$, and $\rho(\xi)=\xi$, there exists a unique assortative CoSESI that solves eq. (\ref{eq:assort}):
    \begin{align}\label{eq:assortcosesi}
    \normalfont    \theta^{(\rho)}_{n,\text{\tiny MLE}}=\frac{5-\frac{1}{n}-\sqrt{\frac{1}{n^2}-\frac{2}{n}+17}}{2\big(1-\frac{1}{n}\big)}\in[0,1],
    \end{align}
    for all $n$ (see, below). When agents acquire only one signal ($n=1$ in eq. (\ref{eq:assortcosesi})), we get $\theta^{(\rho)}_{1,\text{\tiny MLE}}=\frac{1}{2}$, which corresponds to NE and SESI because $c$ is linear. However, as $n$ increases,  the effect of correlated sorting neglect starts to play a role (eq. (\ref{eq:assort})). Specifically, fewer agents take action $A$ in this assortative CoSESI than in both rational expectations and independent-sampling environments as $n$ grows. These dynamics eventually converge to $\theta^{(\rho)}_{\infty,\text{\tiny MLE}}=\frac{5-\sqrt{17}}{2}\approx0.44$. Thus, even with infinite data, about $6\%$ of the population mistakenly take the objectively inferior action $(B)$ at equilibrium. The variance of the sample mean is $\text{var}_{\mu^{\rho(\xi)}_{\infty}}\big(z\big|\theta^{(\rho)}_{\infty,\text{\tiny MLE}}\big)\approx\xi/4$  (Remark \ref{rem:overdis}), which increases by type $\xi\in[0,1]$. Therefore, higher types have noisier beliefs, so they are more likely to make mistakes. \hfill
\end{ex}

To summarize, this appendix has shown that correlated sorting neglect increases the likelihood that some types of agents will mistakenly take objectively inferior actions.
\subsubsection*{Derivations for Example \ref{thm:assortcosesi}}
    Let $c(\theta)=\theta$, $G=\text{MLE}$, and $\rho(\xi)=\xi$. Then, it follows that $C_{n}=c(\theta)$ by MLE, so $\mathcal{B}_n(\theta;C_{n})=c(\theta)$, and $\varLambda_{C_{n}}=\int_{C_n}^1\xi\hspace{0.03in}d\xi=(1-C^2_n)/2$. Thus, eq. (\ref{eq:assor}) becomes
    \begin{align*}
        1-\theta&=\mathcal{B}_n(\theta;C_{n})+\mathcal{B}_n(\theta;\varLambda_{C_{n}})-\mathbb{B}_n(\theta;\varLambda_{C_{n}})=\theta+\Bigg[\frac{1}{2}-\frac{1}{2}\mathcal{B}_n(\theta;C^2_{n})\Bigg]-\Bigg[\frac{1}{2}-\frac{1}{2}\mathbb{B}_n(\theta;C^2_{n})\Bigg]\\
        &=\theta-\frac{1}{2}\mathcal{B}_n(\theta;\theta^2)+\frac{1}{2}\mathbb{B}_n(\theta;\theta^2)=\theta-\frac{1}{2}\Bigg[\theta^2+\frac{1}{n}\theta(1-\theta)\Bigg]+\frac{1}{2}\theta,
    \end{align*}
    where the fourth equality used the fact that $\mathcal{B}_n(\theta;\theta^2)=\theta^2+\frac{1}{n}\theta(1-\theta)$ \citep[see,][eq. (7.14)]{phil}. Then, after rewriting this, we get
     $2+\big(\frac{1}{n}-5\big)\theta+\big(1-\frac{1}{n}\big)\theta^2=0,$
   which has $\theta^{(\rho)}_{n,\text{\tiny MLE}}$ in eq. (\ref{eq:assortcosesi}) as its unique solution on $[0,1]$ for all $n$.
{

\phantomsection\label{app:hetero}
\subsection*{Online Appendix B.IV: Heterogeneous Agents in CoSESI}
We outline how to extend our CoSESI (Section \ref{sec:cosesi}) to account for heterogeneous subjective models $q$, sample size $n$, and inference procures $F^q$. This is straightforward, so we will keep this analysis brief.
The intuition for this extension is that one could argue that agents may acquire different subjective models in practice because they have access to private ex-ante information. Similarly, agents may have different sample sizes if they have different abilities to access or process data. Agents may also favor different inference procedures based on fundamental statistical approaches (e.g., frequentist vs. Bayesian). 
\par  Fix a true distribution $p$, a continuous function $c$, and a type distribution $\upsilon$. Let $\zeta_{rjk}$ denote
the mass of agents who use the inference procedure $F^{r,j}$ on $\theta$, for $r\in\mathcal{R}=\{1,\dots, R\}$, $j\in\mathcal{J}=\big\{1,\dots,J\big\}$, and acquire $k\in\mathcal{K}=\big\{1,\dots,K\big\}$ samples from eq. (\ref{eq:prho}) such that $\sum_{(r,j,k)\in\mathcal{R}\times\mathcal{J}\times\mathcal{K}}\zeta_{rjk}=1$. Thus, a CoSESI with respect to the distribution $\m\zeta=(\zeta_{111},\dots,\zeta_{RJK})$ of  subjective models $\mathcal{Q}=\{q_1,\dots,q_R\}$, inference procedures $\mathcal{F}^q=\{F^{q,1},F^{q,2},\dots, F^{q,j}\}$ and samples $\mathcal{K}=\big\{1,\dots,K\big\}$, is a number $\theta_{\m\zeta}(p)\in[0,1]$ such that a fraction $\theta_{\m\zeta}(p)$ of agents
choose action $A$ when, for every $(r,j,k)\in\mathcal{R}\times \mathcal{J}\times\mathcal{K}$, each agent with subjective model $q_r\in\mathcal{Q}$ acquires $k$ signals from the joint distribution in eq. (\ref{eq:prho}) while using $F^{q,j}\in\mathcal{F}^q$, and a fraction $\zeta_{rjk}$ of agents best responds by choosing an action. 
\par We can write the equilibrium equation of CoSESI in the Observation \ref{thm:cosesichar} explicitly as 
\begin{align*}
    T(\theta|q,F^q,n)=\sum_{s=0}^n\mathscr{L}_{n}(p)(s|\theta)\hspace{0.03in}\sigma(s/n|q,F^q),
\end{align*}
where $\mathscr{L}_{n}(p)(s|\theta)=\Proba_p\big(\sum_{i=1}^nx_i=s\big|\theta\big)$ and $\sigma(s/n|q,F^q)$ represents the fraction of agents who choose action $A$ after observing a sample mean $s/n$ and using the subjective model $q$ and inference procedure $F^q$. In the heterogeneous case, the equilibrium equation becomes
\begin{align*}
    T_{\m{\zeta}}(\theta):=\E_{\m{\zeta}}\big[T(\theta|q,F^q,n)\big]=\sum_{(r,j,k)\in\mathcal{R}\times \mathcal{J}\times\mathcal{K}}\zeta_{rjk}\Bigg[\sum_{s=0}^k\mathscr{L}_{k}(p)(s|\theta)\hspace{0.03in}\sigma(s/k|q_r,F^{q_r,j})\Bigg].
\end{align*}
Since the outer sum is finite, all CoSESI's existence results apply in this setting. 
}

\phantomsection\label{app:ABEE}
\subsection*{Online Appendix B.V: ABEE and CoSESI}

We extend our baseline framework to allow a finite set of payoff-relevant states and define \citeauthor{jehiel}'s (\citeyear{jehiel}) Analogy-Based Expectation Equilibrium (ABEE) adapted to this setting.

\subsection*{States and analogy classes}

Let $S$ be a finite set of payoff-relevant states and let $\pi \in \Delta(S)$ be commonly known. At the beginning of the interaction, a state $s \in S$ is drawn according to $\pi$.

In each state $s \in S$, the stage game is exactly as in the baseline framework:
\begin{itemize}
    \item there is a unit mass of agents;
    \item each agent chooses an action in $\{A,B\}$;
    \item types are given by $\xi \in \Xi \subseteq [-1,1]$ with $\xi \sim \upsilon$;
    \item the utility from taking action $A$ in state $s$ is
    $u_\gamma(\xi,\theta(s))  =  \xi + \gamma c(\theta(s)),$
    where $\theta(s) \in [0,1]$ denotes the fraction of agents taking action $A$ in state $s$.
\end{itemize}

An \emph{analogy partition} is a partition $\mathcal{P} = \{\mathscr{C}_1,\dots,\mathscr{C}_K\}$ of $S$. For each $s \in S$, we denote by $\mathscr{C}_s \in \mathcal{P}$ the unique element of $\mathcal{P}$ that contains $s$.

\subsection*{Strategies and aggregates}

A (pure) strategy is a measurable function $\sigma : S \times \Xi \to \{A,B\}.$
Given a strategy profile $\sigma$ and a state $s \in S$, the induced fraction of agents taking action $A$ in state $s$ is
\[
    \theta(s) 
     :=  \int_{\Xi} \mathds{1}\{\sigma(s,\xi) = A\}   d\upsilon(\xi).
\]

For each $\mathscr{C} \in \mathcal{P}$, the \emph{within-class average aggregate} induced by $\sigma$ is
\[
    \overline{\theta}(\mathscr{C})
     :=  \sum_{s \in \mathscr{C}} \pi(s \mid \mathscr{C}) \theta(s),
    \qquad
    \pi(s \mid \mathscr{C}) := \frac{\pi(s)}{\sum_{t \in \mathscr{C}} \pi(t)}.
\]
We are ready to define \citeauthor{jehiel}'s (\citeyear{jehiel}) ABEE in this setting.
\begin{definition}[ABEE]
\label{def:ABEE}
Given an analogy partition $\mathcal{P}$ of $S$, an ABEE is a pair $(\sigma,\hat{\theta})$:
\begin{itemize}
    \item $\sigma : S \times \Xi \to \{A,B\}$ is a strategy profile,
    \item $\hat{\theta} : \mathcal{P} \to [0,1]$ assigns to each analogy class $\mathscr{C} \in \mathcal{P}$ a \emph{class-level expectation} $\hat{\theta}(\mathscr{C})$,
\end{itemize}
such that:
\begin{enumerate}
    \item \textit{Analogy-based best responses.}  
    For all $s \in S$ and all $\xi \in \Xi$,
    \begin{equation}
        \sigma(s,\xi) = A
        \quad\Longleftrightarrow\quad
        u_\gamma\big(\xi, \hat{\theta}(\mathscr{C}_s)\big)  \geq  0
        \quad\Longleftrightarrow\quad
        \xi  \geq  -\gamma c\big(\hat{\theta}(\mathscr{C}_s)\big).
        \label{eq:ABEE_BR}
    \end{equation}
    That is, in state $s$, agents best respond to the \emph{class-level} expectation $\hat{\theta}(\mathscr{C}_s)$ instead of the true aggregate $\theta(s)$.

    \item \textit{Class-level consistency.}  
    For every $\mathscr{C} \in \mathcal{P}$,
    \begin{equation}
        \hat{\theta}(\mathscr{C})  =  \overline{\theta}(\mathscr{C})
         =  \sum_{s \in \mathscr{C}} \pi(s \mid \mathscr{C}) \theta(s),
        \label{eq:ABEE_consistency}
    \end{equation}
    where $\theta(s)$ is the aggregate defined above under the strategy profile $\sigma$.
\end{enumerate}
\end{definition}

An ABEE is a profile in which (i) agents best respond to beliefs that depend only on the analogy class $\mathscr{C}_s$ of the state, and (ii) the belief associated with each class $C$ coincides with the average aggregate outcome within $C$ induced by the strategy profile.

\subsection*{Embedding ABEE into CoSESI}

We now show that any ABEE can be represented as a CoSESI in subjective games.

Fix a partition $\mathcal{P}$ and an ABEE $(\sigma,\hat{\theta})$ as in Definition \ref{def:ABEE}. For each state $s \in S$, we construct a subjective game
$\Gamma^s_n:=\langle u_\gamma, \upsilon, p^s, F^{q^s} \rangle_n$
where the true distribution $p^s$, the subjective model $q^s$, and the inference procedure $F^{q^s}$ are defined below.

\subsection*{True joint distribution $p^s$}

Given the ABEE $(\sigma,\hat{\theta})$, the induced aggregate in state $s$ is $\theta(s)$. For each $s \in S$, pick any joint distribution of actions $p^s \in \bar{\Delta}_n$ such that
\begin{equation}
    \mathbb{P}_{p^s}(x_i = 1) = \theta(s)
    \quad\text{for all } i = 1,\dots,n.
    \label{eq:p_s_success_prob}
\end{equation}

\subsection*{Subjective model $q^s$}

For each $s \in S$, the subjective model $q^s$ is given by the independent Bernoulli model $p_0$, i.e., 
$q^s := p_0 \in \bar{\Delta}_n$ $\forall s\in S$.
Thus, in every state, agents suffer from correlation neglect.

\subsection*{Inference procedure $F^{q^s}$}

For each state $s \in S$, we define an inference procedure $F^{q^s}$ by specifying, for each sample $(n,z)$, a distribution $F^{q^s}_{n,z}$ over $\theta$. Consider the following \textit{dogmatic} estimate
\begin{equation}
    F^{q^s}_{n,z}(\theta)  :=  \mathds{1}_{\theta \ge \hat{\theta}(\mathscr{C}_s)}
    \quad\text{for all } z \in [0,1],
    \label{eq:ABEE_posterior}
\end{equation}
which is reminiscent of the MLE monotone inference procedure $G_{n,z}=\mathds{1}\{\theta\geq z\}$, where $z$ is replaced with $\hat{\theta}(\mathscr{C}_s)$. Here is an intuition for this. An agent may have already formed an analogy-based expectation $\hat{\theta}(\mathscr{C}_s)$ from past data, so in the current one-shot decision problem, she does not update that expectation based on the noisy finite sample she obtained in this period. The associated expected social incentive function becomes $C^{q,s}_{n}
    := \int_0^1 c(\theta)dF^{q^s}_{n,z}(\theta) 
    = c\big(\hat{\theta}(\mathscr{C}_s)\big),$
which does not depend on $z$.

\subsection*{Best responses in subjective games}

In the subjective game $\Gamma^s_n = \langle u_\gamma,\upsilon,p^s,F^{q^s}\rangle_n$, the expected utility from taking action $A$ for a type $\xi$ in state $s$ is $\xi + \gamma C^{q,s}_{n}
     =  \xi + \gamma c\big(\hat{\theta}(\mathscr{C}_s)\big),$
while the utility of action $B$ is $0$. Thus, in $\Gamma^s_n$ a type $\xi$ follows the best-response rule
\begin{equation}
    \text{type }\xi\text{ chooses } A \quad\Longleftrightarrow\quad
    \xi  \ge  -\gamma c\big(\hat{\theta}(\mathscr{C}_s)\big).
    \label{eq:CoSESI_BR_s}
\end{equation}

Comparing \eqref{eq:CoSESI_BR_s} with the ABEE best-response condition \eqref{eq:ABEE_BR}, we get
\[
    \sigma(s,\xi) = A
    \quad\Longleftrightarrow\quad
    \xi  \geq  -\gamma c\big(\hat{\theta}(\mathscr{C}_s)\big).
\]
Hence, under $\Gamma^s_n$, if agents follow the CoSESI decision rule, the resulting action profile in state $s$ coincides with the ABEE strategy $\sigma(s,\cdot)$.

Consequently, the aggregate fraction of agents taking action $A$ in $\Gamma^s_n$ is
\begin{align*}
    \theta^{\text{CoSESI}}(s)
    := \int_{\Xi} \mathds{1}\{\xi \ge -\gamma c(\hat{\theta}(\mathscr{C}_s))\}   d\upsilon(\xi) = \int_{\Xi} \mathds{1}\{\sigma(s,\xi) = A\}   d\upsilon(\xi) = \theta(s).
\end{align*}

\par We can now state and prove the precise relationship between ABEE and CoSESI.

\begin{proposition}
\label{prop:ABEE_CoSESI}
Consider the multi-state large population game described above and an analogy partition $\mathcal{P}$ of $S$. Let $(\sigma,\hat{\theta})$ be an ABEE with respect to $\mathcal{P}$. For each state $s \in S$, define the subjective game $\Gamma^s_n := \langle u_\gamma,\upsilon,p^s,F^{p^s_0}\rangle_n.$
Then, for each $s \in S$, the ABEE aggregate $\theta(s)$ is a CoSESI of the subjective game $\Gamma^s_n$, i.e.
\[
    \theta_{n,F^{q^s}}(p^s) = \theta(s),
\]
and the CoSESI behavior in state $s$ coincides with the ABEE strategy $\sigma(s,\cdot)$.
\end{proposition}

In summary, the construction in Proposition \ref{prop:ABEE_CoSESI} shows explicitly how the components of ABEE map into the CoSESI primitives:
\begin{itemize}
    \item The analogy partition $\mathcal{P}$ is encoded in the \emph{state-dependent} inference procedures $F^{q^s}$, which assign the same belief $\hat{\theta}(\mathscr{C})$ to all states $s \in \mathscr{C}$.
    \item The misspecification in ABEE---pooling states within an analogy class---is captured by a combination of correlation neglect in the subjective model ($q^s = p_0$) and a coarse inference rule $F^{q^s}$ that ignores state-specific information and treats each state $s$ as if its aggregate were $\hat{\theta}(\mathscr{C}_s)$.
    \item Since CoSESI allows for arbitrary subjective models and inference procedures, ABEE appears as a specific parametric restriction where (i) the true distributions $p^s$ are chosen so that their success probabilities match the ABEE aggregates, and (ii) the inference procedures are dogmatic at the analogy-based expectations $\hat{\theta}(\mathscr{C})$.
\end{itemize}

\phantomsection\label{app:proof}
\section*{Online Appendix C: Omitted Proofs}

\subsection*{Proof of Observations \ref{thm:beta2}--\ref{thm:beta}}

\begin{proof}[Proof of Observation \ref{thm:beta2}]
We apply Definition~\ref{def:infproc} to Case~(1), where the agent is correctly specified ($q=p_{\rho}$). The parameter vector is partitioned as $\vartheta_1=\rho$ (known) and $\vartheta_2=\varnothing$. The prior on $\theta$ is $\pi(\theta)=\text{\normalfont Beta}(\alpha,\beta)$ with density
\[
\pi(\theta)\propto \theta^{\alpha-1}(1-\theta)^{\beta-1},\qquad \theta\in(0,1).
\]
Let $y=\sum_{i=1}^n x_i\in\{0,1,\dots,n\}$ denote the number of successes and $z:=y/n$ the sample mean. By \eqref{eq:distpositive}, the likelihood of $y$ under the correlated binomial distribution is
\[
\mu^{\rho}_n(y|\theta)
=
(1-\rho)\hspace{0.02in}\underbrace{\binom{n}{y}\theta^y(1-\theta)^{n-y}}_{\mu^{0}_n(y|\theta)}
+
\rho\hspace{0.02in}\underbrace{\Big(\theta\mathds{1}_{\{y=n\}}+(1-\theta)\mathds{1}_{\{y=0\}}\Big)}_{\mu^{1}_n(y|\theta)}.
\]

By Definition~\ref{def:infproc}, the posterior density of $\theta$ conditional on $\rho$ and the sufficient statistic $z$ is $\pi^q_{n,z}(\theta|\rho)\propto \pi(\theta)\hspace{0.02in}\mu^{\rho}_n(y|\theta)
\propto \theta^{\alpha-1}(1-\theta)^{\beta-1}\hspace{0.02in}\mu^{\rho}_n(y|\theta).$
We distinguish interior counts ($0<y<n$) and extreme counts ($y=0$ or $y=n$).

\medskip
\noindent\emph{Case 1: $0<y<n$ (equivalently, $0<z<1$).}
In this case the “correlated” component vanishes, so $\mu^{\rho}_n(y|\theta)
=
(1-\rho)\binom{n}{y}\theta^y(1-\theta)^{n-y},$
and hence
\[
\pi^q_{n,z}(\theta|\rho)
\propto
\theta^{\alpha-1}(1-\theta)^{\beta-1}
\cdot
\theta^y(1-\theta)^{n-y}
\propto
\theta^{\alpha+y-1}(1-\theta)^{\beta+n-y-1}.
\]
Since $y=nz$, this is the kernel of a $\text{\normalfont Beta}(\alpha+nz,\beta+n(1-z))$ distribution. The multiplicative constant $(1-\rho)\binom{n}{y}$ does not depend on $\theta$, so it is absorbed into the normalizing constant. Thus, $F^{q}_{n,z}(\theta|\rho)
=
\text{\normalfont Beta}\big(\alpha+nz,\beta+n(1-z)\big),$ for $0<z<1.$
For such $z$, we have $\mathds{1}_{z\in\{0,1\}}=0$, so $\eta_z(\rho)
=
(1-0)+0=1,$
and the unified formula in the statement reduces to the above expression, with $1-\eta_z(\rho)=0$.

\medskip
\noindent\emph{Case 2: $y=0$ or $y=n$ (i.e., $z\in\{0,1\}$).}
Consider first $y=0$ (so $z=0$). From \eqref{eq:distpositive},
$\mu^{\rho}_n(0|\theta)
=
(1-\rho)(1-\theta)^n+\rho(1-\theta),$
and therefore
\begin{align*}
\pi^q_{n,0}(\theta|\rho)
\propto
\theta^{\alpha-1}(1-\theta)^{\beta-1}\big[(1-\rho)(1-\theta)^n+\rho(1-\theta)\big]=
(1-\rho)\hspace{0.02in}\theta^{\alpha-1}(1-\theta)^{\beta+n-1}
+\rho\hspace{0.02in}\theta^{\alpha-1}(1-\theta)^{\beta}.
\end{align*}
Define the two kernels $K_0(\theta):=\theta^{\alpha-1}(1-\theta)^{\beta+n-1},$ and $K_1(\theta):=\theta^{\alpha-1}(1-\theta)^{\beta},$
so that $\pi^q_{n,0}(\theta|\rho)\propto (1-\rho)K_0(\theta)+\rho K_1(\theta).$
The normalizing constants are $\int_0^1 K_0(\theta)\hspace{0.02in}d\theta = B(\alpha,\beta+n)$ and $\int_0^1 K_1(\theta)\hspace{0.02in}d\theta = B(\alpha,\beta+1).$
If $f_{\alpha,\beta+n}$ and $f_{\alpha,\beta+1}$ denote the densities of $\text{\normalfont Beta}(\alpha,\beta+n)$ and $\text{\normalfont Beta}(\alpha,\beta+1)$, we have $K_0(\theta)=B(\alpha,\beta+n)\hspace{0.02in}f_{\alpha,\beta+n}(\theta)$ and $
K_1(\theta)=B(\alpha,\beta+1)\hspace{0.02in}f_{\alpha,\beta+1}(\theta).$
The overall normalizing constant is $C_0(\rho)
:=
(1-\rho)B(\alpha,\beta+n)+\rho B(\alpha,\beta+1),$
so the posterior density can be written as
\[
\pi^q_{n,0}(\theta|\rho)
=
\frac{(1-\rho)B(\alpha,\beta+n)}{C_0(\rho)}\hspace{0.02in}f_{\alpha,\beta+n}(\theta)
+
\frac{\rho B(\alpha,\beta+1)}{C_0(\rho)}\hspace{0.02in}f_{\alpha,\beta+1}(\theta).
\]
Thus $F^{q}_{n,0}(\cdot|\rho)$ is a mixture of $\text{\normalfont Beta}(\alpha,\beta+n)$ and $\text{\normalfont Beta}(\alpha,\beta+1)$ with mixing weight
\[
\eta_0(\rho)
=
\frac{(1-\rho)B(\alpha,\beta+n)}
     {(1-\rho)B(\alpha,\beta+n)+\rho B(\alpha,\beta+1)}.
\]
This coincides with the expression for $\eta_z(\rho)$ in the statement when $z=0$, since then $\mathds{1}_{\{z\in\{0,1\}\}}=1$ and $nz=0$, $1-z=1$.

For $y=n$ (so $z=1$), we have $\mu^{\rho}_n(n|\theta)
=
(1-\rho)\theta^n+\rho\theta,$
and
\begin{align*}
\pi^q_{n,1}(\theta|\rho)
\propto
\theta^{\alpha-1}(1-\theta)^{\beta-1}\big[(1-\rho)\theta^n+\rho\theta\big] =
(1-\rho)\hspace{0.02in}\theta^{\alpha+n-1}(1-\theta)^{\beta-1}
+\rho\hspace{0.02in}\theta^{\alpha}(1-\theta)^{\beta-1}.
\end{align*}
Define $\widetilde K_0(\theta):=\theta^{\alpha+n-1}(1-\theta)^{\beta-1},$ and $\widetilde K_1(\theta):=\theta^{\alpha}(1-\theta)^{\beta-1}.$
Then, $\pi^q_{n,1}(\theta|\rho)\propto (1-\rho)\widetilde K_0(\theta)+\rho \widetilde K_1(\theta),$
with normalizing constants $\int_0^1 \widetilde K_0(\theta)\hspace{0.02in}d\theta = B(\alpha+n,\beta)$ and $
\int_0^1 \widetilde K_1(\theta)\hspace{0.02in}d\theta = B(\alpha+1,\beta).$
Let $\widetilde f_{\alpha+n,\beta}$ and $\widetilde f_{\alpha+1,\beta}$ denote the densities of $\text{\normalfont Beta}(\alpha+n,\beta)$ and $\text{\normalfont Beta}(\alpha+1,\beta)$, respectively. Then, $\widetilde K_0(\theta)=B(\alpha+n,\beta)\hspace{0.02in}\widetilde f_{\alpha+n,\beta}(\theta)$ and $\widetilde K_1(\theta)=B(\alpha+1,\beta)\hspace{0.02in}\widetilde f_{\alpha+1,\beta}(\theta).$
The normalizing constant is $C_1(\rho)
:=
(1-\rho)B(\alpha+n,\beta)+\rho B(\alpha+1,\beta)$
and
\[
\pi^q_{n,1}(\theta|\rho)
=
\frac{(1-\rho)B(\alpha+n,\beta)}{C_1(\rho)}\hspace{0.02in}\widetilde f_{\alpha+n,\beta}(\theta)
+
\frac{\rho B(\alpha+1,\beta)}{C_1(\rho)}\hspace{0.02in}\widetilde f_{\alpha+1,\beta}(\theta).
\]
Hence $F^{q}_{n,1}(\cdot|\rho)$ is a mixture of $\text{\normalfont Beta}(\alpha+n,\beta)$ and $\text{\normalfont Beta}(\alpha+1,\beta)$ with mixing weight
\[
\eta_1(\rho)
=
\frac{(1-\rho)B(\alpha+n,\beta)}
     {(1-\rho)B(\alpha+n,\beta)+\rho B(\alpha+1,\beta)}.
\]
This again coincides with the formula for $\eta_z(\rho)$ in the statement when $z=1$, since then $\mathds{1}_{\{z\in\{0,1\}\}}=1$ and $nz=n$, $1-z=0$.

Combining the three cases yields the unified expression in the statement.
\end{proof}

\begin{proof}[Proof of Observation \ref{thm:beta}]
We apply Definition~\ref{def:infproc} to Case~(3), where the agent estimates the correlation ($q=p_{\hat{\rho}}$). The parameter partition is $\vartheta_1=\varnothing$ and $\vartheta_2=\hat{\rho}$ (unknown). The joint prior is $\pi(\theta,\hat{\rho})\propto \theta^{\alpha-1}(1-\theta)^{\beta-1},$
which implies that $\theta\sim\text{\normalfont Beta}(\alpha,\beta)$ and $\hat{\rho}\sim\mathcal{U}[0,1]$ are independent.

Let $y=\sum_{i=1}^n x_i$ and $z:=y/n$ as before. Given $(\theta,\hat{\rho})$, the likelihood of $y$ is $\mu^{\hat{\rho}}_n(y|\theta)$ from \eqref{eq:distpositive}. Thus the joint posterior density of $(\theta,\hat{\rho})$ is
\[
\pi^q_{n,z}(\theta,\hat{\rho}|\varnothing)\propto \pi(\theta,\hat{\rho})\hspace{0.02in}\mu^{\hat{\rho}}_n(y|\theta)
\propto \theta^{\alpha-1}(1-\theta)^{\beta-1}\hspace{0.02in}\mu^{\hat{\rho}}_n(y|\theta).
\]
By Definition~\ref{def:infproc}, the marginal posterior of $\theta$ is obtained by integrating out $\hat{\rho}$:
\[
\pi^q_{n,z}(\theta|\varnothing)
\propto
\int_0^1 \theta^{\alpha-1}(1-\theta)^{\beta-1}\hspace{0.02in}\mu^{\hat{\rho}}_n(y|\theta)\hspace{0.02in}d\hat{\rho}.
\]

\medskip
\noindent\emph{Case 1: $0<y<n$ (equivalently, $0<z<1$).}
From \eqref{eq:distpositive},
\[
\mu^{\hat{\rho}}_n(y|\theta)
=
(1-\hat{\rho})\binom{n}{y}\theta^y(1-\theta)^{n-y},
\qquad 0<y<n,
\]
because the “correlated” component is zero for interior counts. Therefore
\begin{align*}
\pi^q_{n,z}(\theta|\varnothing)
\propto
\theta^{\alpha-1}(1-\theta)^{\beta-1}
\binom{n}{y}\theta^y(1-\theta)^{n-y}
\int_0^1 (1-\hat{\rho})\hspace{0.02in}d\hat{\rho} \propto
\theta^{\alpha+y-1}(1-\theta)^{\beta+n-y-1}.
\end{align*}
Since $y=nz$, this is the kernel of a $\text{\normalfont Beta}(\alpha+nz,\beta+n(1-z))$ distribution. Hence
\[
F^{q}_{n,z}(\theta)
=
\text{\normalfont Beta}\big(\alpha+nz,\beta+n(1-z)\big),
\qquad 0<z<1.
\]
For such $z$, we have $\mathds{1}_{z\in\{0,1\}}=0$, so $\tilde{\eta}_z
=
(1-0)+0=1,$
and the unified expression in the statement reduces to this Beta posterior, with $1-\tilde{\eta}_z=0$.

\medskip
\noindent\emph{Case 2: $y=0$ or $y=n$ (i.e., $z\in\{0,1\}$).}
Consider first $y=0$ (so $z=0$). From \eqref{eq:distpositive},
$\mu^{\hat{\rho}}_n(0|\theta)
=
(1-\hat{\rho})(1-\theta)^n+\hat{\rho}(1-\theta).$
Thus,
\begin{align*}
\pi^q_{n,0}(\theta|\varnothing)
&\propto
\theta^{\alpha-1}(1-\theta)^{\beta-1}
\int_0^1\big[(1-\hat{\rho})(1-\theta)^n+\hat{\rho}(1-\theta)\big]\hspace{0.02in}d\hat{\rho} \\
&=
\theta^{\alpha-1}(1-\theta)^{\beta-1}
\left[
(1-\theta)^n\int_0^1(1-\hat{\rho})\hspace{0.02in}d\hat{\rho}
+
(1-\theta)\int_0^1\hat{\rho}\hspace{0.02in}d\hat{\rho}
\right] \\
&=
\frac12\hspace{0.02in}\theta^{\alpha-1}(1-\theta)^{\beta+n-1}
+
\frac12\hspace{0.02in}\theta^{\alpha-1}(1-\theta)^{\beta}.
\end{align*}
Define $K_0(\theta):=\theta^{\alpha-1}(1-\theta)^{\beta+n-1}$ and $K_1(\theta):=\theta^{\alpha-1}(1-\theta)^{\beta}.$
Then, $\pi^q_{n,0}(\theta|\varnothing)\propto \tfrac12 K_0(\theta)+\tfrac12 K_1(\theta).$
The normalizing constants are $\int_0^1 K_0(\theta)\hspace{0.02in}d\theta = B(\alpha,\beta+n)$
and $\int_0^1 K_1(\theta)\hspace{0.02in}d\theta = B(\alpha,\beta+1).$
Let $f_{\alpha,\beta+n}$ and $f_{\alpha,\beta+1}$ be the densities of $\text{\normalfont Beta}(\alpha,\beta+n)$ and $\text{\normalfont Beta}(\alpha,\beta+1)$, respectively. Then, $K_0(\theta)=B(\alpha,\beta+n)\hspace{0.02in}f_{\alpha,\beta+n}(\theta)$ and $K_1(\theta)=B(\alpha,\beta+1)\hspace{0.02in}f_{\alpha,\beta+1}(\theta).$
The overall normalizing constant is $\widetilde C_0
:=
\tfrac12 B(\alpha,\beta+n)+\tfrac12 B(\alpha,\beta+1),$
so we can rewrite
\[
\pi^q_{n,0}(\theta|\varnothing)
=
\frac{\tfrac12 B(\alpha,\beta+n)}{\widetilde C_0}\hspace{0.02in}f_{\alpha,\beta+n}(\theta)
+
\frac{\tfrac12 B(\alpha,\beta+1)}{\widetilde C_0}\hspace{0.02in}f_{\alpha,\beta+1}(\theta).
\]
Hence $F^{q}_{n,0}(\cdot)$ is a mixture of $\text{\normalfont Beta}(\alpha,\beta+n)$ and $\text{\normalfont Beta}(\alpha,\beta+1)$ with mixing weight
\[
\tilde{\eta}_0
=
\frac{B(\alpha,\beta+n)}{B(\alpha,\beta+n)+B(\alpha,\beta+1)}.
\]
This coincides with the formula for $\tilde{\eta}_z$ in the statement when $z=0$, since then $\mathds{1}_{\{z\in\{0,1\}\}}=1$ and $nz=0$, $1-z=1$.

For $y=n$ (so $z=1$), we have $\mu^{\hat{\rho}}_n(n|\theta)
=
(1-\hat{\rho})\theta^n+\hat{\rho}\theta,$
and
\begin{align*}
\pi^q_{n,1}(\theta|\varnothing)
&\propto
\theta^{\alpha-1}(1-\theta)^{\beta-1}
\int_0^1\big[(1-\hat{\rho})\theta^n+\hat{\rho}\theta\big]\hspace{0.02in}d\hat{\rho} \\
&=
\theta^{\alpha-1}(1-\theta)^{\beta-1}
\left[
\theta^n\int_0^1(1-\hat{\rho})\hspace{0.02in}d\hat{\rho}
+
\theta\int_0^1\hat{\rho}\hspace{0.02in}d\hat{\rho}
\right] \\
&=
\frac12\hspace{0.02in}\theta^{\alpha+n-1}(1-\theta)^{\beta-1}
+
\frac12\hspace{0.02in}\theta^{\alpha}(1-\theta)^{\beta-1}.
\end{align*}
Define $\widetilde K_0(\theta):=\theta^{\alpha+n-1}(1-\theta)^{\beta-1}$ and $\widetilde K_1(\theta):=\theta^{\alpha}(1-\theta)^{\beta-1}.$
Then, $\pi^q_{n,1}(\theta|\varnothing)\propto \tfrac12 \widetilde K_0(\theta)+\tfrac12 \widetilde K_1(\theta),$
with normalizing constants
$\int_0^1 \widetilde K_0(\theta)\hspace{0.02in}d\theta = B(\alpha+n,\beta)$ and $\int_0^1 \widetilde K_1(\theta)\hspace{0.02in}d\theta = B(\alpha+1,\beta).$
Let $\widetilde f_{\alpha+n,\beta}$ and $\widetilde f_{\alpha+1,\beta}$ be the densities of $\text{\normalfont Beta}(\alpha+n,\beta)$ and $\text{\normalfont Beta}(\alpha+1,\beta)$, respectively. Then, $\widetilde K_0(\theta)=B(\alpha+n,\beta)\hspace{0.02in}\widetilde f_{\alpha+n,\beta}(\theta),$ and $\widetilde K_1(\theta)=B(\alpha+1,\beta)\hspace{0.02in}\widetilde f_{\alpha+1,\beta}(\theta).$
The normalizing constant is $\widetilde C_1
:=
\tfrac12 B(\alpha+n,\beta)+\tfrac12 B(\alpha+1,\beta),$
and
\[
\pi^q_{n,1}(\theta|\varnothing)
=
\frac{\tfrac12 B(\alpha+n,\beta)}{\widetilde C_1}\hspace{0.02in}\widetilde f_{\alpha+n,\beta}(\theta)
+
\frac{\tfrac12 B(\alpha+1,\beta)}{\widetilde C_1}\hspace{0.02in}\widetilde f_{\alpha+1,\beta}(\theta).
\]
Hence $F^{q}_{n,1}(\cdot)$ is a mixture of $\text{\normalfont Beta}(\alpha+n,\beta)$ and $\text{\normalfont Beta}(\alpha+1,\beta)$ with mixing weight
\[
\tilde{\eta}_1
=
\frac{B(\alpha+n,\beta)}{B(\alpha+n,\beta)+B(\alpha+1,\beta)}.
\]
This agrees with the formula for $\tilde{\eta}_z$ in the statement when $z=1$, since then $\mathds{1}_{z\in\{0,1\}}=1$ and $nz=n$, $1-z=0$. Combining the interior and extreme cases, and noticing that $\tilde{\eta}_z=\eta_z(1/2)$ in Observation \ref{thm:beta2} yields the unified expression in the statement.
\end{proof}

\subsection{Proofs of Lemmas \ref{thm:sesigen}--\ref{lem:unique_generic_center}}
\begin{proof}[Proof of Lemma \ref{thm:sesigen}]
    Let $H_n(\theta) := 1 - \theta - \mathcal{B}_n(\theta; c)$ for every $\theta\in[0,1]$. SESI $\theta_n:=\theta_{n,G^{\mathrm{MLE}}}(p_0))$ is the root $H_n(\theta_n) = 0$, and NE $\theta^*:=\theta_{\text{\normalfont\tiny NE}} = 1/2$ solves $1 - \theta^* - c(\theta^*) = 0$.

    \textit{Step 1: Convergence to NE.} 
    Since $G^{\mathrm{MLE}}_{n,z}(\theta)=\mathds{1}_{\theta\geq z}$ for all $(n,z)$ and $\theta$,  \citet[][Observation 2]{stat} implies that SESI converges to NE: $\theta_n \to \theta^*$ as $n \to \infty$.

    \textit{Step 2: Taylor Expansion.}
    Since $c$ is twice continuously differentiable, $H_n(\theta)$ is differentiable. We expand $H_n(\theta_n)$ around $\theta^*$: $0 = H_n(\theta_n) = H_n(\theta^*) + H_n'(\zeta_n) (\theta_n - \theta^*),$
    where $\zeta_n$ lies strictly between $\theta_n$ and $\theta^*$. Since $\theta_n \to \theta^*$, it follows that $\zeta_n \to \theta^*$ as $n\rightarrow\infty$.

    \textit{Step 3: Analysis of $|H_n'(\zeta_n)|$.}
    The derivative is $H_n'(\theta) = -1 - \mathcal{B}_n'(\theta; c)$.
    Since $c$ is increasing, in \citet[][Property 1]{stat} implies that $\mathcal{B}_n(\theta; c)$ is increasing in $\theta$ for all $n$, so $\mathcal{B}_n'(\theta; c) \geq 0$. Consequently, $H_n'(\theta) \leq -1$ for all $\theta \in [0,1]$, ensuring that $|H_n'(\zeta_n)|$ is non-zero and invertible for all $n$. Moreover, since $c$ is continuously differentiable,  \citet[Theorem 7.1.6]{phil} yields
    $\lim_{n \to \infty} \sup_{\theta \in [0,1]} |\mathcal{B}_n'(\theta; c) - c'(\theta)| = 0.$
    Thus, as $n \to \infty$, $H_n'(\zeta_n) \to -1 - c'(\theta^*)$.

    \textit{Step 4: Analysis of $|H_n(\theta^*)|$.}
    We evaluate $H_n(\theta^*)$:
    \[
    H_n(\theta^*) = (1 - \theta^*) - \mathcal{B}_n(\theta^*; c) = c(\theta^*) - \mathcal{B}_n(\theta^*; c).
    \]
    Since $c$ is twice continuously differentiable, \citet[][Theorem 7.1.10]{phil} yields:
    \[
    \lim_{n \to \infty} n \left[ \mathcal{B}_n(\theta^*; c) - c(\theta^*) \right] = \frac{\theta^*(1-\theta^*)}{2} c''(\theta^*).
    \]
    Thus, $|H_n(\theta^*)| = O(1/n)$ and $|\theta_n - \theta^*|=\frac{|H_n(\theta^*)|}{|H_n'(\zeta_n)|} = O(1/n)$, so $|\theta_n - \theta^*|=o(1/\sqrt{n})$.
    \par It is straightforward to verify that $c(\theta)=\theta^2/2+3/8$ satisfies all the assumptions: it is a polynomial, increasing, bounded on $[0,1]$, and its unique fixed point is $1/2$.
\end{proof}

\begin{proof}[Proof of Lemma \ref{lem:unique_generic_center}]
Fix $\epsilon>0$. Define the events
\[
A_n(\epsilon):=\Big\{p\in\Delta_n: d_{TV}\big(\mathscr{L}_n(p),B_n\big)<\epsilon\Big\},
\qquad
D_n(\epsilon):=\Big\{p\in\Delta_n: d_{TV}\big(\mathscr{L}_n(p),F_n\big)<\epsilon\Big\}.
\]
By assumption, $\gamma_n(A_n(\epsilon)^c)\to 0$ and $\gamma_n(D_n(\epsilon)^c)\to 0$, hence
\[
\gamma_n\big(A_n(\epsilon)\cap D_n(\epsilon)\big)
\ge 1-\gamma_n\big(A_n(\epsilon)^c\big)-\gamma_n\big(D_n(\epsilon)^c\big)\to 1.
\]
In particular, for all sufficiently large $n$, the intersection $A_n(\epsilon)\cap D_n(\epsilon)$ is nonempty.
Fix such an $n$ and choose any $p_n\in A_n(\epsilon)\cap D_n(\epsilon)$. Then
\[
d_{TV}\big(\mathscr{L}_n(p_n),B_n\big)<\epsilon
\qquad\text{and}\qquad
d_{TV}\big(\mathscr{L}_n(p_n),F_n\big)<\epsilon.
\]
By the triangle inequality for $d_{TV}$,
\[
d_{TV}(B_n,F_n)\le d_{TV}\big(B_n,\mathscr{L}_n(p_n)\big)+d_{TV}\big(\mathscr{L}_n(p_n),F_n\big)<2\epsilon.
\]
Thus $\limsup_{n\to\infty} d_{TV}(B_n,F_n)\le 2\epsilon$. Since $\epsilon>0$ is arbitrary, $d_{TV}(B_n,F_n)\to 0$.
\end{proof}

\subsection*{Proof of Lemma \ref{lem:symmetric_network}}
\begin{proof}
In a symmetric profile with $\sigma_i=\sigma$ for all $i$, expected degree is $d(\sigma)=(N-1)\sigma^2$. For such a profile, agent $i$'s payoff is
\[
U_i(\sigma,\sigma_{-i})
=B\big(\sigma(N-1)\sigma\big)-D\big(\sigma(N-1)\sigma\big)
=B\big((N-1)\sigma^2\big)-D\big((N-1)\sigma^2\big).
\]
The derivative with respect to $\sigma$ at a symmetric profile is
\[
\frac{\partial U_i}{\partial\sigma}
=2(N-1)\sigma\big(B'(d)-D'(d)\big),
\]
where $d=(N-1)\sigma^2$. For $\sigma>0$, the first-order condition $\partial U_i/\partial\sigma=0$ is equivalent to $B'(d)=D'(d)$, i.e., to \eqref{eq:FOC_network}. By Assumption \ref{ass:BC}, the function $d\mapsto B'(d)-D'(d)$ is continuous, strictly decreasing, positive at $d=0$, and negative for all sufficiently large $d$. Hence there exists a unique $d^*>0$ such that $B'(d^*)=D'(d^*)$. Setting $d^*_N:=d^*$ and solving $d^*_N=(N-1)(\sigma^*_N)^2$ yields a unique symmetric $\sigma^*_N\in(0,1)$.

To see that $(\sigma^*_N,\dots,\sigma^*_N)$ is a NE, note that for fixed $\sigma_{-i}$, $d_i(\sigma_i,\sigma_{-i})$ is strictly increasing in $\sigma_i$, so $U_i$ as a function of $\sigma_i$ is strictly concave by Assumption \ref{ass:BC}. Thus, any interior critical point is the unique best response. At the symmetric profile, $\partial U_i/\partial\sigma_i=0$ with $\sigma_i=\sigma^*_N$, so the profile is a NE. Uniqueness of $d^*$ implies uniqueness of the symmetric equilibrium. Finally, conditional on $\sigma^*_N$, link formation is independent across unordered pairs with probability $\sigma^*_N\sigma^*_N=(\sigma^*_N)^2=d^*_N/(N-1)$, so $\mathbb{G}_N$ is an Erd\H{o}s--R\'enyi graph $K_{N,p_N}$ with $p_N=d^*_N/(N-1)$.
\end{proof}

\subsection*{Proof of Lemma \ref{lem:qs}}
\begin{proof}
Fix $s\ge2$ and suppose all agents other than $j$ mix independently, choosing $\mathrm{K}$ with probability $q_s\in[0,1]$. Then the number of other agents choosing $\mathrm{K}$ is
\[
M_{-j}:=\sum_{\ell\in C\setminus\{j\}}\mathds{1}_{a_\ell=\mathrm{K}}
\sim\text{Bin}(s-1,q_s).
\]
If $j$ chooses $\mathrm{K}$, the total number of $\mathrm{K}$-choosers is $m=M_{-j}+1$, so the expected payoff is
\[
\mathbb{E}\big[u_j(\mathrm{K},a_{-j})\big]
=\mathbb{E}\left[\beta\frac{M_{-j}+1}{s}-c\right]
=\beta\frac{\mathbb{E}[M_{-j}]+1}{s}-c
=\beta\frac{(s-1)q_s+1}{s}-c.
\]
If $j$ chooses $\mathrm{I}$, the total number of $\mathrm{K}$-choosers remains $m=M_{-j}$, so the expected payoff is
\[
\mathbb{E}\big[u_j(\mathrm{I},a_{-j})\big]
=\mathbb{E}\left[\beta\frac{s-M_{-j}}{s}\right]
=\beta\frac{s-\mathbb{E}[M_{-j}]}{s}
=\beta\frac{s-(s-1)q_s}{s}.
\]
In a symmetric mixed equilibrium, $j$ must be indifferent between $\mathrm{K}$ and $\mathrm{I}$, so
\[
\beta\frac{(s-1)q_s+1}{s}-c
=\beta\frac{s-(s-1)q_s}{s}.
\]
Multiplying both sides by $s/\beta$ and rearranging, $(s-1)q_s+1-\frac{cs}{\beta}=s-(s-1)q_s,$
so that $2(s-1)q_s=s-1+\frac{cs}{\beta}.$
Solving for $q_s$ yields \eqref{eq:qs}. This equilibrium is unique because, for fixed $a_{-j}$, $u_j$ is affine in $q_s$ on each action and hence the indifference condition can only hold at a unique $q_s$. Suitable choices of $\beta$ and $c$ (e.g., small $c/\beta$) ensure $q_s\in(0,1)$.
\end{proof}

\subsection*{Proof of Lemma \ref{lem:equiv_prho}}

\begin{proof}
Let $S=(S_1,\dots,S_n)$ denote the random vector of signals generated by the two-step process. Fix $\theta$ and $s$. Conditional on $W_C=0$, the coordinates of $S$ are independent Bernoulli$(\theta)$, so for any vector $x=(x_i)_{i=1}^n\in\{0,1\}^n$ with $k$ ones and $n-k$ zeros,
\[
\Proba(S=x\mid W_C=0,\theta,|C|=s)=\theta^k(1-\theta)^{n-k}.
\]
Conditional on $W_C=1$, all coordinates of $S$ are equal, so
\[
\Proba(S=\m{1}_n\mid W_C=1,\theta,|C|=s)=\theta,\qquad
\Proba(S=\m{0}_n\mid W_C=1,\theta,|C|=s)=1-\theta,
\]
and $\Proba(S=x\mid W_C=1,\theta,|C|=s)=0$ when $x$ has both zeros and ones.

Unconditionally on $W_C$, but conditional on $\theta$ and $|C|=s$, we have for $0<k<n$,
\[
\Proba(S=x\mid\theta,|C|=s)
=(1-k(s))\theta^k(1-\theta)^{n-k}
\]
when $x$ has $k$ ones and $n-k$ zeros. This matches the third line of eq. (\ref{eq:prho}) with $\rho=k(s)$. For $x=\m{1}_n$ we obtain
$\Proba(S=\m{1}_n\mid\theta,|C|=s)
=(1-k(s))\theta^n+k(s)\theta$
and for $x=\m{0}_n$ we obtain
\[
\Proba(S=\m{0}_n\mid\theta,|C|=s)
=(1-k(s))(1-\theta)^n+k(s)(1-\theta),
\]
which match the first and second lines of eq. (\ref{eq:prho}) with $\rho=k(s)$. This proves the claim.
\end{proof}

\subsection*{Proof of Lemma \ref{lem:Y_condSN}}

\begin{proof}
Conditional on $(\theta,S_N=s)$, Lemma \ref{lem:equiv_prho} implies that the joint distribution of $(x_i)_{i=1}^n$ is $p_{k(s)}(\cdot|\theta)$. By construction of the correlated-signal process,
\[
(x_i)_{i=1}^n=
\begin{cases}
(x_i^0)_{i=1}^n & \text{with probability }1-k(s),\\
(x_i^1)_{i=1}^n & \text{with probability }k(s),
\end{cases}
\]
where $(x_i^0)_{i=1}^n$ are independent Bernoulli$(\theta)$ variables and $(x_i^1)_{i=1}^n$ are all equal to a single Bernoulli$(\theta)$ variable. Let
$Y^0:=\sum_{i=1}^n x_i^0$ and $Y^1:=\sum_{i=1}^n x_i^1.$
Then $Y^0\sim\text{Bin}(n,\theta)$ with distribution $\mu^0_n(\cdot|\theta)$, and $Y^1$ takes value $0$ with probability $1-\theta$ and $n$ with probability $\theta$, with distribution $\mu^1_n(\cdot|\theta)$. It follows that
\begin{align*}
\Proba(Y=y\mid\theta,S_N=s)
&=(1-k(s))\Proba(Y^0=y\mid\theta)+k(s)\Proba(Y^1=y\mid\theta)\\
&=(1-k(s))\mu^0_n(y|\theta)+k(s)\mu^1_n(y|\theta),
\end{align*}
for all $y\in\{0,\dots,n\}$. By eq. (\ref{eq:distpositive}), this is exactly $\mu_n^{\rho}(y|\theta)$ with $\rho=k(s)$.
\end{proof}

\subsection*{Proof of Proposition \ref{prop:network_micro}}

\begin{proof}
By law of total probability (for first equality) and Lemma \ref{lem:Y_condSN} (for second equality),
\begin{align*}
\Proba(Y=y\mid\theta, S_N\geq n)
&=\sum_{s\ge n}\frac{\Proba(S_N=s)}{\Proba(S_N\geq n)} \Proba(Y=y\mid\theta,S_N=s)\\
&=\sum_{s\ge n}\frac{\Proba(S_N=s)}{\Proba(S_N\geq n)}\Big[(1-k(s)) \mu^0_n(y|\theta)+k(s) \mu^1_n(y|\theta)\Big]\\
&=\Bigg[\sum_{s\ge n}(1-k(s))\frac{\Proba(S_N=s)}{\Proba(S_N\geq n)}\Bigg]\mu^0_n(y|\theta)
+\Bigg[\sum_{s\ge n}k(s)\frac{\Proba(S_N=s)}{\Proba(S_N\geq n)}\Bigg]\mu^1_n(y|\theta).
\end{align*}
Define $\rho_N$ as in \eqref{eq:rhoN}. Since $\sum_{s\ge n}\frac{\Proba(S_N=s)}{\Proba(S_N\geq n)}=1$, we have
\[
\sum_{s\ge n}(1-k(s))\frac{\Proba(S_N=s)}{\Proba(S_N\geq n)}=1-\rho_N.
\]
Thus, $\Proba(Y=y\mid\theta, S_N\geq n)
=(1-\rho_N)\mu^0_n(y|\theta)+\rho_N\mu^1_n(y|\theta).$
Since $k(s)\in[0,1)$ for all $s$, $\rho_N\in[0,1)$, so we get $\mu_n^{\rho_N}(y|\theta)$ in eq. (\ref{eq:distpositive}).
\end{proof}

\subsection*{Proofs of Propositions \ref{prop:market_eq}--\ref{prop:symmetric_invest} and Lemma \ref{lem:concavity}}
\begin{proof}[Proof of Proposition \ref{prop:market_eq}]
    The first-order condition for coalition $\omega$ yields $x_\omega(P) = B_\omega / P$. Market clearing requires $\sum_\omega x_\omega = 1$, which implies $\sum_\omega (B_\omega/P) = 1$, so $P^* = \sum_\omega B_\omega$. Substituting $P^*$ back into the demand function yields the result.
\end{proof}
\begin{proof}[Proof of Lemma \ref{lem:concavity}]
    Let $D(\alpha_\omega) = A_{-\omega} + a + \alpha_\omega$ denote the total shape parameter of the Dirichlet distribution. The first derivative of the revenue term $g(\alpha_\omega) = (a+\alpha_\omega)/D(\alpha_\omega)$ is $g'(\alpha_\omega) = A_{-\omega}/D(\alpha_\omega)^2 > 0$. The second derivative is $g''(\alpha_\omega) = -2 A_{-\omega} / D(\alpha_\omega)^3 < 0$. Since the cost function is strictly convex, the total payoff $\Pi_\omega$ is strictly concave.
\end{proof}

\begin{proof}[Proof of Proposition \ref{prop:symmetric_invest}]
    The first-order condition for an interior maximum is $\frac{\partial \Pi_\omega}{\partial \alpha_\omega} = 0$. Using the derivative derived in Lemma \ref{lem:concavity}:
    $\pi \frac{A_{-\omega}}{D(\alpha_\omega)^2} - \kappa \alpha_\omega = 0.$
    In a symmetric equilibrium candidate, we set $\alpha_j = \alpha$ for all $j \in \Omega$. Recall that the cardinality of the outcome space is $R=2^n$. Thus, $A_{-\omega} = (R-1)(a+\alpha)$ and $D(\alpha) = R(a+\alpha)$. Substituting these into the FOC yields:
    \[
    \pi \frac{(R-1)(a+\alpha)}{ [R(a+\alpha)]^2 } = \kappa \alpha 
    \quad \iff \quad 
    \kappa  \alpha (a+\alpha) = \frac{\pi(R-1)}{R^2}.
    \]
    The left-hand side is strictly increasing in $\alpha$ (from 0 to $\infty$ for $\alpha \ge 0$), while the right-hand side is a positive constant. By the intermediate value theorem, there exists a unique solution $\alpha^* > 0$. By Lemma \ref{lem:concavity}, this critical point is a global maximum.
\end{proof}


\subsection*{Proof of Observation \ref{thm:dyndist}}
\begin{proof}
    Given the definition of $\varLambda_t$ and the transition matrix $\varSigma$ in eq. (\ref{eq:transmat}), we can derive the distribution of actions at time $t$ as follows. Let $y_t=\sum_{i=1}^nx_{it}$, where $x_{it}\sim\text{Bern}(\theta_t)$, $\text{corr}(x_{it},x_{jt})=\rho_t\in[0,1]$, for any pair $i\neq j$, and no higher-order correlations. When $\varLambda_t=0$, agents' signals $\{x_{it}\}_{i=1}^n$ appear mutually independent at time $t$. Otherwise, when $\varLambda_t=1$, an information cascade forms in the population at time $t$, so with probability $\theta_t$ all agents observe only successes and with probability $1-\theta_t$ they observe only failures. Then, the distribution of $y_t$ can be expanded as follows
    \begin{align*}
        \mu_{\rho_t}(y_t|\theta_t)&=\Proba\big(y_t|\theta_t,\rho_t\big)\\
        &=\Proba(\varLambda_t=0)\Proba\Big(y_t\Big|\theta_t,\rho_t,\varLambda_t=0\Big)+\Proba(\varLambda_t=1)\Proba\Big(y_t\Big|\theta_t,\rho_t,\varLambda_t=1\Big)\\
        &=(1-\rho_t)\Proba\Big(y_t\Big|\theta_t,\rho_t,\varLambda_t=0\Big)+\rho_t\Proba\Big(y_t|\theta_t,\rho_t,\varLambda_t=1\Big)\\
        &=(1-\rho_t)\underbrace{\binom{n}{y_t}\theta_t^{y_t}(1-\theta_t)^{n-y_t}}_{\mu^{0}_n(y_t|\theta_t)}+\rho_t\underbrace{\Big[ \theta_t\mathds{1}_{y_t=n}+(1-\theta_t)\mathds{1}_{y_t=0}\Big]}_{\mu^{1}_n(y_t|\theta_t)},
    \end{align*}
    where the law of total probability is used and recalling that $\varLambda_t\sim\text{Bern}(\rho_t)$. Here, $y_t|\{\varLambda_t=0\}\sim\mu^{0}_n:=\text{Bin}(n,\theta_t)$, and $y_t|\{\varLambda_t=1\}\sim\mu^{1}_n:=\text{modified-Bernoulli}(\theta_t)$, which puts unit mass at either $y_t=n$ or $y_t=0$ at time $t$ with probability $\theta_t$ or $1-\theta_t$, respectively.
\end{proof}

\subsection*{Proof of Proposition \ref{thm:maindynamic}}
\begin{proof}
We prove the convergence of the joint system $(\rho_t, \theta_t)$.

\medskip
\noindent\textit{Step 1: Convergence of $\rho_t$.}
The variable $\rho_t$ represents the probability distribution of the state $\varLambda_t$ at time $t$. Since $\varLambda_t$ is an irreducible and aperiodic finite Markov chain, the standard convergence theorem for Markov chains applies. Specifically, for any initial $\rho_0$,
\[ |\rho_t - \rho^*| \leq K b^t, \]
for some constants $K > 0$ and $b \in (0,1)$, where $\rho^* = \frac{\phi_{\varLambda}}{\phi_{\varLambda}+\varphi_{\varLambda}}$ is the stationary probability of state 1 \citep[e.g., see,][Theorem 4.9]{expo17}. Thus, $\rho_t \to \rho^*$ exponentially fast.

\medskip
\noindent\textit{Step 2: Convergence of $\theta_t$.}
Let $H_t(\theta) := (1-\delta)\theta + \delta(1 - \varPsi_n(\theta, \rho_{t-1}; C_n))$ be the time-dependent update map, and let $H^*(\theta) := (1-\delta)\theta + \delta(1 - \varPsi_n(\theta, \rho^*; C_n))$ be the limit map. The dynamic process is $\theta_t = H_{t-1}(\theta_{t-1})$.

First, we establish that $H^*$ is a contraction mapping for small $\delta$. The function $\varPsi_n(\theta, \rho^*; C_n)$ is continuously differentiable on $[0,1]$ (it is a polynomial in $\theta$). Let $M = \max_{\theta \in [0,1]} |\frac{\partial}{\partial \theta}\varPsi_n(\theta, \rho^*; C_n)|$. The derivative of the limit map is:
\[ \frac{d}{d\theta}H^*(\theta) = (1-\delta) - \delta \frac{\partial}{\partial \theta}\varPsi_n(\theta, \rho^*; C_n). \]
By Lemma \ref{thm:cosesilemma}.1, $\varPsi_n$ is increasing in $\theta$ because $c$ is increasing, so $\frac{\partial}{\partial \theta}\varPsi_n \geq 0$. Thus:
\[ 1 - \delta(1+M) \leq \frac{d}{d\theta}H^*(\theta) \leq 1-\delta. \]
For $H^*$ to be a contraction, it must be that $|\frac{d}{d\theta}H^*(\theta)| < 1$. The upper bound $1-\delta < 1$ holds for all $\delta \in (0,1]$. For the lower bound, we require $1 - \delta(1+M) > -1$, which implies $\delta < \frac{2}{1+M}$. Let $\bar{\delta} = \frac{2}{1+M}$. For any $\delta < \bar{\delta}$, there exists a Lipschitz constant $L \in [0,1)$ such that $|H^*(x) - H^*(y)| \leq L|x-y|$. Since $[0,1]$ is compact, $H^*$ has a unique globally stable fixed point $\theta^{(\rho^*)}$ by the Banach fixed-point theorem.

Next, we bound the distance between the trajectories. Using the triangle inequality:
\begin{align*}
    |\theta_t - \theta^{(\rho^*)}| = |H_{t-1}(\theta_{t-1}) - H^*(\theta^{(\rho^*)})| \leq |H_{t-1}(\theta_{t-1}) - H^*(\theta_{t-1})| + |H^*(\theta_{t-1}) - H^*(\theta^{(\rho^*)})|.
\end{align*}
Consider the first term. Recall $\varPsi_n(\theta, \rho; C_n) = (1-\rho)\mathcal{B}_n + \rho\mathbb{B}_n$. This function is linear in $\rho$. Thus:
\[ |H_{t-1}(\theta) - H^*(\theta)| = \delta \left| \varPsi_n(\theta, \rho_{t-1}) - \varPsi_n(\theta, \rho^*) \right| = \delta |\rho_{t-1} - \rho^*| \cdot |\mathbb{B}_n - \mathcal{B}_n|. \]
Since $\mathcal{B}_n, \mathbb{B}_n \in [0,1]$, $|\mathbb{B}_n - \mathcal{B}_n| \leq 1$. Therefore, $|H_{t-1}(\theta_{t-1}) - H^*(\theta_{t-1})| \leq \delta |\rho_{t-1} - \rho^*|$.
Using the contraction property for the second term:
\[ |\theta_t - \theta^{(\rho^*)}| \leq \delta |\rho_{t-1} - \rho^*| + L |\theta_{t-1} - \theta^{(\rho^*)}|. \]
From Step 1, $|\rho_{t-1} - \rho^*| \leq K b^{t-1}$. Let $e_t = |\theta_t - \theta^{(\rho^*)}|$. We have the recurrence:
\[ e_t \leq L e_{t-1} + \delta K b^{t-1}. \]
Since $L < 1$ and $b < 1$, $e_t$ converges to 0 as $t \to \infty$. Thus, $\theta_t \to \theta^{(\rho^*)}$.
\end{proof}

\subsection*{Proofs of Proposition \ref{prop:multi_exist} and Corollary \ref{cor:multi_coscont}}

\begin{proof}[Proof of Proposition \ref{prop:multi_exist}]
We construct a continuous map on \(\Delta(\mathcal{A})\) whose fixed points are
exactly the multi-action CoSESIs and apply Brouwer's fixed-point theorem.

\medskip\noindent
\emph{Step 1: Aggregate best-response map.}
Fix \(\boldsymbol{\theta}\in\Delta(\mathcal{A})\). When signals are drawn from
\(p_{\boldsymbol{\theta}}\), the probability that the empirical share vector of a
sample equals \(\boldsymbol{z}\in Z_n\) is \(\mu_n(\boldsymbol{z}\mid\boldsymbol{\theta})\) in eq. (\ref{eq:distsum}).

Conditional on the event \(\boldsymbol{z}(x)=\boldsymbol{z}\), the inference procedure
\(G_{n,\boldsymbol{z}}\) induces a distribution over \(\boldsymbol{\theta}\), and hence
an expected payoff from each action \(a\in\mathcal{A}\). Given the payoff primitives
of Section \ref{sec:cosesi} and the type distribution \(\upsilon\), agents best respond
to this estimate. Let \(\sigma_{n,\boldsymbol{z}}(\xi)\in\mathcal{A}\) be a measurable
best-response function mapping each type \(\xi\) to an action when the empirical share
vector is \(\boldsymbol{z}\) (Note: as in the main text, ties can be broken arbitrarily on a
set of types of \(\upsilon\)-measure zero).

Define the induced aggregate action profile given \(\boldsymbol{z}\) by
\[
S_a(\boldsymbol{z})
:=\int_{\Xi}\mathds{1}\{\sigma_{n,\boldsymbol{z}}(\xi)=a\} d\upsilon(\xi),
\qquad
\boldsymbol{S}(\boldsymbol{z}):=\bigl(S_a(\boldsymbol{z})\bigr)_{a\in\mathcal{A}}.
\]
By construction, \(\boldsymbol{S}(\boldsymbol{z})\in\Delta(\mathcal{A})\) for each
\(\boldsymbol{z}\in Z_n\). Moreover, the map
\(\boldsymbol{z}\mapsto\boldsymbol{S}(\boldsymbol{z})\) depends only on the payoff
primitives, \(\upsilon\), and the inference procedure \(G\), but not on the true joint
distributions \(p_{\boldsymbol{\theta}}\). The \emph{unconditional} fraction of agents who choose action \(a\) when the success
probability vector is \(\boldsymbol{\theta}\) is given by
\[
T_a(\boldsymbol{\theta})
:=\sum_{\boldsymbol{z}\in Z_n}
\mu_n(\boldsymbol{z}\mid\boldsymbol{\theta}) S_a(\boldsymbol{z}),
\]
and let $\boldsymbol{T}(\boldsymbol{\theta})
:=\bigl(T_a(\boldsymbol{\theta})\bigr)_{a\in\mathcal{A}}.$ We claim that \(\boldsymbol{T}:\Delta(\mathcal{A})\to\Delta(\mathcal{A})\) is 
continuous. First, for every \(\boldsymbol{\theta}\in\Delta(\mathcal{A})\) and
each \(a\in\mathcal{A}\), $T_a(\boldsymbol{\theta})\geq0$ since $\mu_n(\boldsymbol{z}\mid\boldsymbol{\theta})\geq0$ \text{and} $S_a(\boldsymbol{z})\geq0.$
Moreover,
\begin{align*}
\sum_{a\in\mathcal{A}}T_a(\boldsymbol{\theta})
=\sum_{a\in\mathcal{A}}\sum_{\boldsymbol{z}\in Z_n}
\mu_n(\boldsymbol{z}\mid\boldsymbol{\theta}) S_a(\boldsymbol{z})=\sum_{\boldsymbol{z}\in Z_n}
\mu_n(\boldsymbol{z}\mid\boldsymbol{\theta})
\sum_{a\in\mathcal{A}}S_a(\boldsymbol{z})=\sum_{\boldsymbol{z}\in Z_n}
\mu_n(\boldsymbol{z}\mid\boldsymbol{\theta})=1,
\end{align*}
because \(\boldsymbol{S}(\boldsymbol{z})\in\Delta(\mathcal{A})\) and
\(\mu_n(\cdot\mid\boldsymbol{\theta})\) is a probability distribution on the finite set
\(Z_n\). Hence \(\boldsymbol{T}(\boldsymbol{\theta})\in\Delta(\mathcal{A})\). Second, for each fixed \(\boldsymbol{z}\in Z_n\) and \(a\in\mathcal{A}\), the function
\[
\Delta(\mathcal{A})\ni\boldsymbol{\theta}
\mapsto\mu_n(\boldsymbol{z}\mid\boldsymbol{\theta})\cdot S_a(\boldsymbol{z})
\]
is continuous by Assumption \ref{ass:multi_cont}, because
\(\mu_n(\boldsymbol{z}\mid\boldsymbol{\theta})\) is continuous and \(S_a(\boldsymbol{z})\)
is constant. Since \(Z_n\) is finite, \(T_a(\boldsymbol{\theta})\) is a finite sum of
continuous functions and is therefore continuous. It follows that
\(\boldsymbol{T}:\Delta(\mathcal{A})\to\Delta(\mathcal{A})\) is continuous.

\medskip\noindent
\emph{Step 2: Existence of a fixed point.}
The simplex \(\Delta(\mathcal{A})\) is nonempty, compact, and convex. By Brouwer's
fixed-point theorem, the continuous map \(\boldsymbol{T}:\Delta(\mathcal{A})\to\Delta(\mathcal{A})\) has at least one fixed
point. Let \(\boldsymbol{\theta}^*\in\Delta(\mathcal{A})\) be such that $\boldsymbol{T}(\boldsymbol{\theta}^*)=\boldsymbol{\theta}^*.$

\medskip\noindent
\emph{Step 3: Connection to CoSESI.}
Consider the subjective game in which each agent's \(n\) signals are drawn from the
true joint distribution \(p_{\boldsymbol{\theta}^*}\). When agents follow the statistical
decision rule described above (sampling from \(p_{\boldsymbol{\theta}^*}\), inferring via
\(G\), and best responding), the fraction of agents choosing action \(a\) is exactly
\(T_a(\boldsymbol{\theta}^*)\). Since \(\boldsymbol{\theta}^*\) is a fixed point, 
\[
T_a(\boldsymbol{\theta}^*)=\theta^*_a
\qquad\text{for all }a\in\mathcal{A}.
\]
Thus, the aggregate distribution of actions coincides with the success probability
vector of the true joint distribution, so \(\boldsymbol{\theta}^*\) satisfies the
conditions of Definition \ref{def:multi_CoSESI} and is therefore a multi-action
CoSESI, i.e., \(\boldsymbol{\theta}^*=\boldsymbol{\theta}_{n,G}(p)\).
\end{proof}

\begin{proof}[Proof of Corollary \ref{cor:multi_coscont}]
Fix \(\boldsymbol{z}\in Z_n\). The set of action profiles \(x\in\mathcal{A}^n\) such
that \(\boldsymbol{z}(x)=\boldsymbol{z}\) is finite. By definition,
\[
\mu_n(\boldsymbol{z}\mid\boldsymbol{\theta})
=\sum_{x\in\mathcal{A}^n: \boldsymbol{z}(x)=\boldsymbol{z}}
p_{\boldsymbol{\theta}}(x).
\]
Each summand \(p_{\boldsymbol{\theta}}(x)\) is continuous in \(\boldsymbol{\theta}\) by
hypothesis, so \(\mu_n(\boldsymbol{z}\mid\boldsymbol{\theta})\) is a finite sum of
continuous functions, hence continuous. Thus, Assumption \ref{ass:multi_cont}
holds, and Proposition \ref{prop:multi_exist} implies the existence of a multi-action
CoSESI \(\boldsymbol{\theta}_{n,G}(p)\).
\end{proof}

\subsection*{Proof of Proposition \ref{thm:multinom_approx}}

\begin{proof}[Proof of Proposition \ref{thm:multinom_approx}]
    The proof proceeds in five steps. We first represent the uniform distribution on the simplex using independent exponential random variables, then partition the sample space by count vectors to characterize the induced distribution.

    \textit{Step 1: Uniform distribution via exponentials.}
    Let $R := K^n$ be the total number of possible signal realizations in the sample space $\Omega_n = \mathcal{A}^n$. Index the elements of $\Omega_n$ as $\{\omega_1, \dots, \omega_R\}$.
    Let $Z_1, \dots, Z_R$ be independent exponential random variables, $\text{Exp}(1)$, with joint density $f_Z(z_1, \dots, z_R) = \exp\big(-\sum_{i=1}^R z_i\big)\mathds{1}_{\{z_i>0\}}.$
    Define the random sum $T := \sum_{i=1}^R Z_i$ and the random vector $P := (P_1, \dots, P_R)$ where $P_i := Z_i/T$.

    \begin{lemma}\label{lem:uniform_rep}
        The random vector $P$ is uniformly distributed on $\Delta_n^K$ (i.e., it is distributed according to $\lambda_{n,K}$).
    \end{lemma}
    \begin{proof}
        Consider the change of variables $(z_1, \dots, z_R) \mapsto (t, p_1, \dots, p_R)$ where $z_i = t p_i$, subject to constraints $t>0$, $p_i \ge 0$, and $\sum p_i = 1$. The Jacobian determinant of this transformation is $t^{R-1}$. The joint density of $(T, P)$ is:
        $f_{T,P}(t,p) = \exp(-t) t^{R-1} \mathds{1}_{\{t>0\}} \mathds{1}_{\{p \in \Delta_n^K\}}.$
        This factors as:
        \[
        f_{T,P}(t,p) = \underbrace{\frac{1}{\Gamma(R)} t^{R-1} e^{-t}}_{\text{Gamma}(R,1) \text{ density in } t} \cdot \underbrace{\Gamma(R) \mathds{1}_{\{p \in \Delta_n^K\}}}_{\text{constant density in } p}.
        \]
        Thus, $T$ and $P$ are independent, and the marginal density of $P$ is constant on the simplex $\Delta_n^K$.
    \end{proof}
    Thus, sampling a joint distribution $p$ from $\lambda_{n,K}$ is equivalent to setting $p_{\omega_i} = P_i = Z_i/T$.

    \textit{Step 2: Count vectors.}
    For each sample realization $\omega = (\omega_1, \dots, \omega_n) \in \Omega_n$, we define the \emph{count vector} $c(\omega) \in \mathbb{N}^K$ as:
    \[
    c(\omega) := (N_1(\omega), \dots, N_K(\omega)), \quad \text{where } N_j(\omega) := \#\{1 \le t \le n : \omega_t = j\}.
    \]
    Let $\mathcal{C}_{n,K}$ be the set of all count vectors summing to $n$:
    \[
    \mathcal{C}_{n,K} := \left\{ \boldsymbol{y} = (y_1, \dots, y_K) \in \mathbb{N}^K : \sum_{j=1}^K y_j = n \right\}.
    \]
    For each count vector $\boldsymbol{y} \in \mathcal{C}_{n,K}$, define the subset of sample realizations $\mathcal{G}_{\boldsymbol{y}} \subset \Omega_n$ that generate this specific count:
    $\mathcal{G}_{\boldsymbol{y}} := \{ \omega \in \Omega_n : c(\omega) = \boldsymbol{y} \}.$
    The cardinality of this set is given by the multinomial coefficient:
    $m(\boldsymbol{y}) := |\mathcal{G}_{\boldsymbol{y}}| = \frac{n!}{y_1! y_2! \cdots y_K!}.$
    Since the sets $\{\mathcal{G}_{\boldsymbol{y}}\}_{\boldsymbol{y} \in \mathcal{C}_{n,K}}$ form a partition of the sample space $\Omega_n$, we can define the \textit{block sums} of the exponential variables corresponding to each count vector:
    \[
    Y_{\boldsymbol{y}} := \sum_{\omega \in \mathcal{G}_{\boldsymbol{y}}} Z_\omega, \qquad Q_{\boldsymbol{y}} := \frac{Y_{\boldsymbol{y}}}{T} = \sum_{\omega \in \mathcal{G}_{\boldsymbol{y}}} P_\omega.
    \]
    By definition, the induced probability of observing count vector $\boldsymbol{y}$ under the random distribution $p$ is $\mathscr{L}_n^K(p)(\boldsymbol{y}) = Q_{\boldsymbol{y}}$.

    \begin{lemma}\label{lem:gamma_blocks}
        For each $\boldsymbol{y} \in \mathcal{C}_{n,K}$, $Y_{\boldsymbol{y}} \sim \text{Gamma}(m(\boldsymbol{y}), 1)$. The family of random variables $\{Y_{\boldsymbol{y}}\}_{\boldsymbol{y} \in \mathcal{C}_{n,K}}$ is mutually independent.
    \end{lemma}
    \begin{proof}
        For a fixed $\boldsymbol{y}$, $Y_{\boldsymbol{y}}$ is the sum of $m(\boldsymbol{y})$ independent $\text{Exp}(1)$ variables, which follows a Gamma distribution with shape parameter $m(\boldsymbol{y})$ and scale 1. Since the sets $\mathcal{G}_{\boldsymbol{y}}$ are disjoint, the corresponding collections of $Z_\omega$ are disjoint, which ensures independence across $\boldsymbol{y}$.
    \end{proof}

    \textit{Step 3: Moments of $Q_{\boldsymbol{y}}$.}
    We use the Beta-Gamma algebra to find the moments. Let $\alpha_{\boldsymbol{y}} := m(\boldsymbol{y})$ and let $\alpha_{*} := \sum_{\boldsymbol{y} \in \mathcal{C}_{n,K}} m(\boldsymbol{y}) = |\Omega_n| = K^n$. Note that $T = \sum_{\boldsymbol{y}} Y_{\boldsymbol{y}}$.
    
    \begin{lemma}\label{lem:beta_gamma}
        Let $U \sim \text{Gamma}(\alpha, 1)$ and $V \sim \text{Gamma}(\beta, 1)$ be independent. Then, $R = \frac{U}{U+V}$ follows a $\text{Beta}(\alpha, \beta)$ distribution, and $\mathbb{E}[R] = \frac{\alpha}{\alpha+\beta}$ and $\text{\normalfont var}(R) = \frac{\alpha\beta}{(\alpha+\beta)^2(\alpha+\beta+1)}$.
    \end{lemma}
    \begin{proof}
        Standard change of variables $(u,v) \mapsto (s,r)$ with $s=u+v$ and $r=u/(u+v)$ yields the joint density $f_{S,R}(s,r)$. Factoring shows $S$ and $R$ are independent, with $R \sim \text{Beta}(\alpha,\beta)$.
    \end{proof}

    Apply Lemma \ref{lem:beta_gamma} with $U = Y_{\boldsymbol{y}}$ and $V = T - Y_{\boldsymbol{y}}$. Note that $V \sim \text{Gamma}(\alpha_{*} - \alpha_{\boldsymbol{y}}, 1)$.
    Thus, $Q_{\boldsymbol{y}} = U/(U+V)$ follows a Beta distribution with parameters $\alpha_{\boldsymbol{y}}$ and $\alpha_{*} - \alpha_{\boldsymbol{y}}$.
    The expectation is: $\mathbb{E}[Q_{\boldsymbol{y}}] = \frac{\alpha_{\boldsymbol{y}}}{\alpha_{*}} = \frac{m(\boldsymbol{y})}{K^n} = M_{n,K}(\boldsymbol{y}).$
    The variance satisfies:
    \[
    \text{var}(Q_{\boldsymbol{y}}) = \frac{\alpha_{\boldsymbol{y}}(\alpha_{*}-\alpha_{\boldsymbol{y}})}{\alpha_{*}^2(\alpha_{*}+1)} \le \frac{\alpha_{\boldsymbol{y}}\alpha_{*}}{\alpha_{*}^3} = \frac{\alpha_{\boldsymbol{y}}}{\alpha_{*}^2}.
    \]
    Summing the variances over all possible count vectors $\boldsymbol{y} \in \mathcal{C}_{n,K}$ yields:
    \begin{equation}\label{eq:sum_var}
        \sum_{\boldsymbol{y} \in \mathcal{C}_{n,K}} \text{var}(Q_{\boldsymbol{y}}) \le \frac{1}{\alpha_{*}^2} \sum_{\boldsymbol{y} \in \mathcal{C}_{n,K}} \alpha_{\boldsymbol{y}} = \frac{1}{\alpha_{*}} = \frac{1}{K^n}.
    \end{equation}

    \textit{Step 4: Sup-norm bound (part 1).}
    For any count vector $\boldsymbol{y} \in \mathcal{C}_{n,K}$, Chebyshev's inequality gives:
    \[
    \lambda_{n,K}\Bigl(|Q_{\boldsymbol{y}} - M_{n,K}(\boldsymbol{y})| > \epsilon\Bigr) \le \frac{\text{var}(Q_{\boldsymbol{y}})}{\epsilon^2}.
    \]
    Using the union bound over all $\boldsymbol{y} \in \mathcal{C}_{n,K}$:
    \[
    \lambda_{n,K}\Bigl(\|\mathscr{L}_n^K(p)-M_{n,K}\|_\infty > \epsilon\Bigr) 
    \le \sum_{\boldsymbol{y} \in \mathcal{C}_{n,K}} \lambda_{n,K}\Bigl(|Q_{\boldsymbol{y}} - M_{n,K}(\boldsymbol{y})| > \epsilon\Bigr)
    \le \sum_{\boldsymbol{y} \in \mathcal{C}_{n,K}} \frac{\text{var}(Q_{\boldsymbol{y}})}{\epsilon^2} \le \frac{1}{\epsilon^2 K^n},
    \]
    where the last step uses eq. (\ref{eq:sum_var}). This proves part 1.

    \textit{Step 5: Total Variation bound (part 2).}
    Let $N^* := |\mathcal{C}_{n,K}| = \binom{n+K-1}{K-1}$ be the number of weak compositions. By the Cauchy-Schwarz inequality:
    \[
    \|\mathscr{L}_n^K(p)-M_{n,K}\|_{TV}^2 = \left(\frac{1}{2}\sum_{\boldsymbol{y} \in \mathcal{C}_{n,K}} |Q_{\boldsymbol{y}} - \mathbb{E}[Q_{\boldsymbol{y}}]|\right)^2 \le \frac{1}{4} N^* \sum_{\boldsymbol{y} \in \mathcal{C}_{n,K}} (Q_{\boldsymbol{y}} - \mathbb{E}[Q_{\boldsymbol{y}}])^2.
    \]
    Taking expectations with respect to the random measure $p$:
    \[
    \mathbb{E}\left[\|\mathscr{L}_n^K(p)-M_{n,K}\|_{TV}^2\right] \le \frac{1}{4} N^* \sum_{\boldsymbol{y} \in \mathcal{C}_{n,K}} \text{var}(Q_{\boldsymbol{y}}) \le \frac{1}{4 K^n} \binom{n+K-1}{K-1}.
    \]
    Now apply Markov's inequality to the nonnegative random variable $X := \|\mathscr{L}_n^K(p)-M_{n,K}\|_{TV}^2$:
    \[
    \lambda_{n,K}\Bigl(\|\mathscr{L}_n^K(p)-M_{n,K}\|_{TV} > \epsilon\Bigr) = \mathbb{P}\left(X > \epsilon^2\right) \le \frac{\mathbb{E}[X]}{\epsilon^2} \le \frac{1}{4 \epsilon^2 K^n} \binom{n+K-1}{K-1}.
    \]
    This proves part 2.
\end{proof}

\subsection*{Proof of Proposition \ref{thm:select} and Corollaries \ref{thm:rho1}--\ref{thm:var}}
\begin{proof}
     Since $\varPsi_n(0,\rho;c)=c(0)>0$ and $\varPsi_n(1,\rho;c)=c(1)<1$ for all $n$ and $\rho\in[0,1]$ (Lemma \ref{thm:cosesilemma}.2), the function $\varPsi_n(\theta,\rho;c)$ crosses the $45^{\circ}$ line at least once and thus there always exists at least one simple CoSESI.

    \par For Corollary \ref{thm:rho1}, let $\rho=1$. Then, $\varPsi_n(\theta,1;c)=\mathbb{B}_n(\theta;c)=\theta c(1)+(1-\theta)c(0)$ is the line with slope $c(1)-c(0)\in(0,1)$ and intercept $c(0)>0$, and hence must cross the $45^{\circ}$ line exactly once. Since the $45^{\circ}$ line is steeper than $\varPsi_n(\theta,1;c)$ and has a lower intercept, its lowest intersection with $c(\theta)$ is lower than with $\varPsi_n(\theta,1;c)$, and its second-lowest intersection with $c(\theta)$ is higher than with $\varPsi_n(\theta,1;c)$. Thus, when $\rho=1$, the unique simple CoSESI is located between the smallest and second-largest NE.
    \par When $\rho=0$, simple CoSESI is SESI. \citet[][Theorem 5]{stat} shows that there is at least one and at most three SESIs for any given $n$. Thus, as $\rho$ goes from 0 to 1, the number of simple CoSESIs decreases from (at least one and) at most three to exactly one.
    \par Corollary \ref{thm:var} follows from \citet[][Proposition A1]{stat} because in the $\rho$-weighted Bernstein polynomial $\varPsi_n(\theta,\rho;c)=(1-\rho)\mathcal{B}_n(\theta;c)+\rho\mathbb{B}_n(\theta;c)$, $\mathbb{B}_n(\theta;c)$ is linear and therefore does not affect the curvature of $\varPsi_n$ (Lemma \ref{thm:cosesilemma}.3). Thus, the curvature of $\varPsi_n$ is the same as the curvature of the Bernstein polynomial $\mathcal{B}_n(\theta;c)$ for all $n$ and $\rho<1$.
\end{proof}

\subsection*{Proof of Proposition \ref{thm:assortcoss}}
It will be useful in some proofs to observe that the pure best response correspondence is
\begin{align}\label{eq:bestrep}
    b^{\xi}_G(n,z)=\underset{s\in\{A,B\}}{\text{ arg max }}\int_0^1u_{s}(\xi,\theta) \hspace{0.03in}dG_{n,z}(\theta),
\end{align}
where the utility of actions $A$ and $B$ are $u_{A}(\xi,\theta)=\xi-c(\theta)$ and $u_{B}(\xi,\theta)=0$, respectively. 
\begin{proof}
    We fix a sample size $n$, monotone inference procedure $G$, and any (continuous) matching technology $\rho(\xi)\in[0,1]$, for $\xi\sim\mathcal{U}[0,1]$. Recall the pure best-response correspondence $b^{\xi}_{G}(n,z)$ in eq. (\ref{eq:bestrep}), which satisfies $b^{\xi}_{G}(n,z)\in\{A,B\}$. Then, we define \begin{align}\label{eq:index}
    M^{\xi}_G(z)=\begin{cases}
        1 & \text{ if }b^{\xi}_{G}(n,z)=A\\
        0 & \text{ if }b^{\xi}_{G}(n,z)=B.
    \end{cases}
    \end{align} 
    An assortative CoSESI $\theta^{(\rho)}_{n,G}$ is then the value of $\theta$ that satisfies
    \begin{align}
        \theta&=\int_0^1\sum_{y=0}^n\mu\big(y|\theta,\rho(\xi)\big)M^{\xi}_G(y/n)\hspace{0.03in}d\xi\nonumber=\int_0^1\sum_{y=0}^n\Bigg\{(1-\rho(\xi))\mu_{0}(y|\theta)+\rho(\xi)\mu_{1}(y|\theta)\Bigg\}M^{\xi}_G(y/n)\hspace{0.03in}d\xi\nonumber\\
   &=\underbrace{\sum_{y=0}^n\mu_{0}(y|\theta)\int_0^1M^{\xi}_G(y/n)\hspace{0.03in}d\xi}_{1-\mathcal{B}_n(\theta;C_{n})}+\sum_{y=0}^n\Big[\mu_{1}(y|\theta)-\mu_{0}(y|\theta)\Big]\int_0^1\rho(\xi)M^{\xi}_G(y/n)\hspace{0.03in}d\xi\nonumber\\
   &=1-\mathcal{B}_n(\theta;C_{n})+\underbrace{\sum_{y=0}^n\mu_{1}(y|\theta)\varLambda_{C_{n,y/n}}}_{\mathbb{B}_n(\theta;\varLambda_{C_{n}})}-\underbrace{\sum_{y=0}^n\mu_{0}(y|\theta)\varLambda_{C_{n,y/n}}}_{\mathcal{B}_n(\theta;\varLambda_{C_{n}})}\nonumber\\
   &=1-\mathcal{B}_n(\theta;C_{n})+\mathbb{B}_n(\theta;\varLambda_{C_{n}})-\mathcal{B}_n(\theta;\varLambda_{C_{n}}),\label{eq:assor}
    \end{align}
    where $\int_0^1M^{\xi}_G\big(\frac{y}{n}\big)\hspace{0.03in}d\xi=\int_{C_{n,y/n}}^1d\xi=1-C_{n,y/n}$, and therefore $\varLambda_{C_{n,y/n}}=\int_0^1\rho(\xi)M^{\xi}_G\big(\frac{y}{n}\big)\hspace{0.03in}d\xi=\int_{C_{n,y/n}}^1\rho(\xi)d\xi,$ which follows by definition of $M^{\xi}_G\big(\frac{y}{n}\big)$ in eq. (\ref{eq:index}). There exists a solution to eq. (\ref{eq:assor}) by continuity of both $c(.)$ and $\rho(.)$.
\end{proof}

\phantomsection\label{ex:assor}

\subsection*{Proof of Proposition \ref{prop:ABEE_CoSESI}}

\begin{proof}
Fix an arbitrary state $s \in S$ and its corresponding analogy class $\mathscr{C}_s \in \mathcal{P}$.

\medskip\noindent
\textit{Step 1: CoSESI best responses coincide with ABEE best responses.}

By the construction of the inference procedure $F^{q^s}$, for any sample $(n,z)$, the estimate $F^{q^s}_{n,z}$ is degenerate at $\hat{\theta}(\mathscr{C}_s)$, so $C^{q,s}_{n}
     :=  \int_0^1 c(\theta) dF^{q^s}_{n,z}(\theta)
     =  c\big(\hat{\theta}(\mathscr{C}_s)\big).$
Therefore, in the subjective game $\Gamma^s_n$, a type $\xi$ chooses $A$ if and only if
\[
    \xi + \gamma c(\hat{\theta}(\mathscr{C}_s))  \geq  0
    \quad\Longleftrightarrow\quad
    \xi  \geq  -\gamma c(\hat{\theta}(\mathscr{C}_s)).
\]
This is exactly the ABEE best-response condition \eqref{eq:ABEE_BR} in state $s$:
\[
    \sigma(s,\xi) = A
    \quad\Longleftrightarrow\quad
    \xi  \geq  -\gamma c(\hat{\theta}(\mathscr{C}_s)).
\]
Thus, if agents in $\Gamma^s_n$ follow the CoSESI decision rule, then for every type $\xi$ their action in state $s$ coincides with $\sigma(s,\xi)$.

\medskip\noindent
\textit{Step 2: Aggregates coincide.}

Since the CoSESI decision rule in $\Gamma^s_n$ yields the same action profile as $\sigma(s,\cdot)$, the aggregate fraction of agents taking $A$ in state $s$ in $\Gamma^s_n$ is
\begin{align*}
    \theta^{\text{CoSESI}}(s)
    &:= \int_{\Xi} \mathds{1}\big\{\text{CoSESI rule chooses $A$ in state $s$ for type $\xi$}\big\}   d\upsilon(\xi) \\
    &= \int_{\Xi} \mathds{1}\{\sigma(s,\xi) = A\}   d\upsilon(\xi) = \theta(s).
\end{align*}

\medskip\noindent
\textit{Step 3: CoSESI fixed point condition.}

By construction of $p^s$, the success probability of the Bernoulli signals in $\Gamma^s_n$ is
\[
    \mathbb{P}_{p^s}(x_i = 1) = \theta(s), \quad \forall i.
\]
In a CoSESI, the success probability of the joint distribution $p^s$ from which agents sample equals the fraction of agents taking $A$ under the induced decision rule.

\par From Step 2 we know that, under the CoSESI decision rule constructed in Step 1, the fraction of agents taking $A$ is $\theta(s)$. Since the success probability of $p^s$ is also $\theta(s)$, the number $\theta(s)$ satisfies the CoSESI fixed point condition in Definition \ref{def:CoSESI}. Therefore, $\theta_{n,F^{q^s}}(p^s)$ exists and must equal $\theta(s)$.

\par Since $s \in S$ was arbitrary, the same conclusion holds for each state. Hence, the ABEE profile $(\sigma,\hat{\theta})$ is represented as a family of CoSESI $\{\theta_{n,F^{q^s}}(p^s)\}_{s \in S}$ that reproduces exactly the same aggregate behavior and analogy-based beliefs in each state.
\end{proof}
\end{document}